\documentclass[11pt]{article}
\usepackage{graphicx}
\usepackage{amssymb}
\usepackage{amscd}
\usepackage{mathrsfs}
\usepackage{longtable,lscape}
\usepackage{amsthm}
\usepackage{amsfonts}
\usepackage{amsmath}
\usepackage{bbm}
\usepackage{float}
\usepackage{url}

\newcommand{\compl}{{\mathbb C}}
\newcommand{\real}{{\mathbb R}}
\newcommand{\captionfonts}{\footnotesize}
\makeatletter  % Allow the use of @ in command names
\long\def\@makecaption#1#2{%
  \vskip\abovecaptionskip
  \sbox\@tempboxa{{\captionfonts #1: #2}}%
  \ifdim \wd\@tempboxa >\hsize
    {\captionfonts #1: #2\par}
  \else
    \hbox to\hsize{\hfil\box\@tempboxa\hfil}%
  \fi
  \vskip\belowcaptionskip}
\makeatother 
\begin{document}
\title{The Extended Bloch Representation of Quantum Mechanics. Explaining Superposition, Interference and Entanglement}
\author{Diederik Aerts$^1$ and Massimiliano Sassoli de Bianchi$^{2}$ \vspace{0.5 cm} \\ 
        \normalsize\itshape
        $^1\!\,$Center Leo Apostel for Interdisciplinary Studies \\
        \normalsize\itshape
        Brussels Free University, 1050 Brussels, Belgium \\ 
         \normalsize
        E-Mail: \url{diraerts@vub.ac.be}
          %\vspace{0.5 cm} 
          \\ 
        \normalsize\itshape
        $^2\!\,$Laboratorio di Autoricerca di Base \\ 
        \normalsize\itshape
         6914 Lugano, Switzerland \\
        \normalsize
        E-Mail: \url{autoricerca@gmail.com} \\
              }
\date{}
\maketitle

\begin{abstract}
\noindent 
The extended Bloch representation of quantum mechanics was recently derived to offer a (hidden-measurement) solution to the measurement problem. In this article we use it to investigate the geometry of superposition and entangled states, explaining the interference effects, and the entanglement correlations, in terms of the different orientations that a state-vector can take within the generalized Bloch sphere. We also introduce a tensorial determination of the generators of $SU(N)$, particularly suitable to describe multipartite systems, from the viewpoint of the sub-entities. We then use it to show that non-product states admit a general description in which the sub-entities can always remain in well-defined states, even when they are entangled. Therefore, the completed version of quantum mechanics provided by the extended Bloch representation, in which the density operators are also representative of pure states, allows to solve not only the well-known measurement problem, but also the lesser-known entanglement problem. This because we no longer need to give up the general physical principle saying that a composite entity exists, and therefore is in a pure state, if and only if its components also exist, and therefore are in well-defined pure states. 
\end{abstract}
\medskip
{\bf Keywords}: Extended Bloch representation, Hidden-measurement interpretation, Superposition, Interference effects, Entanglement, Correlations, Bell's inequalities, $SU(N)$, Non-spatiality.

\section{Introduction}
\label{Introduction}

The theory of the so-called \emph{Poincar\'e sphere} dates back to 1892, when the French physicist and mathematician Henri Poincar\'e discovered that a surprisingly simple representation of the polarization states of electromagnetic radiation could be obtained by representing the polarization ellipse on a complex plane, and then further projecting such plane onto a sphere~\cite{Poincare1892}. In 1946, this representation was adapted by the Swiss physicist Felix Bloch to represent the states of two-level quantum systems, like spin-${1\over 2}$ entities, in what is today known as the \emph{Bloch sphere}~\cite{Bloch1946}.

Forty years later, one of us showed that the Bloch sphere could be used not only as a valuable tool for visualizing the states of a two-level (qubit) system, but also for modeling its measurements, providing a possible explanation for the origin of the Born rule~\cite{Aerts1986,Aerts1987}. This preliminary 1986 study generated over the years a number of works, further exploring the explicative power contained in this modelization, in what is today known as the \emph{hidden-measurement interpretation} of quantum mechanics (see~\cite{AertsSassoli2014c} and the references cited therein). According to this interpretation, a quantum measurement is an experimental context characterized by a situation of lack of knowledge not about the state of the measured entity, but about the measurement-interaction between the apparatus and the entity.

For some time this hidden-measurement extension of the Bloch sphere, called the \emph{$\epsilon$-model}, or the \emph{sphere model}~\cite{{Aerts1998b,Aerts1999,AertsSassoli2014c}}, was only taken into consideration by a small number of physicists working on the foundations of physical theories, whereas in general it was mostly considered as a mathematical curiosity. This in particular because of the existence of the so-called no-go theorems, like those of Gleason~\cite{Gleason1957} and Kochen-Specker~\cite{Kochen1967}, which were known to be  valid only for $N$-dimensional Hilbert spaces with $N>2$, hence not for the special case of two-level systems, which in that sense were considered to be pathological. But this was a misconception, as the hidden-measurement interpretation had little to do with a classical hidden-variable theory, considering that the lack of knowledge was not associated with the state of the measured entity, but with its measurement-interactions, so that the no-go theorems did not apply. 

Also, contrary to the initial prejudice that no general hidden-measurement modelization could be given for dimensions $N>2$, new results became available over the years, showing that the hidden-measurement mechanism was by no means restricted to two-dimensional situations~\cite{Aertsetal1997, Coecke1995aa, Coecke1995bb}. However, these promising results were not totally convincing, as what was still lacking was a natural generalization of the Bloch sphere representation, beyond the two-dimensional situation. Things changed in more recent times, when this much-sought generalized Bloch representation was obtained, using the properties of the so-called generators of $SU(N)$, the special unitary group of degree $N$~\cite{Arvind1997, Kimura2003, Byrd2003, Kimura2005, Bengtsson2006, Bengtsson2013}. Thanks to these results, the representation could recently be further extended, to also include the description of the measurements, providing what we believe is a general and convincing solution to the measurement problem~\cite{AertsSassoli2014c}.

The present article can be considered to be the continuation of the investigation that we have started in~\cite{AertsSassoli2014c}, further exploring the explanatory power contained in the extended Bloch model. It consists in three logically distinct parts. In the first part (Sec.~\ref{General measurement}), we provide an introduction to the essentials of the model, explaining how the notion of hidden-measurement can be used to derive, in a non-circular way, the Born rule. We will not give all the mathematical details of the derivation, but only those results and ideas that are necessary to understand the following of our analysis (we refer the reader to~\cite{AertsSassoli2014c} for a more general and systematic presentation; see also~\cite{AertsSassoli2014d,AertsSassoli2014e}). 

In the second part (Sections~\ref{Superposition states}, \ref{Entanglement} and~\ref{More general superpositions}), we use the model to analyze the origin of the interference effects, when measurements are performed on superposition states. More specifically, we show that superposition states can be represented as states moving on specific circles within the Bloch sphere, and that these circular movements explain the effects of overextension and underextension with respect to the classical probabilities, produced by the interference phenomena. We also investigate the situation where more than two (orthogonal) states are superposed, showing that the complexity of the representation rapidly increases, because of the interplay between the different phase factors. Entangled states, which are a special class of superposition states, are also considered. The extended Bloch model can then be used to illustrate the process of \emph{creation of correlations}  taking place when ``coincidence counts'' measurements are carried out, on  joint entities in non-product states. A simple macroscopic simulation of these measurements, using a breakable elastic band, is also proposed. 

In the third part of the article (Sections~\ref{Multipartite systems} and~\ref{Completed quantum mechanics}), we introduce a new basis of generators of $SU(N)$, constructed using the properties of the tensor product. This will allow us to characterize, directly within the extended Bloch representation, product, separable and entangled states, and study the unfolding of the measurement processes also from the viewpoint of the sub-entities. In particular, we will show that coincidence spin measurements on singlet states are equivalent to sequential spin measurements performed on the two individual sub-entities, when they are connected through a ``rigid-rod mechanism,'' which is broken during the measurement, creating in this way the correlations. More importantly, our general description will allow us to offer a solution to an old (often overlooked)  problem about entanglement, which we will formulate in a clear way in Sec.~\ref{Completed quantum mechanics}, by showing that the existence of a joint entangled entity is not incompatible with the existence of its sub-systems, which although entangled can nevertheless always remain in well defined states. Finally, in Sec.~\ref{Conclusion}, we conclude by summarizing the obtained results.

\section{Modeling a quantum measurement}
\label{General measurement}

In this section we explain how a general $N$-outcome measurement can be described, and explained, in the extended Bloch model. We consider an entity whose Hilbert state is ${\cal H}_N=\compl^N$. Its states are described by  positive semidefinite Hermitian operators $D$, of unit trace, i.e., operators such that: $D^\dagger = D$, ${\rm Tr}\, D=1$, and  $\langle\psi |D|\psi\rangle \geq 0$, $\forall |\psi\rangle\in {\cal H}_N$. When the state is of the form $D=|\psi\rangle\langle\psi|$, i.e., is a rank-one projection operator, we also have $D=D^2$, and therefore ${\rm Tr}\, D^2=1$. In standard quantum mechanics, we then say that the operator describes a pure state. On the other hand, when $D$ is not idempotent, ${\rm Tr}\, D^2 \leq 1$ (the minimum possible value being ${1\over N}$), and according to the standard interpretation $D$ describes a  classical mixture of states (and is usually called a \emph{density matrix}), in accordance with the fact that $D$ can always be written as a convex linear combination of orthogonal projection operators. 

However, this interpretation is not without difficulty, considering that a same density operator can have infinitely many representations  as  mixtures of one-dimensional projection operators~\cite{Hughston1993}, and that, as we will show in the following, density matrices also admit an interpretation as pure states. More precisely, they correspond to those states that are necessary to complete the quantum formalism in order to also obtain a description of the measurement processes. Therefore, in the present work we shall not a priori distinguish, in an ontological sense, states described by one-dimensional orthogonal projection operators, which we will call \emph{vector-states}, from density operators, i.e., convex linear combinations of vector-states, which  we will call \emph{operator-states} (a vector-state being of course a special case of an operator-state).  

The genralized Bloch representation (which we will extend to also include the measurements) is based on the observation that one can find a basis for the linear operators acting in $\compl^N$, made of $N^2$ orthogonal (in the Hilbert-Schmidt sense) operators. One of them is the identity matrix ${\mathbb I}$, and the other $N^2-1$ ones correspond to a determination of the generators of $SU(N)$, the \emph{special unitary group of degree $N$}, which are traceless and orthogonal self-adjoint matrices $\Lambda_i$, $i=1,\dots, N^2-1$. More precisely, choosing the normalization ${\rm Tr}\, \Lambda_i\Lambda_j=2\delta_{ij}$, we can always express an operator-state $D$ in a given basis $\{{\mathbb I},\Lambda_1,\dots,\Lambda_{N^2-1}\}$, by writing: 
\begin{equation}
D({\bf r}) = {1\over N}\left(\mathbb{I} +c_N\, {\bf r}\cdot\mbox{\boldmath$\Lambda$}\right) = {1\over N}\left(\mathbb{I} + c_N\sum_{i=1}^{N^2-1} r_i \Lambda_i\right),
\label{formulaNxN}
\end{equation}
where for convenience we have introduced the dimensional constant $c_N\equiv \sqrt{N(N-1)\over 2}$. 

In other terms, we can always establish a correspondence between operator-states $D\equiv D({\bf r})$ and $(N^2-1)$-dimensional real vectors  ${\bf r}=(r_1,\dots, r_{N^2-1})^\top$. Also, taking the trace of $D^2({\bf r}) $, and using the orthogonality of the $\Lambda_i$, after a simple calculation one finds that ${\rm Tr}\, D^2({\bf r})={1\over N} + (1-{1\over N})\|{\bf r}\|^2$, and since an operator-state generally obeys ${1\over N}\leq {\rm Tr}\, D^2({\bf r})\leq 1$, it follows that $0\leq \|{\bf r}\|^2\leq 1$, i.e., the representative vectors ${\bf r}$ belongs to a $(N^2-1)$-dimensional unit ball $B_1(\real^{N^2-1})$. Clearly, $\|{\bf r}\|=1$ iff $D^2({\bf r})=D({\bf r})$, i.e., iff $D({\bf r})$ is a one-dimensional projection operator (a pure state, in the standard quantum terminology). 

The main advantage of representing states as real vectors in the unit ball $B_1(\real^{N^2-1})$, is that this will allow us to obtain a description also of the quantum measurements, by representing them, within $B_1(\real^{N^2-1})$, as $(N-1)$-simplexes having some very special properties. But before explaining how all this works, let us introduce a convenient determination of the $SU(N)$ generators, which is the following~\cite{Hioe1981, Alicki1987, Mahler1995}: $\{\Lambda_i\}_{i=1}^{N^2-1}=\{U_{jk},V_{jk},W_{l}\}$, with:
\begin{eqnarray}
\label{rgeneratorsN}
&&U_{jk}=|b_j\rangle\langle b_k| + |b_k\rangle\langle b_j|, \quad V_{jk}=-i(|b_j\rangle\langle b_k| - |b_k\rangle\langle b_j|),\\
&&W_l =\sqrt{2\over l(l+1)}\left(\sum_{j=1}^l |b_j\rangle\langle b_j|-l|b_{l+1}\rangle\langle b_{l+1}|\right),\label{generatorsW}\label{W}\\
&&1\leq j < k\leq N, \quad 1\leq l\leq N-1,
\end{eqnarray}
where $\{|b_1\rangle,\dots,|b_N\rangle\}$ is an arbitrary orthonormal basis of $\compl^N$. 

It is instructive to write down explicitly the generators for the simplest situations. The $N=2$ case corresponds to the standard three-dimensional Bloch representation, with the generators corresponding to the well-known Pauli matrices (usually denoted $\sigma_1$, $\sigma_2$ and $\sigma_3$):
 \begin{equation}
\Lambda_1=
\left[ \begin{array}{cc}
0 & 1 \\
1 & 0 \end{array} \right],\quad
\Lambda_2=
\left[\begin{array}{cc}
0 & -i \\
i & 0 \end{array} \right],\quad 
\Lambda_3=
\left[ \begin{array}{cc}
1 & 0 \\
0 & -1 \end{array} \right].
\end{equation}
 For the $N=3$ case, the generators correspond to the so-called \emph{Gell-Mann} matrices (usually denoted $\lambda_i$, $i=1,\dots,8$):
\begin{eqnarray}
&&\Lambda_1 = \begin{bmatrix} 0 & 1 & 0 \\ 1 & 0 & 0 \\ 0 & 0 & 0 \end{bmatrix},\quad \Lambda_2 = \begin{bmatrix} 0 & -i & 0 \\ i & 0 & 0 \\ 0 & 0 & 0 \end{bmatrix},\quad
\Lambda_3 =\begin{bmatrix} 1 & 0 & 0 \\ 0 & -1 & 0 \\ 0 & 0 & 0 \end{bmatrix},\quad \Lambda_4 = \begin{bmatrix} 0 & 0 & 1 \\ 0 & 0 & 0 \\ 1 & 0 & 0 \end{bmatrix},\nonumber \\
&&\Lambda_5 = \begin{bmatrix} 0 & 0 & -i \\ 0 & 0 & 0 \\ i & 0 & 0 \end{bmatrix},\quad
\Lambda_6 = \begin{bmatrix} 0 & 0 & 0 \\ 0 & 0 & 1 \\ 0 & 1 & 0 \end{bmatrix},\quad \Lambda_7 = \begin{bmatrix} 0 & 0 & 0 \\ 0 & 0 & -i \\ 0 & i & 0 \end{bmatrix},\quad
\Lambda_8 = \frac{1}{\sqrt{3}} \begin{bmatrix} 1 & 0 & 0 \\ 0 & 1 & 0 \\ 0 & 0 & -2 \end{bmatrix}.
\label{generatorsN=3}
\end{eqnarray}
The $N=2$ and $N=3$ describe qubit and qutrit systems, like spin-${1\over 2}$ and spin-$1$ entities, respectively. Since in Sec.~\ref{tensorial basis} we will introduce an additional representation of the generators, particularly suitable for the description of composite systems, let us also give, for later comparison, their explicit form of the $N=4$ case: 
\begin{eqnarray}
&&\Lambda_1 = \begin{bmatrix} 0 & 1 & 0 &0  \\ 1 & 0 & 0 & 0 \\ 0 & 0 & 0 & 0 \\0 & 0 & 0 & 0 \end{bmatrix},\quad \Lambda_2 = \begin{bmatrix} 0 & -i & 0 &0  \\ i & 0 & 0 & 0 \\ 0 & 0 & 0 & 0 \\0 & 0 & 0 & 0 \end{bmatrix}, \quad \Lambda_3 = \begin{bmatrix} 1 & 0 & 0 &0  \\ 0 & -1 & 0 & 0 \\ 0 & 0 & 0 & 0 \\0 & 0 & 0 & 0 \end{bmatrix},\nonumber\\
&&\Lambda_4 = \begin{bmatrix} 0 & 0 & 1 &0  \\ 0 & 0 & 0 & 0 \\ 1 & 0 & 0 & 0 \\0 & 0 & 0 & 0 \end{bmatrix},\quad  \Lambda_5 = \begin{bmatrix} 0 & 0 & -i &0  \\ 0 & 0 & 0 & 0 \\ i & 0 & 0 & 0 \\0 & 0 & 0 & 0 \end{bmatrix},\quad \Lambda_6 = \begin{bmatrix} 0 & 0 & 0 &0  \\ 0 & 0 & 1 & 0 \\ 0 & 1 & 0 & 0 \\0 & 0 & 0 & 0 \end{bmatrix},\nonumber
\end{eqnarray}
 \vspace{-0.75cm}
\begin{eqnarray}
&&\Lambda_7 = \begin{bmatrix} 0 & 0 & 0 &0  \\ 0 & 0 & -i & 0 \\ 0 & i & 0 & 0 \\0 & 0 & 0 & 0 \end{bmatrix},\quad \Lambda_8 = {1\over\sqrt{3}}\begin{bmatrix} 1 & 0 & 0 &0  \\ 0 & 1 & 0 & 0 \\ 0 & 0 & -2 & 0 \\0 & 0 & 0 & 0 \end{bmatrix},\quad \Lambda_9 = \begin{bmatrix} 0 & 0 & 0 & 1  \\ 0 & 0 & 0 & 0 \\ 0 & 0 & 0 & 0 \\1 & 0 & 0 & 0 \end{bmatrix},\nonumber\\
&&\Lambda_{10} = \begin{bmatrix} 0 & 0 & 0 & -i  \\ 0 & 0 & 0 & 0 \\ 0 & 0 & 0 & 0 \\i & 0 & 0 & 0 \end{bmatrix},\quad \Lambda_{11} = \begin{bmatrix} 0 & 0 & 0 &0  \\ 0 & 0 & 0 & 1 \\ 0 & 0 & 0 & 0 \\0 & 1 & 0 & 0 \end{bmatrix},\quad \Lambda_{12} = \begin{bmatrix} 0 & 0 & 0 &0  \\ 0 & 0 & 0 & -i \\ 0 & 0 & 0 & 0 \\0 & i & 0 & 0 \end{bmatrix},\nonumber
\end{eqnarray}
 \vspace{-0.75cm}
\begin{eqnarray}
&&\Lambda_{13} = \begin{bmatrix} 0 & 0 & 0 &0  \\ 0 & 0 & 0 & 0 \\ 0 & 0 & 0 & 1 \\0 & 0 & 1 & 0 \end{bmatrix},\quad F_{14} = \begin{bmatrix} 0 & 0 & 0 &0  \\ 0 & 0 & 0 & 0 \\ 0 & 0 & 0 & -i \\0 & 0 & i & 0 \end{bmatrix},\quad  F_{15} = {1\over\sqrt{6}}\begin{bmatrix} 1 & 0 & 0 &0  \\ 0 & 1 & 0 & 0 \\ 0 & 0 & 1 & 0 \\0 & 0 & 0 & -3 \end{bmatrix}
\label{F15}.
\end{eqnarray} 

We can immediately see that since the factor $c_N$ appearing in (\ref{formulaNxN}) has values $c_2=1$, $c_3=\sqrt{3}$, and $c_4= \sqrt{12}$,  and considering that some of the generators have negative eigenvalues, when they are multiplied by $c_N$, apart the special $N=2$ case, they can become strictly less than $-1$, so that the identity matrix in (\ref{formulaNxN}) will not be able  to compensate them and ensure the positiveness of $D({\bf r})$. More precisely, considering for instance the vector ${\bf r}=(0,0,0,0,0,0,0,1)^\top$ in the $N=3$ case, the associated operator is: $D({\bf r}) ={1\over 3}(\mathbb{I} +\sqrt{3}\Lambda_8)$, and one of its eigenvalues is $-{1\over 3}$. This means that not all vectors ${\bf r}$ within the Bloch ball $B_1(\real^{N^2-1})$ are representative of bona fide states, if $N>2$, as not all vectors can be associated with a positive semidefinite $D({\bf r})$. 

This is the main and important difference between the standard Bloch representation, where the three-dimensional Bloch sphere is filled with states, and its $N>2$ generalization, were the $(N^2-1)$-dimensional sphere is only partially filled with states. But although the shape of the region containing the states is rather complex, and depends on the chosen determination of the generators $\Lambda_i$, it has the nice property of being a closed convex region. This follows from the well-known fact that a convex linear combination of operator-states is again an operator-state. Thus, if ${\bf r} = a\, {\bf s} + b\, {\bf t}$, with $a+b=1$, $a, b\geq0$, and ${\bf s}$ and ${\bf t}$ are representative of two operator-states, from (\ref{formulaNxN}) we immediately obtain that $D({\bf r})=aD({\bf s})+bD({\bf t})$, being a convex linear combination of operator-states is again an operator-state. Thus, a vector ${\bf r}$ which is a convex linear combination of two vectors ${\bf s}$ and ${\bf t}$, both representative of good states, is also representative of a good state $D({\bf r})$. 

To explain how the Born rule can be derived by means of a hidden-measurement mechanism, let us first show how the rule is formulated within the Blochean representation. The standard formulation of the Born rule is that, given an operator-state $D({\bf r})$, and an arbitrary observable $A=\sum_{i=1}^N a_i P_{a_i} = \sum_{i=1}^N a_i  |a_i\rangle\langle a_i|$ (which for simplicity is assumed to be non-degenerate), then the probability of observing the eigenvalue $a_i$ corresponds to the probability of the transition $D({\bf r})\to P_{a_i}$, and is given by the trace: ${\cal P}(D({\bf r})\to P_{a_i}) = {\rm Tr}\, D({\bf r}) P_{a_i}$. Let us denote ${\bf n}_i$ the unit vectors representative of the vector-eigenstates $P_{a_i}\equiv P({\bf n}_i)$, $i=1,\dots,N$. We then have: 
\begin{eqnarray}
\lefteqn{{\cal P}(D({\bf r}) \to P_{a_i}) = {\rm Tr}\, D({\bf r}) P({\bf n}_i) = {\rm Tr}\, {1\over N^2}\left(\mathbb{I} +c_N\, {\bf r}\cdot\mbox{\boldmath$\Lambda$}\right)\left(\mathbb{I} + c_N\,{\bf n}_i\cdot\mbox{\boldmath$\Lambda$}\right)}\nonumber\\
&=& {\rm Tr}\, {1\over N^2}\left[\mathbb{I} + c_N\,({\bf r}\cdot\mbox{\boldmath$\Lambda$}+{\bf n}_i\cdot\mbox{\boldmath$\Lambda$}) +c_N^2\,({\bf r}\cdot\mbox{\boldmath$\Lambda$})({\bf n}_i\cdot\mbox{\boldmath$\Lambda$})\right] = {1\over N} +{c_N^2\over N^2}\,{\rm Tr}\,({\bf r} \cdot\mbox{\boldmath$\Lambda$})\, ({\bf n}_i\cdot\mbox{\boldmath$\Lambda$})\nonumber\\
&=& {1\over N} \left[1+ (N-1)\,{\bf r}\cdot {\bf n}_i\right]={1\over N} \left[1+ (N-1)\, r \cos\theta\right],
\label{transitiongeneralNxN}
\end{eqnarray}
where $\theta\equiv \theta({\bf r},{\bf n}_i)$ is the angle between ${\bf r}$ and ${\bf n}_i$, and of course $r\equiv \| {\bf r}\|=1$ if $D({\bf r})$ is a vector-state.

If the initial state is the eigenstate $P_{a_j}$, i.e., ${\bf r} = {\bf n}_j$, we know that ${\cal P}(P_{a_j} \to P_{a_i}) = \delta_{ji}$, and it follows from (\ref{transitiongeneralNxN}) that $\cos\theta({\bf n}_j,{\bf n}_i)=-{1\over N-1}$, that is: $\theta({\bf n}_j,{\bf n}_i) = \theta_N\equiv \cos^{-1} (-{1\over N-1})$, for all $i\neq j$. This means that the $N$ unit vectors ${\bf n}_i$, representative of the eigenstates $P_{a_i}$, $i=1\dots,N$, are the vertices of a $(N-1)$-dimensional \emph{simplex} $\triangle_{N-1}$, inscribed in the unit ball, with edges of length $\|{\bf n}_i -{\bf n}_j\| = \sqrt{2(1-\cos\theta_N)} = \sqrt{2N\over N-1}$. Also, considering that $\triangle_{N-1}$ is a convex set of vectors, it immediately follows that all its points are representative of operator-states, in accordance with the fact that the states in $B_1(\mathbb{R}^{N^2-1})$ form a closed convex subset. For $N=2$, $\theta_1 = \pi$, and $\triangle_{1}$ is a line segment of length $2$, inscribed in a $3$-dimensional ball. For $N=3$, $\theta_2 = {\pi\over 3}$, and $\triangle_{2}$ is an equilateral triangle of area ${3\sqrt{3}\over 4}$. For $N=4$, $\theta_3 \approx 0.6\,\pi$, and $\triangle_{3}$ is a tetrahedron of volume ${1\over 3}({4\over 3})^{3\over 2}$. For $N=5$, $\theta_4 \approx 0.58\,\pi$, and $\triangle_{4}$ is a pentachoron; and so on. 

We can write (\ref{transitiongeneralNxN}) in a more compact form by expressing ${\bf r}$ as the sum ${\bf r} = {\bf r}^\perp + {\bf r}^\parallel$, where ${\bf r}^\parallel$ is the vector obtained by orthogonally projecting ${\bf r}$ onto $\triangle_{N-1}$. Since by definition ${\bf r}^\perp\cdot {\bf n}_i =0$, for all $i=1,\dots, N$, (\ref{transitiongeneralNxN}) becomes ${\cal P}(D({\bf r}) \to P_{a_i}) = {1\over N} \left[1+ (N-1)\,{\bf r}^\parallel\cdot {\bf n}_i\right]$. Also, as ${\bf r}^\parallel\in\triangle_{N-1}$, by definition of a simplex it can be uniquely written as a convex linear combination of the $N$ vertex vectors ${\bf n}_i$:
\begin{equation}
{\bf r}^\parallel =\sum_{i=1}^{N} r^\parallel_i \,{\bf n}_i, \quad \sum_{i=1}^{N} r^\parallel_i=1, \quad r^\parallel_i\geq 0, \,\, i=1\dots,N.
\label{n-parallelexpansion}
\end{equation}
Since ${\bf n}_i \cdot {\bf n}_j= -{1\over N-1}$, for $i\neq j$, we have: ${\bf r}^\parallel \cdot {\bf n}_i ={1\over N-1}(Nr^\parallel_i -1)$, and (\ref{transitiongeneralNxN}) becomes:
\begin{equation}
{\cal P}(D({\bf r}) \to P_{a_i}) = r^\parallel_i.
\label{trans-general}
\end{equation}

Having expressed the Born rule in terms of the geometry of the vectors representative of the states in the generalized Bloch sphere, we now want to use the model to derive the rule in a non-circular way, obtaining in this way a solution to the measurement problem, in the sense of providing a consistent mechanism explaining the emergence, both qualitatively and quantitatively, of the quantum probabilities. We start by exploring the simple $N=2$ case.

\subsection{The $N=2$ case}

For $N=2$, (\ref{formulaNxN}) becomes $D({\bf r})={1\over 2}(\mathbb{I} + {\bf r}\cdot \mbox{\boldmath$\Lambda$})$, with $\mbox{\boldmath$\Lambda$}\equiv(\sigma_1,\sigma_2,\sigma_3)^\top$. If we write ${\bf r}$ in spherical coordinates, we have: ${\bf r}=(r\sin\theta\cos\phi, r\sin\theta\sin\phi, r\cos\theta)^\top$, and $D({\bf r})\equiv D(r,\theta,\phi)$ takes the explicit form: 
\begin{equation}
D(r,\theta,\phi)=
{1\over 2}\left[ \begin{array}{cc}
1 +r\cos\theta & r\sin\theta\, e^{-i\phi} \\
r\sin\theta\, e^{i\phi} & 1-r\cos\theta \end{array} \right] 
\label{spherical-coordinates}
\end{equation}
To keep the discussion simple, we consider the measurement of the observable $\sigma_3=P({\bf n}_1)-P({\bf n}_2)$, where ${\bf n}_2 = - {\bf n}_1$, and the eigenvector-states are given by:
\begin{equation}
P({\bf n}_1)=
\left[ \begin{array}{cc}
1 &0 \\
0 & 0 \end{array} \right], \quad P({\bf n}_2) = 
\left[ \begin{array}{cc}
0 & 0 \\
0 & 1 \end{array} \right].
\end{equation}
According to the Born rule, ${\cal P}(D({\bf r}) \to  P({\bf n}_i) = {\rm Tr}\, D({\bf r}) P({\bf n}_i)$, $i=1,2$, so that: 
\begin{equation}
{\cal P}(D({\bf r}) \to  P({\bf n}_1)) = {1\over 2}(1 +r\cos \theta), \quad 
{\cal P}(D({\bf r}) \to  P({\bf n}_2)) = {1\over 2}(1 +r\cos (\pi +\theta)).
\label{prob2dim}
\end{equation}
We can observe that (\ref{prob2dim}) is exactly of the form (\ref{transitiongeneralNxN}), with $\theta$ the angle between ${\bf r}$ and ${\bf n}_1$, and $\pi + \theta$ the angle between ${\bf r}$ and ${\bf n}_2$. Also, when $r=1$ (the initial state is a vector-state), we obtain the simpler formulae: 
\begin{equation}
{\cal P}(D({\bf r}) \to  P({\bf n}_1)) = \cos^2 {\theta\over 2}, \quad 
{\cal P}(D({\bf r}) \to  P({\bf n}_2)) = \sin^2 {\theta\over 2}.
\label{prob2dim-bis}
\end{equation}

We now explain how the measurement of $\sigma_3$, producing the probabilities (\ref{prob2dim}), can be modeled within the three-dimensional Bloch sphere $B_1(\real^3)$, thus extending (and completing) the Bloch representation~\cite{AertsSassoli2014c}. Since the Bloch sphere is a sphere of states, to geometrically represent measurements inside of it we have to consider the associated eigenstates. The observable $\sigma_3$ is characterized by the two eigenstates $P({\bf n}_1)$ and $P({\bf n}_2)=P(-{\bf n}_1) $, whose representative vectors on the sphere are associated with the zenith and nadir points ${\bf n}_1= (0,0,1)^\top$ and ${\bf n}_2 =(0,0,-1)^\top$, respectively. In addition to these two antipodal points, we also consider the intermediary points going from ${\bf n}_2$ to ${\bf n}_1$, forming the 1-simplex $\triangle_{1}$, of length 2, which is the vertical diameter of the sphere. 

\begin{figure}[!ht]
\centering
\includegraphics[scale =.6]{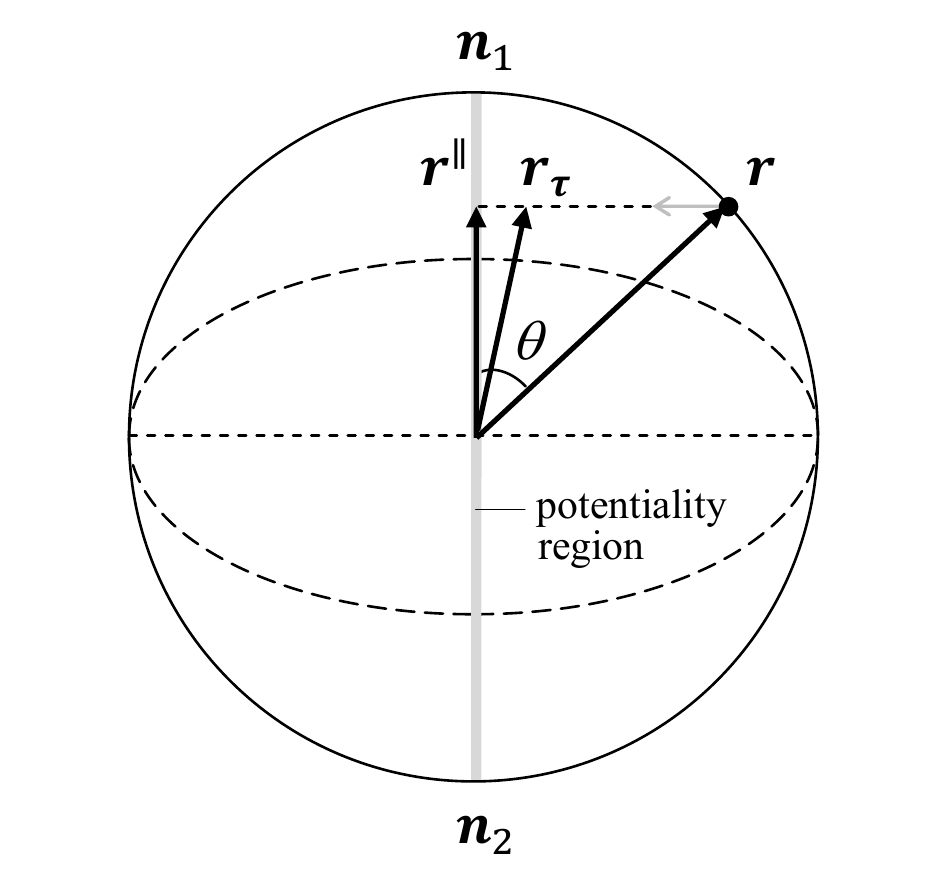}
\caption{The orthogonal path ${\bf r}_\tau$, $0\leq \tau\leq 1$, followed by the point particle representative of the measured entity in the Bloch sphere, going from the initial position ${\bf r}_0={\bf r}$, here considered to be at the surface (i.e., ${\bf r}$ is representative of a vector-state), to the position ${\bf r}_1={\bf r}^\parallel$, onto the 1-simplex (in gray color) associated with the measurement of  $\sigma_3$.
\label{Bloch-sphere2}}
\end{figure}

This 1-simplex describes the `region of potentiality' associated with $\sigma_3$, responsible for the indeterministic ``collapse'' of the state during the measurement, as we will see in a moment. For this to occur, the entity has first to enter into contact with $\triangle_{1}$, which means that the measurement process has to involve a preparatory \emph{deterministic} phase, during which the pre-measurement state $D({\bf r})$ approaches $\triangle_{1}$ and enters into contact with it. This process can be viewed as the immersion of a point particle associated with the vector ${\bf r}$ (which lies at the surface of the sphere, or is located inside of it, depending on whether it describes a vector-state or a more general operator-state), to reach $\triangle_{1}$ along an orthogonal path (see Fig.~\ref{Bloch-sphere2}). The point ${\bf r}^\parallel$ on $\triangle_{1}$ obtained in this way is then ${\bf r}^\parallel=({\bf r}\cdot {\bf n}_1)\,{\bf n}_1 = r\cos \theta\, {\bf n}_1$, and we can describe this deterministic movement of approach of the potentiality region by means of a parameter $\tau$, varied from 0 to 1: ${\bf r}_\tau = (1-\tau)\,{\bf r} + \tau\, r\cos \theta\, {\bf n}_1$. Clearly, ${\bf r}_0 = {\bf r}$ is the initial position, and ${\bf r}_1={\bf r}^\parallel$ the final position on the potentiality region $\triangle_{1}$. 

Before describing the purely indeterministic part of the measurement, it is instructive to write the operator-state $D({\bf r}_\tau)$ in explicit terms. We have ${\bf r}_\tau=((1-\tau)r\sin\theta\cos\phi, (1-\tau)r\sin\theta\sin\phi, r\cos\theta)^\top$, so that $D({\bf r}_\tau)\equiv D(\tau,r,\theta,\phi)$ takes the explicit form: 
\begin{equation}
D(\tau,r,\theta,\phi)=
{1\over 2}\left[ \begin{array}{cc}
1 +r\cos\theta & (1-\tau)r\sin\theta\, e^{-i\phi} \\
(1-\tau)r\sin\theta\, e^{i\phi} & 1-r\cos\theta \end{array} \right].
\label{decoherence}
\end{equation}
As we can see on the above expression, the deterministic approach of the entity's state towards $\triangle_{1}$ corresponds to a \emph{decoherence-like} process, causing the off-diagonal elements of the operator-state $D(\tau,r,\theta,\phi)$ to gradually vanish, as $\tau\to 1$, so that the ``on-potentiality region'' state takes the form of a fully reduced density operator: 
\begin{equation}
D({\bf r}^\parallel) = D(1,r,\theta,\phi)= {1\over 2}(1 +r\cos\theta) \, P({\bf n}_1) + {1\over 2}(1 -r\cos\theta) \, P({\bf n}_2).
\label{reduced}
\end{equation}

\begin{figure}[!ht]
\centering
\includegraphics[scale =.6]{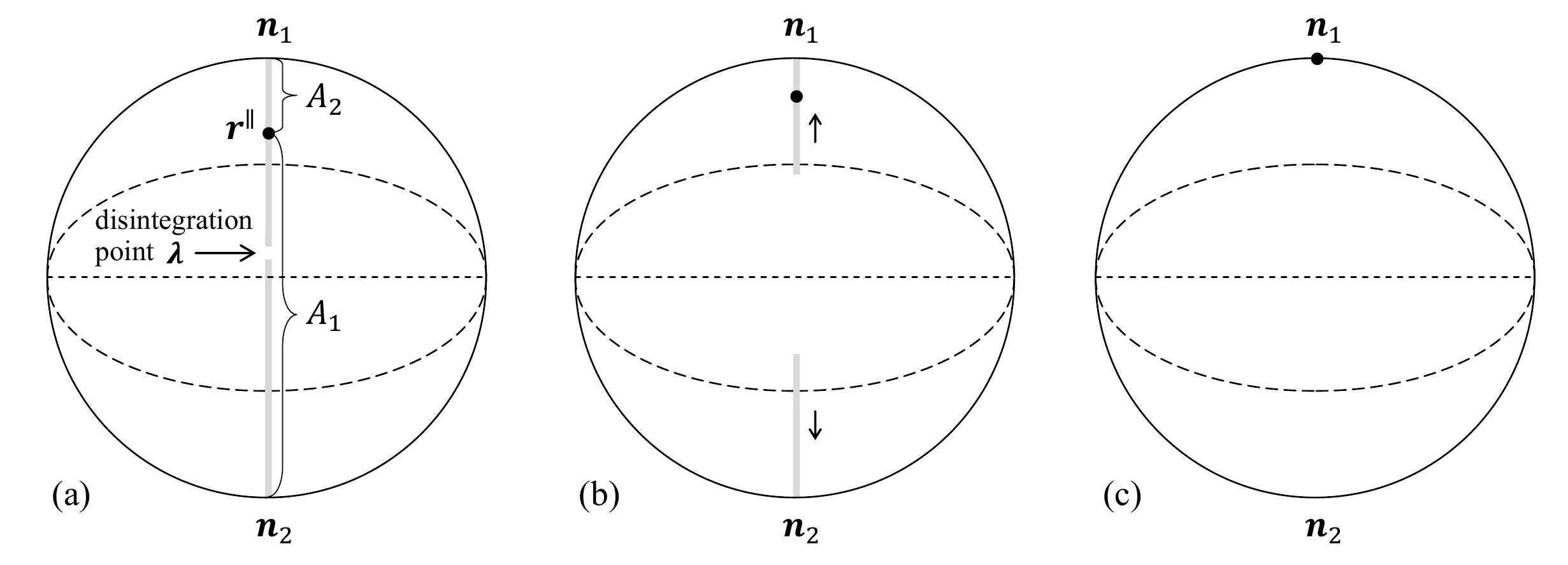}
\caption{The unfolding of the indeterministic part of the measurement of observable $\sigma_3$: (a) the point particle representative of the state reaches the potentiality region (the one-dimensional elastic band represented here in gray color) at point ${\bf r}^\parallel$, so defining two line segments $A_1$ and $A_2$. The substance of the elastic band then disintegrates at some unpredictable point \mbox{\boldmath$\lambda$}, here assumed to be within $A_1$; (b) the elastic substance collapses, drawing the point particle towards one of the two anchor points, here ${\bf n}_1$; (c) the point particle reaches its final destination, here point ${\bf n}_1$, representative of the eigenstate $P({\bf n}_1)$, associated with the eigenvalue $1$.
\label{Bloch-sphere3}}
\end{figure}

We are now in a position to describe the second phase of the measurement process, purely indeterministic, which is responsible for the emergence of the quantum probabilities. For this, we have to think of the potentiality region $\triangle_{1}$ as if it were made of a uniform substance which is not only \emph{attractive} (as it causes the initial state to be orthogonally attracted towards it), but also \emph{unstable} and \emph{elastic}. A simple image we can use is that of a uniform elastic band stretched between the two anchor points ${\bf n}_2$ and ${\bf n}_1$, with the state of the entity represented by a material point particle stuck on it, at point ${\bf r}^\parallel$. The instability of the substance means that the elastic at some moment will break, at some unpredictable point \mbox{\boldmath$\lambda$}, causing it to split into two halves, which will then contract towards the respective anchor points. Depending on which of these two halves the point particle is attached, it will either be drawn to point ${\bf n}_2$, or to point ${\bf n}_1$ (see Fig.~\ref{Bloch-sphere3}).

More precisely, let $A_1$ be the line segment between ${\bf r}^\parallel$ and ${\bf n}_2$, and $A_2$ the line segment between ${\bf n}_1$ and ${\bf r}^\parallel$. Their lengths (Lebesgue measures) are: 
\begin{eqnarray}
&&\mu(A_1)= \|{\bf r}^\parallel - {\bf n}_2\|= \|r\cos \theta\,{\bf n}_1 + {\bf n}_1\| = \|(1+ r\cos \theta)\,{\bf n}_1\| =1+ r\cos \theta,\\
&&\mu(A_2)= \|{\bf n}_1 - {\bf r}^\parallel \|=\| {\bf n}_1 - r\cos \theta\,{\bf n}_1 \| = \|(1- r\cos \theta)\,{\bf n}_1\| =1- r\cos \theta.
\end{eqnarray}
Considering that the first immersive phase of the measurement process is deterministic, it is clear that the transition probability ${\cal P}(D({\bf r})\to P({\bf n}_i))$ is nothing but the probability that the point particle is drawn to point ${\bf n}_i$, i.e., the probability ${\cal P}(\mbox{\boldmath$\lambda$}\in A_i)$ that the disintegration point \mbox{\boldmath$\lambda$} belongs to $A_i$, $i=1,2$. Being the elastic substance, by hypothesis, uniform, and of total length $\mu(A_1)+\mu(A_2)=2$, we thus have: 
\begin{eqnarray}
&&{\cal P}(\mbox{\boldmath$\lambda$}\in A_1) = {1\over 2}\,\mu (A_1) = {1\over 2}(1+ r\cos \theta),\\
&&{\cal P}(\mbox{\boldmath$\lambda$}\in A_2) = {1\over 2}\,\mu (A_2) = {1\over 2}(1- r\cos \theta) ,
\label{PlambdainA+}
\end{eqnarray}
which are precisely the quantum mechanical probabilities (\ref{prob2dim}) predicted by the Born rule. In other terms, the standard Bloch sphere representation can be extended to also include a description of the different possible measurements, in accordance with the predictions of the Born rule.

Before explaining in the next section how this representation can be generalized to an arbitrary number of dimensions, a few remarks are in order. What we have described, by means of the elastic band structure, is clearly a measurement of the \emph{first kind}. Indeed, once the point particle has reached one of the two outcome positions ${\bf n}_i$, if subjected again to the same measurement, being already located in one of the two anchor points of the elastic, we have ${\bf r} = {\bf r}^\parallel = {\bf n}_i$, so that its position cannot be further changed by its collapse. 

The disintegration points \mbox{\boldmath$\lambda$} can be interpreted as variables specifying the \emph{measurement-interactions}. Thus, the model provides a consistent hidden-measurement interpretation of the quantum probabilities, as epistemic quantities characterizing our lack of knowledge regarding the interaction that is actualized between the measured entity and the measuring apparatus, at each run of the experiment. Almost each measurement interaction \mbox{\boldmath$\lambda$} gives rise to a purely deterministic process, changing the state of the entity to either ${\bf n}_1$ or ${\bf n}_2$, depending whether $\mbox{\boldmath$\lambda$}\in A_1$, or $\mbox{\boldmath$\lambda$}\in A_2$. We say `almost' because when $\mbox{\boldmath$\lambda$}={\bf r}^\parallel$, that is, when \mbox{\boldmath$\lambda$} coincides with the point separating $A_1$ from $A_2$, we have a situation of classical unstable equilibrium. This point of classical instability is at the origin of the distinction between the two outcomes, but being of measure zero, it doesn't contribute to the values of the probabilities associated with them. In other terms, the border point $\mbox{\boldmath$\lambda$}={\bf r}^\parallel$ is the ``source of the possibilities,'' but does not contribute to the values of the probabilities that are associated with them.

 \subsection{The $N>2$ case}
 
In the $N=2$ special case we have seen that the measurement results from the interaction between the point particle representative of the state and an attractive, elastic and unstable substance uniformly filling the 1-simplex $\triangle_{1}$, representative of the measurement. In the same way, the measurement context associated with a general $N$-dimensional observable consists of a $(N-1)$-dimensional simplex $\triangle_{N-1}$, uniformly filled with an attractive, elastic and unstable substance, with the point particle representative of the state always orthogonally ``falling'' onto it, and being then drawn to one of its apex points (if the observable is non-degenerate) in an unpredictable way, as a consequence of the disintegration and collapse of said substance. 

\begin{figure}[!ht]
\centering
\includegraphics[scale =.65]{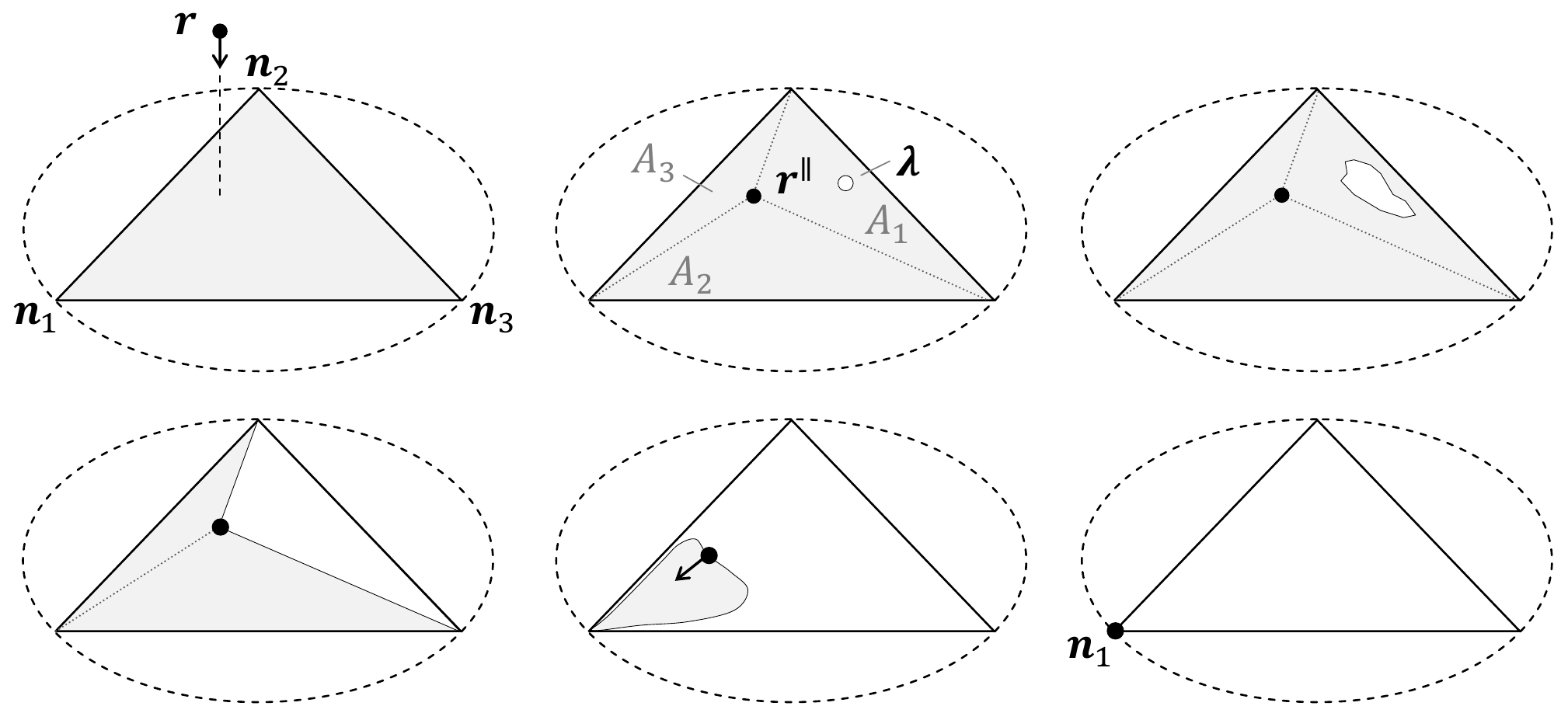}
\caption{The unfolding of a measurement process with three distinguishable outcomes: ${\bf n}_1$, ${\bf n}_2$ and ${\bf n}_3$. The point particle representative of the state, initially located in ${\bf{r}}$, deterministically approaches the triangular elastic membrane along an orthogonal path, reaching the on-membrane point ${\bf r}^\parallel$, defining in this way three sub-regions $A_1$, $A_2$ and $A_3$. The membrane then disintegrates at some unpredictable point \mbox{\boldmath$\lambda$}, producing the complete collapse of the associated sub-region, here $A_1$, causing it to lose two of its anchor points and drawing in this way the point particle to its final location, here ${\bf{n}}_1$.
\label{spinmachine3d}}
\end{figure}

To see how all this works, we only describe here, for simplicity, the $N=3$ situation, as the general situation proceeds according to the same logic and is a straightforward generalization~\cite{AertsSassoli2014c}. So, the measurement context is now represented by a $2$-dimensional triangular elastic membrane inscribed in a 8-dimensional ball, associated with an observable $A=\sum_{i=1}^3 a_i P({\bf n}_i)$, which for the moment we assume to be non-degenerate. We have three possible outcomes, which are the eigenstates $P({\bf n}_i)$, associated with the vertex vectors ${\bf n}_i$, $i=1,2,3$. If the initial, pre-measurement state $D({\bf r})$ is associated with a vector ${\bf r}\in B_1(\real^{8})$, the entity proceeds first with a deterministic movement ${\bf r}_\tau = (1-\tau)\,{\bf r} + \tau\, {\bf r}^\parallel$, $\tau \in [0,1]$, which brings the state of the entity from its initial position ${\bf r}_0 = {\bf r}$ to the on-membrane position ${\bf r}_1={\bf r}^\parallel$, along a path orthogonal to $\triangle_{2}$ (see Fig.~\ref{spinmachine3d}).

Once fixed to membrane, at point ${\bf r}^\parallel$, the particle defines three triangular sub-regions $A_1$, $A_2$ and $A_3$, delineated by the line segments connecting the particle's position with the three vertex points. One should think of these line segments as ``tension lines'' altering the functioning of the membrane, in the sense of making it less easy to disintegrate along them. Then, at some moment, the membrane disintegrates, at some unpredictable point \mbox{\boldmath$\lambda$}, belonging to one of these three sub-regions. The disintegration then propagates, initially inside that specific sub-region, but not into the other two sub-regions, because of the presence of the tension lines. This causes the two anchor points of the disintegrating sub-region to tear away, producing the detachment of the membrane, which being elastic contracts towards the only remaining anchor point, drawing to that position also the particle attached to it, which in this way reaches its final destination, corresponding to the outcome of the measurement (see Fig.~\ref{spinmachine3d}). 

Reasoning in the same way as we did in the $N=2$ case, it is clear that the transition probability ${\cal P}(D({\bf r})\to P_{a_i})$ is given by the probability ${\cal P}(\mbox{\boldmath$\lambda$}\in A_i)$ that the disintegration point \mbox{\boldmath$\lambda$} belongs to the sub-region $A_i$. Considering that $\triangle_{2}$ is an equilateral triangle of area ${3\sqrt{3}\over 4}$, we have ${\cal P}(\mbox{\boldmath$\lambda$}\in A_i)= {4\over 3\sqrt{3}}\mu(A_i)$. Let us consider for instance $A_1$. It is a triangle with vertices ${\bf n}_2$, ${\bf n}_3$ and ${\bf r}^\parallel = \sum_{i=1}^3 r^\parallel_i {\bf n}_i$. Using the explicit coordinates of the three vertices of $A_1$, we can easily calculate its area. For this, we can use a system of coordinates directly in the plane of the triangle, such that ${\bf n}_2=(0,1)\equiv (x_1,x_2)$, ${\bf n}_3=({\sqrt{3}\over 2},-{1\over 2})\equiv (y_1,y_2)$, and ${\bf n}_1=(-{\sqrt{3}\over 2},-{1\over 2})\equiv (z_1,z_2)$. To calculate the area we can use the general formula: $\mu(A_1)={1\over 2}\left|-y_1x_2 +z_1x_2+x_1y_2-z_1y_2-x_1z_2+y_1z_2 \right|$. After a calculation without difficulties, one finds, using (\ref{n-parallelexpansion}): $\mu(A_1)={3\sqrt{3}\over 4}r^\parallel_1$. Doing a similar calculation for $A_2$ and $A_3$, one obtains that $\mu(A_i)={3\sqrt{3}\over 4}r^\parallel_i$, $i=1,2,3$, so that ${\cal P}(\mbox{\boldmath$\lambda$}\in A_i)= {4\over 3\sqrt{3}}\mu(A_i)=r^\parallel_i$, $i=1,2,3$, in accordance with the predictions of the quantum mechanical Born rule (\ref{trans-general}).

The hidden-measurement membrane mechanism that we have here described, for the $N=2$ and $N=3$ case, generalizes in a natural way to an arbitrary number of dimensions $N$, and we refer the reader to~\cite{AertsSassoli2014c} for a general mathematical proof. The membrane mechanism can also be used to describe measurements of degenerate observables. Then, the sub-regions associated with the degenerate eigenvalues are fused together and form bigger composite sub-regions, so that when the initial disintegration point \mbox{\boldmath$\lambda$} takes place inside one of them, the process draws the particle not to a vertex point of $\triangle_{N-1}$, but to one of its sub-simplexes. The collapse of the elastic substance remains compatible with the predictions of the \emph{L\"uders-von Neumann projection formula}, but to complete the process the particle has to re-emerge from the sub-simplex potentiality region, again along an orthogonal path, to deterministically reach its final position~\cite{AertsSassoli2014c}.

In other terms, in the general situation a measurement is to be understood as a tripartite process formed by (1) an initial deterministic \emph{decoherence-like} process, corresponding to the particle reaching the ``on-membrane region of potentiality;'' (2) a subsequent indeterministic \emph{collapse-like} process, corresponding to the disintegration of the elastic substance filling the simplex, with the particle being drawn to some of its peripheral points; and (3) a possible final deterministic \emph{purification-like} process, bringing again the particle to the right distance from the center of the sphere. In this article we will not consider degenerate measurements, and we refer to~\cite{AertsSassoli2014c} for their specific description. In~\cite{AertsSassoli2014c} we also show that it is possible to relax the hypothesis of uniformity of the membranes, by considering a larger uniform average, not only over all the possible measurement-interactions, but also over all the possible non-uniform membranes, and still obtain the Born rule. This means that the latter can describe a very deep condition of lack of knowledge regarding the fluctuations which are present in a measurement context.

\section{Superposition states and interference effects}
\label{Superposition states}

In the previous section we have presented the extended Bloch representation and the associated hidden-measurement mechanism, which allowed us to obtain a complete description of the quantum measurement processes, and consequently an explanation of the possible origin and nature of the quantum probabilities. In the present and following sections we want to use this representation to explore and explain two typical quantum phenomena: \emph{superposition} and \emph{entanglement}. In this section we start by analyzing the typical effects of overextension and underextension of quantum probabilities, when compared to the classical ones, due to the presence of interferences. For this, we consider a state of the form:
\begin{equation}
|\psi \rangle = a_1e^{i\alpha_1}|\varphi_1\rangle + a_2\, e^{i\alpha_2}|\varphi_2\rangle,
\label{superposition}
\end{equation}
with $0\leq a_1,a_2 \leq 1$, $a_1^2+a_2^2 =1$, $\alpha_1,\alpha_2\in \real$, and $\langle \varphi_i|\varphi_j\rangle = \delta_{ij}$, $i,j=1,2$, so that $\langle \psi|\psi\rangle = 1$. The associated projection operator $P_\psi=|\psi\rangle\langle\psi|$ is given by:
\begin{eqnarray}
P_\psi &=&\left( a_1e^{i\alpha_1}\, |\varphi_1\rangle + a_2\, e^{i\alpha_2}|\varphi_2\rangle \right)\left(\langle\varphi_1| a_1e^{-i\alpha_1}+ \langle\varphi_2|a_2\, e^{-i\alpha_2}\right)\nonumber\\
&=& a_1^2 P_1 + a_2^2 P_2 +a_1a_2\left(e^{-i\alpha} |\varphi_1\rangle \langle\varphi_2| + e^{i\alpha}|\varphi_2\rangle \langle\varphi_1|\right),
\label{Ppsi}
\end{eqnarray}
where we have defined $P_i= |\varphi_i\rangle\langle\varphi_i|$, $i=1,2$, and $\alpha = \alpha_2-\alpha_1$. We also consider the two orthonormal vectors $|\chi_\pm\rangle = {1\over\sqrt{2}}( |\varphi_1\rangle \pm |\varphi_2\rangle)$, with associated projection operators $P_\pm=|\chi_\pm\rangle\langle\chi_\pm|$. We then have the transition probabilities: ${\cal P}(P_\psi\to P_i)=a_i^2$, ${\cal P}(P_i\to P_\pm)={1\over 2}$, $i=1,2$, and: 
\begin{eqnarray}
{\cal P}(P_\psi\to P_\pm) &=& {\cal P}(P_\psi\to P_1){\cal P}(P_1\to P_\pm)+{\cal P}(P_\psi\to P_2){\cal P}(P_2\to P_\pm) +I_\pm\label{interference0}\\
&=& {a_1^2\over 2} + {a_2^2\over 2} \pm a_1a_2 \cos\alpha ={1\over 2}(1 +2 I_\pm),
\label{interference}
\end{eqnarray}
where $I_\pm\equiv \pm a_1a_2 \cos\alpha$ is the non-classical interference contribution. 

Projection operators can be associated in quantum mechanics with properties, so that the transition probability ${\cal P}(P_\psi\to P_\pm) = {\rm Tr}\, P_\psi P_\pm$ can be \emph{classically} interpreted as the probability for the entity in state $P_\psi$ to possess the property $P_\pm$, and similarly for the other transition probabilities. This means that, if properties $P_i$ and $P_\pm$ would be compatible, the probability for  $P_\pm$ to be actual would be equal, according to the classical theorem of total probability, to the probability for  ``$P_1$ and $P_\pm$'' to be actual, plus the probability for  ``$P_2$ and $P_\pm$'' to be actual. This corresponds to the first two terms in (\ref{interference0}). Therefore, if the third term in (\ref{interference0}) would be absent, the equality would admit a classical probabilistic interpretation. But because of the presence of the third interference term (oscillating with $\alpha$), we have a violation of the classical theorem of total probability.

Our goal is to understand the origin of this term, within the hidden-measurement paradigm and the extended Bloch representation. For this, we consider an observable $O$ having the two projection operators $P_+$ and $P_-$ in its spectral decomposition. This means that, writing $P_\pm ={1\over N}(\mathbb{I}+c_N \,{\bf n}_\pm \cdot \mbox{\boldmath$\Lambda$})$, the unit vectors ${\bf n}_+$ and ${\bf n}_-$ are two of the $N$ vertex vectors of the measurement simplex $\triangle_{N-1}$, associated with $O$. We also introduce the unit vectors ${\bf n}_1$, ${\bf n}_2$ and ${\bf n}$, representative of $P_1$, $P_2$ and $P_\psi$, respectively, i.e., $P_i ={1\over N}(\mathbb{I}+c_N \,{\bf n}_i \cdot \mbox{\boldmath$\Lambda$})$, $i=1,2$, and $P_\psi ={1\over N}(\mathbb{I}+c_N \,{\bf n}\cdot \mbox{\boldmath$\Lambda$})$.

\subsection{The $N=2$ case}

\begin{figure}[!ht]
\centering
\includegraphics[scale =.6]{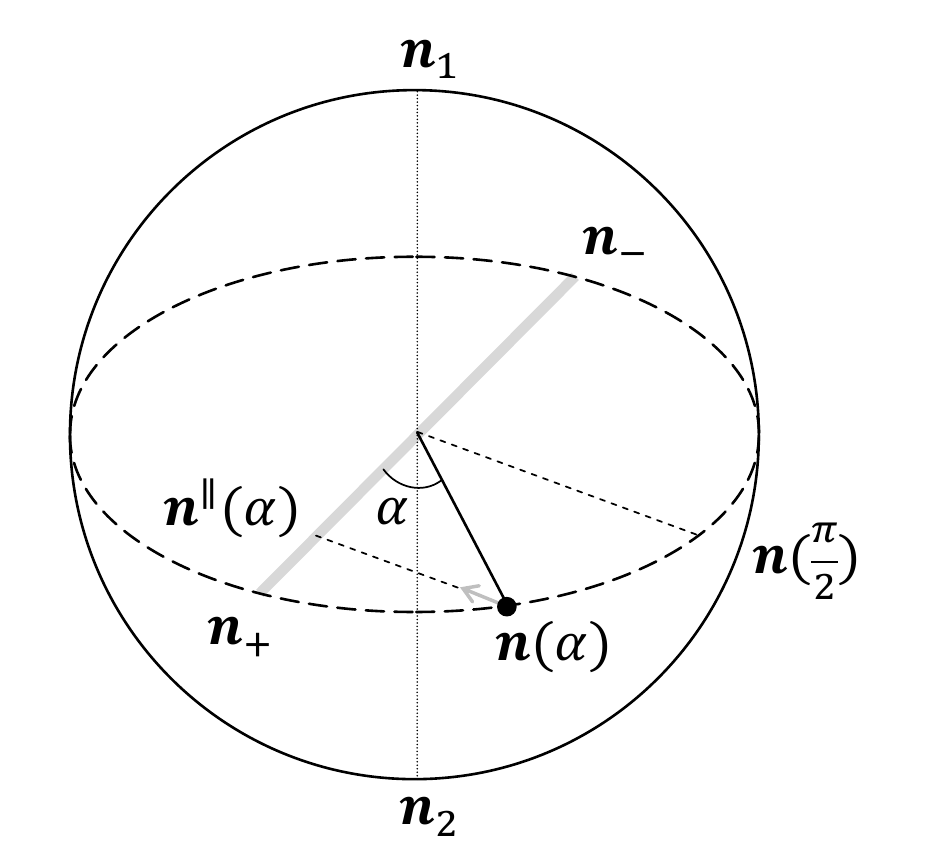}
\caption{The equatorial ``circle of latitude'' which is traveled by the representative vector ${\bf n}({\alpha})$, as the relative phase ${\alpha}$ varies, with the no-interference situation corresponding to the vectors ${\bf n}({\pi\over 2})$ (modulo $\pi$), which exactly ``falls'' onto the middle of the measurement 1-simplex.
\label{figure4}}
\end{figure}

We explore first the $N=2$ case, and to keep the discussion simple we choose $|\varphi_1\rangle$ and $|\varphi_2\rangle$ to be the eigenvectors of $\sigma_3$, for the eigenvalues $+1$ and $-1$, respectively. We also choose $a_1=a_2={1\over\sqrt{2}}$, so that $|\chi_+\rangle$ and $|\chi_-\rangle$ are the eigenvectors of $\sigma_1$, for the eigenvalues $+1$ and $-1$, respectively, and the above mentioned observable $O$ can be taken to be $\sigma_1$. Then, the representative vector ${\bf n}$ of $P_\psi$, within the Bloch sphere, is only a function of the relative phase $\alpha$: ${\bf n}\equiv {\bf n}(\alpha)$, with ${\bf n}(0)={\bf n}_{+}$, and ${\bf n}(\pi)={\bf n}_{-}$. Comparing $P_\psi ={1\over 2}(\mathbb{I}+ n_1\, \sigma_1 + n_2\, \sigma_2 + n_3\, \sigma_3)$, with (\ref{Ppsi}), i.e., with $P_\psi = {1\over 2}(P_1+P_1 + e^{-i\alpha} |\varphi_1\rangle \langle\varphi_2| + e^{i\alpha}|\varphi_2\rangle \langle\varphi_1|)$, one easily finds that (see Fig.~\ref{figure4}): 
\begin{equation}
{\bf n}({\alpha}) =(\cos\alpha, \sin\alpha, 0)^\top,\quad {\bf n}_{\pm}= (\pm 1, 0,0)^\top,\quad {\bf n}_{1\atop 2}= (0, 0, \pm 1)^\top.
\end{equation}
We immediately see that we can write ${\bf n} ={\bf n}^\parallel + {\bf n}^\perp$, with ${\bf n}^\parallel = (\cos\alpha, 0, 0)^\top$ the orthogonal projection of {\bf n} onto the measurement 1-simplex associated with $\sigma_1$, and ${\bf n}^\perp = (0, \sin\alpha, 0)^\top$ the component perpendicular to the latter, so that ${\bf n}^\parallel$ can be written as the convex combination: ${\bf n}^\parallel = n^\parallel_+\,{\bf n}_{+} + n^\parallel_-\,{\bf n}_{-}$, where the components $n^\parallel_\pm={1\over 2}(1\pm \cos\alpha)$, in accordance with (\ref{trans-general}), correspond to the transition probabilities
\begin{equation}
{\cal P}(P_\psi\to P_\pm)={1\over 2}(1 \pm \cos\alpha) ={1\over 2}(1 + 2I_\pm),
\end{equation}
where the term $I_\pm \equiv \pm {1\over 2}\cos\alpha$ is the interference contribution, which is zero for $\alpha={\pi\over 2}, {3\pi\over 2}$ (mod $2\pi$). 

We thus see that the non-classical interferences produced by the superposition can be explained in the extended Bloch representation as follows (see Fig.~\ref{figure4}). When the relative phase $\alpha$ varies, the representative vector ${\bf n}({\alpha})$ moves on the ``circle of latitude'' that is equidistant from the ${\bf n}_{1}$ ``North Pole'' and the ${\bf n}_{2}=-{\bf n}_{1}$ ``South Pole,'' and which contains the two outcome vectors ${\bf n}_{+}$ and ${\bf n}_{-}=-{\bf n}_{+}$. The no-interference situation, compatible with classical probabilities, is when the point particle orthogonally projects exactly at the center of the measurement 1-simplex (i.e., at the center of the Bloch sphere), thus yielding equal probabilities for the two possible transitions. But as $\alpha$ varies from the ${\pi\over 2}$ and ${3\pi\over 2}$ values, the point particle will not  ``fall'' anymore at the center of the 1-simplex, so that the two outcomes ${\bf n}_{\pm}$ will have different probabilities, with the maximum effects of overextension and underextension corresponding to the values $\alpha = 0, \pi$ (modulo $2\pi$), i.e., to the situation when the on-elastic vector coincides with one of the two vertices of the measurement simplex. 

Interference effects can therefore be understood as being the consequence of the fact that superposition states all lie along a circle of latitude within the Bloch sphere (which for the special choice $a_1=a_2={1\over\sqrt{2}}$ corresponds to the equator), and that such circle is traveled when the relative phase $\alpha$ is varied. This alters the way the representative vector ${\bf n}({\alpha})$ projects onto the measurement simplex, so producing a deviation of the probabilities' values with respect to the classical (no-interference) situation. 

We can also observe that if we measure $\sigma_3$ instead of $\sigma_1$, then the projection of ${\bf n}({\alpha})$ onto the associated 1-simplex (an elastic band stretched along the North Pole-South Pole axis) is independent of $\alpha$, and that: ${\bf n}^\parallel({\alpha})={1\over 2}{\bf n}_1+{1\over 2}{\bf n}_2$, for all $\alpha$, in accordance with the fact that ${\cal P}(P_\psi\to P_{1\atop 2})={1\over 2}$.

\subsection{The $N>2$ case}
\label{The general situation}

We want now to investigate the general situation, to see if the simple explanation and illustration of quantum interferences given above, as the result of a movement on a ``circle of latitude'' within the Bloch sphere, can be maintained for $N>2$. For this, we introduce a basis ${\cal B}=\{|b_1\rangle, |b_2\rangle, |b_3\rangle,\dots,|b_N\rangle\}$ of ${\cal H}_N=\compl^N$, such that the first two vectors are precisely: $|b_1\rangle=|\varphi_1\rangle$ and $|b_2\rangle=|\varphi_2\rangle$. We also choose the first two $SU(N)$ generators to be: $\Lambda_1 = |\varphi_1\rangle\langle \varphi_2| + |\varphi_2\rangle\langle \varphi_1|$ and $\Lambda_2 = -i(|\varphi_1\rangle\langle \varphi_2| - |\varphi_2\rangle\langle \varphi_1|)$, and the following $N-1$ ones to be: $\Lambda_{n+2}= W_n$,  $n=1,\dots,N-1$, where the $W_n$ are the diagonal matrices defined in (\ref{W}). Of course, as we have a totality of $N^2-1$ generators, in addition to the above $N+1$ matrices we should also specify the remaining $N^2-N-2$ ones, i.e., the $\Lambda_i$, with $i=N+2,\dots, N^2-1$, but as we will see, we will not need them.

Our next step is to determine the components of the five $(N^2-1)$-dimensional vectors ${\bf n}$, ${\bf n}_1$, ${\bf n}_2$, ${\bf n}_+$ and ${\bf n}_-$. For this, we observe that since the generators are mutually orthogonal, multiplying $P_\psi$ by $\Lambda_i$, then taking the trace, gives: 
\begin{equation}
n_i=e_N{\rm Tr}\, D({\bf r})\Lambda_i, \quad i=1,\dots,N^2-1, \quad e_N\equiv {N\over 2c_N}.
\label{components}
\end{equation}
Since ${\rm Tr}\, P_i\Lambda_1= {\rm Tr}\, P_i\Lambda_2 =0$, ${\rm Tr}\, |\varphi_1\rangle \langle\varphi_2| \Lambda_1={\rm Tr}\, |\varphi_2\rangle \langle\varphi_1| \Lambda_1=1$, ${\rm Tr}\, |\varphi_1\rangle \langle\varphi_2| \Lambda_2= -{\rm Tr}\, |\varphi_2\rangle \langle\varphi_1| \Lambda_2=-i$, ${\rm Tr}\, |\varphi_1\rangle \langle\varphi_2| \Lambda_{n+2}= {\rm Tr}\, |\varphi_2\rangle \langle\varphi_1| \Lambda_{n+2}= 0$, ${\rm Tr}\, P_1\Lambda_{n+2}=\sqrt{2\over n(n+1)}$, ${\rm Tr}\, P_2\Lambda_{n+2}=\sqrt{2\over n(n+1)}(1-2\delta_{1,n})$, for $n=1,\dots,N-1$, and ${\rm Tr}\, P_1\Lambda_i= {\rm Tr}\, P_2\Lambda_i= \sqrt{2\over (i-2)(i-1)}$, for $i=4,\dots,N+1$, we have:
\begin{eqnarray}
&n_1 = 2e_N a_1a_2\cos\alpha, \quad n_2 = 2e_N a_1a_2\sin\alpha,\quad n_3 = e_N (a_1^2 -a_2^2),\\
&n_i = e_N \sqrt{2\over (i-2)(i-1)}, \quad i=4,\dots,N+1.
\end{eqnarray}

To show that there are no other non-zero components, it is sufficient to show that a $(N+1)$-dimensional vector with the above components is of unit length. Considering that $e_N^2={N\over 2(N-1)}$, we have: 
\begin{eqnarray}
\sum_{j=1}^{N+1} n_j^2 &=& e_N^2\left(4a_1^2a_2^2(\cos^2\alpha +\sin^2\alpha) +(a_1^2 -a_2^2)^2 +2\sum_{i=4}^{N+1}{1\over (i-2)(i-1)}\right)\nonumber \\
&=&{N\over 2(N-1)} \left(1 +2\sum_{i=4}^{N+1}{1\over (i-2)(i-1)}\right)= {N\over 2(N-1)} \left(1 +2\sum_{n=2}^{N-1}{1\over n(n+1)}\right)\nonumber\\
&=&{N\over 2(N-1)} \left(1 +2\sum_{n=1}^{N-1}{1\over n(n+1)}-1\right)={N\over N-1} \sum_{n=1}^{N-1}{1\over n(n+1)}=1,
\end{eqnarray}
where for the last equality we have used $\sum_{n=1}^{N-1}{1\over n(n+1)}=\sum_{n=1}^{N-1}{1\over n}-\sum_{n=2}^{N}{1\over n}=1-{1\over N}$. Finally, observing that $P_1$ can be obtained from $P_\psi$ by setting $a_1=1$, $P_2$ by setting $a_1=0$, $P_+$ by setting $a_1= {1\over\sqrt{2}}$ and $\alpha =0$, $P_-$ by setting $a_1= {1\over\sqrt{2}}$ and $\alpha =\pi$, we obtain: 
\begin{eqnarray}
&{\bf n} =e_N(2 a_1a_2\cos\alpha, 2 a_1a_2\sin\alpha, a_1^2 -a_2^2, {1\over\sqrt{3}}, {1\over\sqrt{6}},\dots, \sqrt{2\over N(N-1)}, 0,\dots,0)^\top,\label{gencoord1}\\
&{\bf n}_{\pm}=e_N (\pm 1, 0,0, {1\over\sqrt{3}}, {1\over\sqrt{6}}, \dots, \sqrt{2\over N(N-1)}, 0, \dots, 0)^\top,\label{gencoord2}\\
&{\bf n}_{1\atop 2}=e_N(0, 0, \pm 1, {1\over\sqrt{3}}, {1\over\sqrt{6}}, \dots, \sqrt{2\over N(N-1)}, 0, \dots, 0)^\top.
\label{gencoord3}
\end{eqnarray}

As we said, we are interested in describing the measurement of an observable $O$ having the two projectors $P_+$ and $P_-$ in its spectral decomposition, i.e., such that two of the vertices of the associated simplex are ${\bf n}_+$ and ${\bf n}_-$. For an initial state $P_\psi=|\psi\rangle\langle\psi|$, with $|\psi\rangle$ the superposition state (\ref{superposition}), only the two transitions $P_{\psi}\to P_\pm$ have non-zero probabilities, as is clear from the fact that ${\cal P}(P_{\psi}\to P_+)+{\cal P}(P_{\psi}\to P_-)=1$. This means that when ${\bf n}$ orthogonally ``falls'' onto the measurement simplex $\triangle_{N-1}$, it lands exactly on the edge between ${\bf n}_{+}$ and ${\bf n}_{-}$, of length $\| {\bf n}_{+}-{\bf n}_{-}\|=2e_N$. This can be seen more explicitly by writing: ${\bf n} ={\bf n}^\parallel + {\bf n}^\perp$, where ${\bf n}^\parallel$ is the on-simplex component and ${\bf n}^\perp$ the component perpendicular to it, given by: 
\begin{eqnarray}
&{\bf n}^\parallel =e_N (2 a_1a_2\cos\alpha, 0, 0, {1\over\sqrt{3}}, {1\over\sqrt{6}}, \dots, \sqrt{2\over N(N-1)}, 0, \dots, 0)^\top\\
&{\bf n}^\perp =e_N (0, 2 a_1a_2\sin\alpha, a_1^2 -a_2^2, 0, \dots, 0, \dots, 0)^\top,
\end{eqnarray}
so that ${\bf n}^\parallel$ can be written as the convex linear combination: 
\begin{equation}
{\bf n}^\parallel = {1\over 2}(1+2 a_1a_2\cos\alpha)\,{\bf n}_{+} + {1\over 2}(1-2 a_1a_2\cos\alpha)\,{\bf n}_{-},
\label{convexcomb}
\end{equation}
where in accordance with (\ref{trans-general}), the components $n^\parallel_\pm={1\over 2}(1\pm 2 a_1a_2\cos\alpha)$ are the transition probabilities ${\cal P}(P_{\psi}\to P_\pm)$, given by (\ref{interference}). 

At this point, we observe that only the first three components of ${\bf n}$ depend on the parameters $a_1$, $a_2$ and $\alpha$. This means that the measurement process takes effectively place in the $3$-dimensional sub-space generated by the first three canonical vectors $(1,0,\dots,0)^\top$, $(0,1,0,\dots,0)^\top$ and $(0,0,1,0,\dots,0)^\top$. So, projecting $B_1(\real^{N^2-1})$ onto that subspace, we obtain a 3-dimensional sub-ball of radius $e_N$, in which the measurement simplex $\triangle_{N-1}$ appears as an effective 1-simplex, of length $2e_N$, corresponding to its edge lying between the two apex vectors ${\bf n}_{+}$ and ${\bf n}_{-}$. Then, the three-dimensional projected vectors have the coordinates (we denote these 3-dimensional vectors with a ``tilde,'' not to confuse them with the unit non-projected ones): 
\begin{equation}
\tilde{\bf n} =e_N(2 a_1a_2\cos\alpha, 2 a_1a_2\sin\alpha, 0)^\top,\quad \tilde{\bf n}_{\pm}=e_N (\pm 1, 0,0)^\top,\quad \tilde{\bf n}_{1\atop 2}=e_N(0, 0, \pm 1)^\top,
\label{projectedvectors}
\end{equation}
and (\ref{convexcomb}) remains clearly valid for the tilded vectors.

For the special case $N=2$, $e_N=1$ and the projected vectors (\ref{projectedvectors}) exactly correspond to those of the standard 3-dimensional Bloch representation, with the edge of $\triangle_{N-1}$ reducing to the 1-simplex $\triangle_{1}$ (the one-dimensional elastic band). When $N>2$, the only difference is that the process now unfolds in an effective 3-dimensional sub-ball of radius $e_N$, but apart from that the description we have given for the $N=2$ case, and Fig.~\ref{figure7}, also hold for the general situation of a $N$-dimensional entity. 

More specifically, let us consider the $N=3$ situation. The projected 3-dimensional sub-ball is then of radius ${\sqrt{3}\over 2}$, and the effective one-dimensional elastic band, of length $\sqrt{3}$, corresponds to the edge of a two-dimensional triangular elastic membrane, representing an observable having the three states $P_+$, $P_-$ and $\mathbb{I} - P_+ - P_-$ as its possible outcomes. We can observe that the triangular membrane works as if it was a one-dimensional elastic band, and that one can obtain the outcome probabilities by doing as if the measurement would be governed by a 1-simplex. The reason for this is that the area of a convex region generated by a point particle positioned on one of the edges of an equilateral triangle is simply given by $\mu(A_x)={3\over 4}x$, where $x$ is the length of the segment defined by the particle's position. In other terms, the area is directly proportional to the length $x$. Thus, reasoning in terms of the length $x$ on the edge of the membrane, or of the area $\mu(A_x)$ of the associated convex sub-region, is perfectly equivalent when considering relative quantities, in the sense that ${x\over \sqrt{3}}= {\mu(A_x)\over \mu(\triangle_2)}$, and more generally we also have: ${x\over 2e_N}= {\mu(A_x)\over \mu(\triangle_{N-1})}$.

\section{Entangled states and correlation effects}
\label{Entanglement}

Our analysis of superposition states can be straightforwardly used to also describe measurements on entangled states, as entanglement is a direct consequence of the superposition principle. Consider a Hilbert space of the form ${\cal H}={\cal H}_A\otimes {\cal H}_B$, with ${\cal H}_A=\compl^{N_A}$ describing the first entity, and ${\cal H}_B=\compl^{N_B}$ the second one, so that ${\cal H}$ is isomorphic to $\compl^{N}$, with $N=N_AN_B$. States of the form $ |\psi^A\rangle\otimes |\phi^B\rangle$, $|\psi^A\rangle\in {\cal H}_A$, $|\phi^B\rangle\in {\cal H}_B$, are called product states, and describe a situation where each entity is in a well-defined vector-state. However, if we superpose two (here orthogonal) product states:
\begin{equation}
|\psi\rangle = a_1\, e^{i\alpha_1}|\psi^A\rangle\otimes |\phi^B\rangle + a_2\, e^{i\alpha_2}|\phi^A\rangle\otimes |\psi^B\rangle,
\label{entanglement}
\end{equation}
with $0\leq a_1,a_2 \leq 1$, $a_1^2+a_2^2 =1$, $\alpha_1,\alpha_2\in\real$, $\langle\psi^A|\phi^A\rangle = \langle\psi^B|\phi^B\rangle = 0$, we obtain a non-product, entangled state, where only the joint entity is in a well defined vector-state. 

The two product states $|\psi^A\rangle\otimes |\phi^B\rangle$ and $|\phi^A\rangle\otimes |\psi^B\rangle$ being orthogonal, if we set $|\varphi_1\rangle \equiv |\psi^A\rangle\otimes |\phi^B\rangle$ and $|\varphi_2\rangle \equiv |\phi^A\rangle\otimes |\psi^B\rangle$, we are exactly in the situation (\ref{superposition}), and when the parameter $a_1$ goes from $1$ to $0$, we transition in a continuous way from the product state $e^{i\alpha_1}|\psi^A\rangle\otimes |\phi^B\rangle$ to the product state $e^{i\alpha_2}|\phi^A\rangle\otimes |\psi^B\rangle$, passing through different entangled states. In particular, for the values $a_1={1\over \sqrt{2}}$ and $\alpha_1=\alpha_2 = 0$, we obtain the triplet-like state $|\chi_+\rangle = {1\over \sqrt{2}}( |\varphi_1\rangle +|\varphi_2\rangle)$, and for $a_1={1\over \sqrt{2}}$, $\alpha_1=0$, and $\alpha_2 = \pi$, the singlet-like state: $|\chi_-\rangle = {1\over \sqrt{2}}( |\varphi_1\rangle -|\varphi_2\rangle)$.

Therefore, the calculation of Sec.~\ref{Superposition states} can also be used to describe measurements on entangled states. This time the focus is on the two transitions $P_\psi\to P_1$ and $P_\psi\to P_2$, corresponding to the measurement of a product observable $O^A\otimes O^B$ such that the two states $|\psi^A\rangle$ and  $|\phi^A\rangle$ are among the eigenvectors of $O^A$, and the two states $|\psi^B\rangle$ and $|\phi^B\rangle$ are among the eigenvectors of $O^B$, so that the four product states $|\varphi_1\rangle \equiv |\psi^A\rangle\otimes |\phi^B\rangle$, $|\varphi_2\rangle \equiv |\phi^A\rangle\otimes |\psi^B\rangle$, $|\varphi_3\rangle \equiv |\psi^A\rangle\otimes |\psi^B\rangle$ and $|\varphi_4\rangle \equiv |\phi^A\rangle\otimes |\phi^B\rangle$ are among the eigenvectors of $O^A\otimes O^B$. Of course, only the transitions to the two product states $P_1$ and $P_2$ will have a non-zero probability, which is just another way to state that the entangled state $P_\psi$ contains in potentiality the perfect correlations: ``$|\psi^A\rangle\leftrightarrow |\phi^B\rangle$'' and ``$|\phi^A\rangle\leftrightarrow|\psi^B\rangle$''. 

\begin{figure}[!ht]
\centering
\includegraphics[scale =.6]{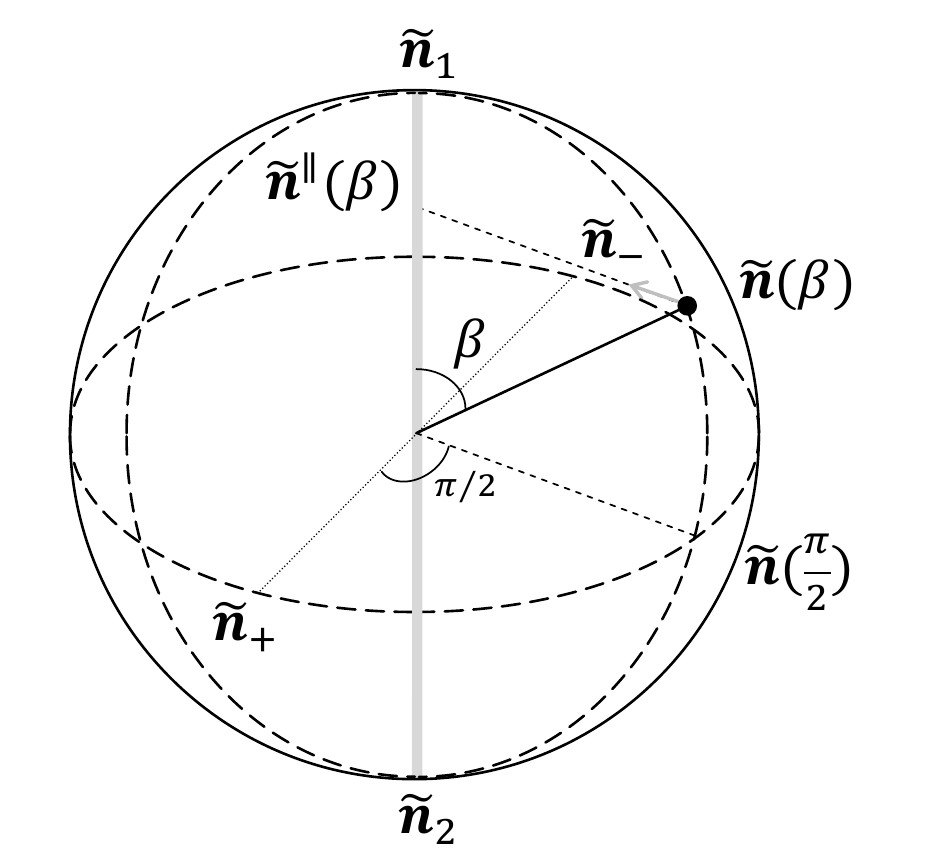}
\caption{The ``circle of longitude'' that is traveled by the representative point particle $\tilde{\bf n}({\beta})$, as the parameter ${\beta}$ varies, here for the choice $\alpha={\pi\over 2}$, with the no-entanglement situation corresponding to $\tilde{\bf n}(0)=\tilde{\bf n}_{1}$ and $\tilde{\bf n}(\pi)=\tilde{\bf n}_{2}$ (modulo $\pi$), whereas the position ${\bf n}({\pi\over 2})$, which projects exactly at the center of the measurement simplex, is representative of the state $|\psi\rangle = {e^{i\alpha_1}\over\sqrt{2}}(|\psi^A\rangle\otimes |\phi^B\rangle + i|\phi^A\rangle\otimes |\psi^B\rangle)$.
\label{figure5}}
\end{figure}

Introducing a real parameter $\beta$, such that $\cos^2 {\beta\over 2}= a_1^2$, and consequently $\sin^2 {\beta\over 2}= a_2^2$, we can write: ${\bf n}^\parallel(\beta) = \cos^2 {\beta\over 2}\,{\bf n}_{1} + \sin^2 {\beta\over 2}\,{\bf n}_{2}$, where ${\bf n}^\parallel(\beta)$ is the vector obtained by orthogonally projecting ${\bf n}$ onto the 1-simplex associated with ${\bf n}_{1}$ and ${\bf n}_{2}$. Of course, the relation remains unchanged if we project these $(N^2-1)$-dimensional vectors onto the subspace generated by the first three canonical vectors, so that we can also write (see the previous section for the notation): $\tilde{\bf n}^\parallel(\beta) = \cos^2 {\beta\over 2}\,\tilde{\bf n}_{1} + \sin^2 {\beta\over 2}\,\tilde{\bf n}_{2}$, where the sub-ball is now of radius $e_N=e_{N_AN_B}$ (see Figure~\ref{figure5}).

So, in accordance with the general discussion of the previous section, we find that entangled states, formed by the superposition of two orthogonal product states, can be described in the extended Bloch representation by unit vectors that are positioned in such a way that, when they orthogonally ``fall'' onto the measurement simplex, they exactly land onto the edge delimited by the two product states forming the entanglement (i.e., the superposition). The process, as we have seen, can be fully represented within the three-dimensional sub-ball generated by the first three components of the Bloch vectors, with the elastic substance filling the measurement simplex  reducing to an effective elastic band, of length $2e_N$. This one-dimensional effective structure conveys the essence of a coincidence measurement on an entangled pair, when two perfectly correlated entities in well-defined vector-states are created, in an unpredictable way.

\subsection{A macroscopic entangled entity}

We can further illustrate this mechanism of ``creation of correlations'' by performing the following macroscopic experiment, which simulates the microscopic process. Two colleagues, Alice ($A$) and Bob ($B$), hold the two ends of a stretched uniform elastic band. A third colleague then draws a black dot somewhere on the stretched elastic, with a marker. Let us assume that the dot is drawn at a distance $L_A$ from Alice, and consequently at a distance $L_B=L-L_A$ from Bob, with $L$ the length of the stretched elastic. The dot is representative of the entangled state $|\psi\rangle = \sqrt{L_A/L}\,|\psi^A\rangle\otimes |\phi^B\rangle + \sqrt{L_B/L}\,|\phi^A\rangle\otimes |\psi^B\rangle$ (we have set the relative phase $\alpha$ equal to zero, as it plays no role here), whereas the stretched elastic between Alice and Bob is representative of the measurement of a product observable having $|\psi^A\rangle\otimes |\phi^B\rangle$ and $|\phi^A\rangle\otimes |\psi^B\rangle$ among its eigenstates. 

\begin{figure}[!ht]
\centering
\includegraphics[scale =.6]{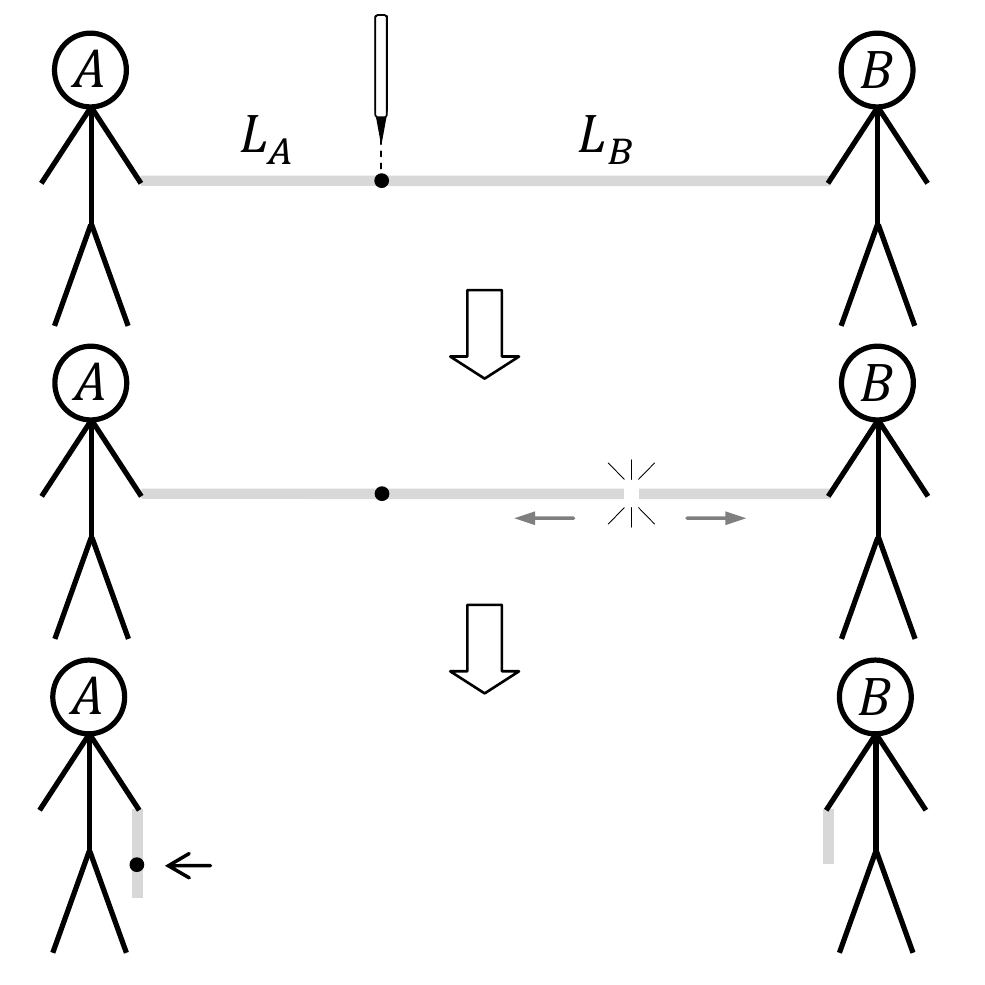}
\caption{A process of creation of correlations, with two possible outcomes: $|\bullet, A\rangle\otimes |\circ, B\rangle$ or $|\circ, A\rangle\otimes |\bullet, B\rangle$ (see the text for the notation). In the figure, a process producing the outcome  $|\bullet, A\rangle\otimes |\circ, B\rangle$ is represented. 
\label{figure6}}
\end{figure}

Once the dot has been drawn, Alice and Bob pull their respective end of the elastic with force, to break it into two separated fragments, which will then collapse into their hands. Alice (resp., Bob) can then check if the dot is in her (resp., his) elastic fragment, and if this is the case $|\psi^A\rangle\otimes |\phi^B\rangle$ (resp., $|\phi^A\rangle\otimes |\psi^B\rangle$) is the outcome of the experiment; otherwise, the outcome is $|\phi^A\rangle\otimes |\psi^B\rangle$ (resp., $|\psi^A\rangle\otimes |\phi^B\rangle$); see Fig.~\ref{figure6}.

It is important not to confuse the two descriptive levels: that of the extended Bloch sphere (which in the case of a singlet state formed by a pair of electrons is 15-dimensional), and that of the above experiment performed by Alice and Bob, which is not only a projection of a higher-dimensional process onto a three-dimensional subspace, but also a ``transliteration'' of a microscopic quantum process by means of an ordinary macroscopic object. Indeed, the position of the point particle, which in the Bloch sphere is representative of the state of the composite system, acquires in the experiment performed by Alice and Bob a different meaning. 

In the Alice and Bob context, the entangled state corresponds to the situation of a black dot that is located somewhere between them, and therefore cannot be attributed to one of them, for as long as the elastic remains unbroken. More precisely, there is a correspondence between the microscopic state $|\psi^A\rangle$ and the macroscopic state ``the black dot is in Alice's hand,'' which we denote $|\bullet, A\rangle$, and between the state $|\phi^A\rangle$ and the situation ``the black dot is not in Alice's hand,'' which we denote $|\circ, A\rangle$. Similarly, $|\psi^B\rangle$ is in a correspondence with $|\bullet, B\rangle$ and $|\phi^B\rangle$ with $|\circ, B\rangle$. The initial entangled state can then be written: $|\psi\rangle = \sqrt{L_A/ L}\,|\bullet, A\rangle\otimes |\circ, B\rangle + \sqrt{L_B/ L}\, |\circ, A\rangle\otimes |\bullet, B\rangle$, and describes a black dot at a distance $L_A$ (resp., $L_B$) form Alice (resp., Bob), so that it has a probability ${L_A/L}$ of being drawn into Alice's hand (resp., ${L_B/ L}$ of being drawn into Bob's hand), when the elastic is torn.

Considering how the experiment is conducted, it is perfectly evident that when Alice observes the state $|\bullet, A\rangle$,  Bob necessarily observes $|\circ, B\rangle$, and vice versa, when Alice observes $|\circ, A\rangle$, Bob necessarily observes $|\bullet, B\rangle$. This because the dot, for obvious reasons, cannot be present (or not present) in both fragments simultaneously, and therefore $|\bullet, A\rangle$ and $|\circ, B\rangle$ describe the same possibility, viewed from the perspective of Alice and Bob, respectively (and similarly for $|\circ, A\rangle$ and $|\bullet, B\rangle$). This also explains why Alice can know the state observed by Bob, without having exchanged with him any information, and vice versa. However, neither Alice nor Bob are able to predict in advance if the dot will end or not in their fragment, being the breaking point unpredictable, so that the correlations ``$|\bullet, A\rangle \leftrightarrow |\circ, B\rangle$'' or ``$|\circ, A\rangle\leftrightarrow|\bullet, B\rangle$'' are not discovered by Alice and Bob, but literally created by them.

In the macroscopic experiment performed by Alice and Bob, the composite entity subjected to the measurement is a pair of entangled \emph{potential} elastic fragments, which when they disentangle, and become \emph{actual} fragments, can only be in two states: with or without a black dot painted on them. However, different from two electrons in a singlet state, an unfragmented elastic possesses the property of \emph{macroscopic wholeness}~\cite{Aerts1991}: it manifests its presence not only in the hands of Alice and Bob, i.e., in the regions of space where Alice and Bob are located, but also in the region between them. This means that the correlations between the elastic fragments that are created by the experiment are the result of their prior \emph{connection through space}. On the other hand, two electrons in a singlet state do not possess this property of macroscopic wholeness, which means that the correlations between the spin orientations that are created by the measurement cannot be attributed to a connection through space. This should not surprise us, as we know that the elastic band is the exemplification of the edge of a  more extended structure, living in a higher-dimensional space, meaning that the connection in question is \emph{non-spatial}, i.e., not representable within our three-dimensional Euclidean space.

\section{Three-state superpositions}
\label{More general superpositions}

\begin{figure}[!ht]
\centering
\includegraphics[scale =.6]{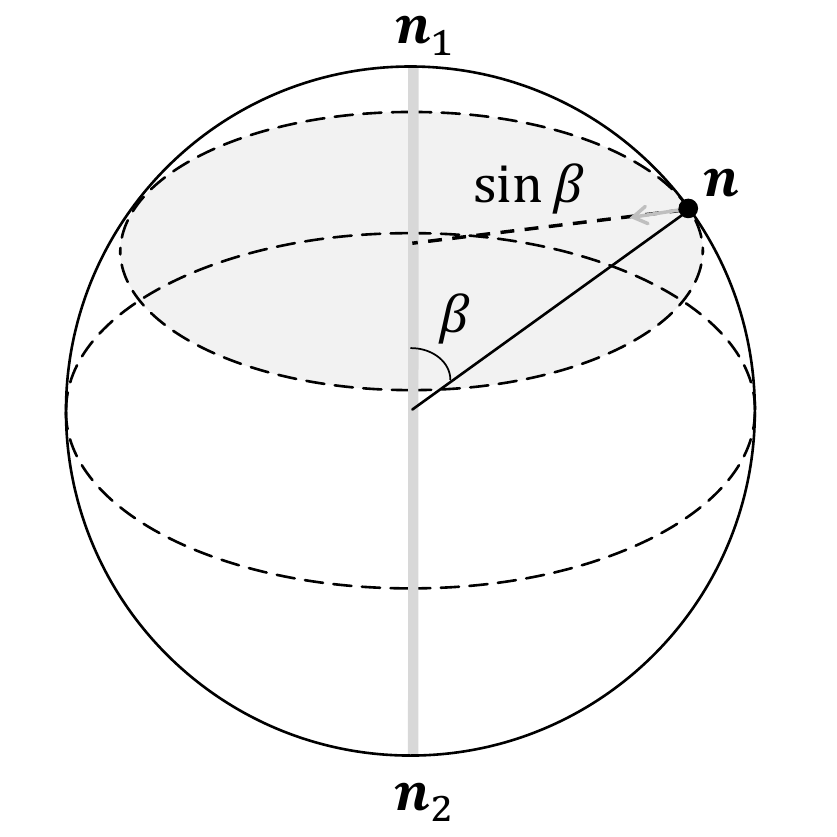}
\caption{The states producing the same transition probabilities as $|\psi \rangle = \cos^2 {\beta\over 2} |\varphi_1\rangle + \sin^2 {\beta\over 2} e^{i\alpha}|\varphi_2\rangle$, with $\alpha\in [0,2\pi]$, are represented, for each $\beta\in [0,\pi]$, as parallel disks within the Bloch sphere. 
\label{figure7}}
\end{figure}

In Sec.~\ref{Superposition states}, we have seen that the two-state superposition (\ref{superposition}), in a two-dimensional Hilbert space, is represented in the three-dimensional Bloch sphere by a unit vector ${\bf n}=(\sin\beta\cos\alpha,\sin\beta\sin\alpha, \cos\beta)^\top$, where $\alpha=\alpha_2-\alpha_1$ and $a_1^2 = \cos^2 {\beta\over 2}$. When we consider this state in relation to the measurement 1-simplex generated by the two outcomes $P_1$ and $P_2$, with representative unit vectors ${\bf n}_1$ and ${\bf n}_2$, the point particle, when plunging into the sphere to reach the simplex, follows the path: ${\bf r}_\tau=((1-\tau)\sin\beta\cos\alpha, (1-\tau)\sin\beta\sin\alpha, \cos\beta)^\top$, with the parameter $\tau$ going from $0$ to $1$. This means that for each $\beta\in [0,\pi]$ and $\tau\in [0,1]$, by varying $\alpha$ the vector ${\bf r}_\tau$ spans a circle of latitude of radius $(1-\tau)\sin\beta$. Therefore, by also varying $\tau$, it fills a disk of radius $\sin\beta$ (see Fig.~\ref{figure7}). In other terms, all points in such disk are  representative of states that give rise to the same transition probabilities ${\cal P}(P_\psi\to P_1)= \cos^2 {\beta\over 2}$ and ${\cal P}(P_\psi\to P_2)= \sin^2 {\beta\over 2}$. And of course, by also varying $\beta$, the ensemble of these disks fills the entire Bloch sphere of states.

We have also seen that a similar picture holds  for $N>2$, if we consider a superposition of only two orthogonal states, so that only three of the $(N^2-1)$ components of the Bloch vector will depend on the parameters $\alpha$ and $\beta$ (and therefore on $\tau$). Consequently, it is still possible to represent the superposition states within an effective three-dimensional ball, of reduced radius, still formed by the juxtaposition of disks characterizing states having the same probabilities for the transitions $P_\psi\to P_1$ and $P_\psi\to P_2$. 

However, the situation becomes rapidly much more complicated if more general superposition states are considered. To see this, we consider the situation of a superposition of three orthogonal states:
\begin{equation}
|\psi\rangle = a_1 \, e^{i\alpha_1} |\varphi_1\rangle + a_2 \, e^{i\alpha_2}|\varphi_2\rangle + a_3 \, e^{i\alpha_3}|\varphi_3\rangle = e^{i\alpha_1}(a_1 |\varphi_1\rangle + a_2 \, e^{i\alpha}|\varphi_2\rangle + a_3 \, e^{i\delta}|\varphi_3\rangle),
\label{superposition-3}
\end{equation}
where the $a_i$ are positive numbers such that: $a_1^2+a_2^2 + a_3^2 =1$, and we have defined: $\alpha \equiv \alpha_2-\alpha_1$, $\delta \equiv \alpha_3-\alpha_1$. Considering also the three orthonormal vectors:
\begin{equation}
|\chi_1\rangle = {1\over\sqrt{3}}( |\varphi_1\rangle + |\varphi_2\rangle +|\varphi_3\rangle ),\,\,\, |\chi_2\rangle = {1\over\sqrt{3}}( |\varphi_1\rangle + \omega|\varphi_2\rangle +\omega^2|\varphi_3\rangle ),\,\,\, 
|\chi_3\rangle = {1\over\sqrt{3}}( |\varphi_1\rangle + \omega^2|\varphi_2\rangle +\omega|\varphi_3\rangle ),
\end{equation}
with $\omega$ obeying: $1+\omega+\omega^2 =0$,  $\omega^*=\omega^2$ (i.e., $\omega=e^{i{2\pi\over 3}}$), and defining: $P_i\equiv |\varphi_i \rangle\langle \varphi_i|$ and $F_i\equiv |\chi_i \rangle\langle \chi_i|$, $i=1,2,3$, we have the transition probabilities: ${\cal P}(P_\psi \to P_{i})= {\rm Tr}\, P_\psi P_i = a_i^2$, ${\cal P}(P_i\to F_j)={\rm Tr}\, P_i F_j = {1\over 3}$, $i,j=1,2,3$, and a simple calculation yields: 
\begin{equation}
{\cal P}(P_\psi \to F_j)=\sum_{i=1}^3{\cal P}(P_\psi \to P_i){\cal P}(P_i \to F_j) + I_j = {1\over 3}(1 + 3I_j),\quad j=1,2,3,
\end{equation}
where the non-classical interference terms $I_j$ are given by:
\begin{eqnarray}
&&I_1= {2\over 3}(a_1a_2\cos\alpha + a_1a_3\cos\delta + a_2a_3\cos\gamma ),\\
&&I_2= {2\over 3}( a_1a_2 \cos(\alpha -{2\pi\over 3}) + a_1a_3 \cos(\delta -{4\pi\over 3}) + a_2a_3 \cos(\gamma - {2\pi\over 3})),\\
&&I_3= {2\over 3}(a_1a_2 \cos(\alpha -{4\pi\over 3}) + a_1a_3 \cos(\delta -{2\pi\over 3}) + a_2a_3 \cos(\gamma - {4\pi\over 3})),
\end{eqnarray}
where $\gamma\equiv \delta-\alpha =\alpha_3-\alpha_2$; and since $\cos\theta +\cos(\theta +{2\pi\over 3}) + \cos(\theta+{4\pi\over 3})=0$, we have $\sum_{i=1}^3 I_j =0$, in accordance with $\sum_{j=1}^3{\cal P}(P_\psi \to F_j)=1$.

As for the two-state superposition, we can use the extended Bloch representation to understand the origin of the above interference contributions. In order to keep the discussion as simple as possible, let us assume that $N=3$, so that $\{|\varphi_1\rangle, |\varphi_2\rangle, |\varphi_3\rangle\}$ and $\{|\chi_1\rangle, |\chi_2\rangle, |\chi_3\rangle\}$ are two (mutually unbiased) bases of $\compl^3$, and the Bloch representation is 8-dimensional. Considering the generators (the Gell-Mann matrices (\ref{generatorsN=3}), here not in the usual order): 
\begin{eqnarray}
&\Lambda_1=|\varphi_1\rangle\langle \varphi_2| + |\varphi_2\rangle\langle \varphi_1|,\quad \Lambda_2=-i(|\varphi_1\rangle\langle \varphi_2| - |\varphi_2\rangle\langle \varphi_1|),\quad \Lambda_3=|\varphi_1\rangle\langle \varphi_3| + |\varphi_3\rangle\langle \varphi_1|,\\
&\Lambda_4 = -i(|\varphi_1\rangle\langle \varphi_3| - |\varphi_3\rangle\langle \varphi_1|),\quad \Lambda_5 =|\varphi_2\rangle\langle \varphi_3| + |\varphi_3\rangle\langle \varphi_2|,\quad \Lambda_6 = -i(|\varphi_2\rangle\langle \varphi_3| - |\varphi_3\rangle\langle \varphi_2|),\\
&\Lambda_7=|\varphi_1\rangle\langle \varphi_1| - |\varphi_2\rangle\langle \varphi_2|,\quad \Lambda_8=(|\varphi_1\rangle\langle \varphi_1| + |\varphi_2\rangle\langle \varphi_2| -2 |\varphi_3\rangle\langle \varphi_3|)/\sqrt{3},
\end{eqnarray}
by performing a calculation similar to that of Sec.~\ref{The general situation}, one obtains for $P_\psi$ the following representative vector:
\begin{eqnarray}
&{\bf n}=\sqrt{3}(a_1a_2\cos\alpha, a_1a_2\sin\alpha, a_1a_3\cos\delta, a_1a_3\sin\delta, a_2a_3\cos \gamma, a_2a_3\sin \gamma, {a_1^2 -a_2^2\over 2}, {a_1^2 +a_2^2 -2a_3^2\over 2\sqrt{3}})^\top,
\label{3statessuperposition}
\end{eqnarray}
whereas the vectors ${\bf n}_i$ and ${\bf m}_i$, representative of the states $P_i$ and $F_i$, respectively, $i=1,2,3$, define the vertices of  two triangular measurement 2-simplexes, given by: 
\begin{eqnarray}
&{\bf n}_1= (0, 0, 0, 0, 0, 0, {\sqrt{3} \over 2}, {1 \over 2})^\top,\quad {\bf n}_2= (0, 0, 0, 0, 0, 0, -{\sqrt{3} \over 2}, {1 \over 2})^\top, \quad {\bf n}_3= (0, 0, 0, 0, 0, 0, 0, -1)^\top, \\
&{\bf m}_{1}= {1\over\sqrt{3}}(1, 0, 1, 0, 1, 0, 0, 0)^\top,\quad {\bf m}_{2}={-1\over 2\sqrt{3}}(1, -\sqrt{3}, 1,\sqrt{3},1,-\sqrt{3}, 0,0)^\top,\nonumber\\
&{\bf m}_{3}={-1\over 2\sqrt{3}}(1, \sqrt{3}, 1,-\sqrt{3}, 1, \sqrt{3}, 0, 0)^\top.
\label{3basevectors}
\end{eqnarray}

According to (\ref{trans-general}), the orthogonal projection of ${\bf n}$ can be written as: 
\begin{equation}
{\bf n}^\parallel = \sum_{i=1}^3 {1\over 3}(1 + 3I_i)\, {\bf m}_{i} = \sum_{i=1}^3 I_i\, {\bf m}_{i},
\end{equation}
where we have used $\sum_{i=1}^3 {\bf m}_{i}=0$. This means that the no-interference condition corresponds to a situation such that the representative vector ${\bf n}$ orthogonally projects exactly at the center of the extended Bloch sphere. For instance, for the special case $a_1=a_2=a_3={1\over\sqrt{3}}$, this happens for the values: $(\alpha,\delta)= (0,{2\pi\over 3}), (0,-{2\pi\over 3}),(-{2\pi\over 3},0),({2\pi\over 3},0),(-{2\pi\over 3},-{2\pi\over 3}),({2\pi\over 3},{2\pi\over 3})$. So, different from the situation of a two-state superposition, where the interferences originated from the movement of ${\bf n}$ along simple circles of latitudes, when three states are superposed, the movement of ${\bf n}$, and consequently its projections onto the measurement simplexes, become much more involved. 

Indeed, considering the explicit form (\ref{3statessuperposition}), we see that when the coefficients $a_1$ and $a_2$ are kept fixed, and the relative phases are varied, the movement of ${\bf n}$ now results from the combination of three different circular movements, of radius $a_1a_2$, $a_1a_3$ and $a_2a_3$, respectively, belonging to three different non-intersecting sub-planes of $B_1(\real^8)$. The first two circular movements are associated with the relative phases $\alpha$ and $\delta$, which can be varied one independently from the other, whereas the movement on the third circle, associated with the relative phase $\gamma=\delta-\alpha$, depends on the other two. More generally, when the four independent parameters $a_1$, $a_2$, $\alpha$ and $\delta$ are varied, ${\bf n}$ travels on a 7-dimensional surface defining a convex region inside $B_1(\real^8)$, which contains all the physical states. The shape of this region is rather complex, and some of its geometrical characteristics have been explored in \cite{Kimura2005, Kimura2003,AertsSassoli2014c}. 

Let us consider more explicitly a measurement with the three outcomes $P_i$, and associated probabilities $a_i^2$, described by the unit vectors ${\bf n}_{i}$, $i=1,2,3$. To reach the measurement triangular membrane, the point particle representative of the state has to follow the orthogonal path: 
\begin{eqnarray}
&{\bf r}_\tau=\sqrt{3} ((1-\tau)a_1a_2\cos\alpha, (1-\tau)a_1a_2\sin\alpha, (1-\tau)a_1a_3\cos\delta, (1-\tau)a_1a_3\sin\delta, \nonumber\\
&(1-\tau)a_2a_3\cos \gamma, (1-\tau)a_2a_3\sin \gamma, {a_1^2 -a_2^2\over 2}, {a_1^2 +a_2^2 -2a_3^2\over 2\sqrt{3}}),
\label{3statessuperposition-path}
\end{eqnarray}
with the parameter $\tau$ going from $0$ to $1$. Different  the two-state superposition, where by varying the only relative phase we obtained a region of states having the shape of a disk (see Fig.~\ref{figure7}), we now have three different disks, located on non-intersecting sub-planes, which generate a 6-dimensional region characterizing the states that all give the same transition probabilities. However, the shape of this 6-dimensional region is not simply given by the Cartesian product of these three orthogonal disks, as only two of them are associated with independent variables. 

Thus, the general explanation of interference phenomena in terms of the deviations of the representative vector ${\bf n}$ with respect to the non-interference orientations, which orthogonally project onto the center of the sphere, remains valid also for superpositions of three orthogonal states, and the present discussion clearly generalizes to more general $n$-state superpositions, with $n\leq N$. However, when $n>2$, a simple geometric characterization of the subsets of superposition states cannot be given anymore, because of the general complexity of the convex region containing the states, in the generalized Bloch sphere.

\section{The description of multipartite systems}
\label{Multipartite systems}

In the present and following sections we continue our investigation of the non-product states, by introducing a new determination of the generators of $SU(N)$, more suitable to discuss multipartite systems (joint entities) from the viewpoint of the sub-systems, which in a sense is complementary to that presented in Sec.~\ref{Superposition states} and \ref{Entanglement}. This new determination uses the tensor product as a means to construct higher order generators from lower order ones, and will allow us to understand how composite systems can be defined directly within the extended Bloch representation, and gain a deeper understanding of the nature of  separable and entangled states.

\subsection{Tensorial determination of the $SU(N)$ generators}
\label{tensorial basis}

We consider a Hilbert space  ${\cal H}={\cal H}_{A_1}\otimes {\cal H}_{A_2} \otimes \cdots \otimes {\cal H}_{A_n}$, with ${\cal H}_{A_i}=\compl^{N_i}$, $i=1,\dots,n$, and ${\cal H}=\compl^{N}$, with $N=N_1\cdots N_n$. We denote $\mathbb{I}^{A_i}$ the $N_i\times N_i$ identity matrix, and  ${\Lambda^{A_i}_j}$, $j=1,\dots,N_i^2-1$, a given determination of the generators of $SU(N_i)$, $i=1,\dots,n$. We assume that the normalization is the standard one: ${\rm Tr}\, ({\Lambda^{A_i}_j})^2=2$, for all  $j=1,\dots,N_i^2-1$, and $i=1,\dots,n$.

We want to construct the generators of $SU(N)$ as tensor products of the generators of $SU(N_i)$, $i=1,\dots,n$. For this, we also introduce the matrices: ${\Lambda^{A_i}_0}\equiv \sqrt{2\over N_i}\,\mathbb{I}^{A_i}$, whose normalization is: ${\rm Tr} \, ({\Lambda^{A_i}_0})^2 =2$. Then, we have $n$ bases: $\{\Lambda^{A_i}_0,\Lambda^{A_i}_1,\dots, \Lambda^{A_i}_{N_i^2-1}\}$, $i=1,\dots,n$, each one made of $N_i^2$ orthogonal $N_i\times N_i$ matrices, all having the same dimension-independent normalization, and we can define the $N^2=N_1^2\cdots N_n^2$ matrices: 
\begin{equation}
\Lambda_{(j_1,\dots,j_n)} \equiv  2^{1-n\over 2}\, \Lambda^{A_1}_{j_1}\otimes\cdots\otimes  \Lambda^{A_n}_{j_n},\quad  j_i=0,\dots,N_i^2-1,\quad i=1,\dots,n.
\label{tensor-generators}
\end{equation}
We observe that: 
\begin{eqnarray}
{\rm Tr}\, \Lambda_{(j_1,\dots,j_n)}\Lambda_{(k_1,\dots,k_n)}&=& 2^{1-n}\, {\rm Tr}\, \Lambda^{A_1}_{j_1} \Lambda^{A_1}_{k_1}\otimes \cdots \otimes \Lambda^{A_n}_{j_n} \Lambda^{A_n}_{k_n} =2^{1-n}\, {\rm Tr}\, \Lambda^{A_1}_{j_1} \Lambda^{A_1}_{k_1}\otimes \cdots \otimes {\rm Tr}\, \Lambda^{A_n}_{j_n} \Lambda^{A_n}_{k_n}\nonumber \\
&=& 2\,  \delta_{j_1,k_1}\dots \delta_{j_n,k_n}=2\, \delta_{(j_1,\dots,j_n),(k_1,\dots,k_n)},
\label{tensor-generatorsN}
\end{eqnarray}
where we have used the fact that the trace of a tensor product is the product of the traces. Thus, the $\Lambda_{(j_1,\dots,j_n)}$ generate a maximal family of orthogonal Hermitian matrices, all having the right normalization. Apart $\Lambda_{(0,\dots,0)}= \sqrt{2\over N}\mathbb{I}$, they are  also traceless matrices, and therefore constitute a specific determination of the $SU(N)$ generators. Let us write them explicitly in the case $N_1=N_2 = n=2$. We have: $\Lambda_{(0,0)}={1\over\sqrt{2}}\mathbb{I}^{A_1}\otimes\mathbb{I}^{A_2}= {1\over\sqrt{2}}\mathbb{I}$, and the 15 generators are: 
\begin{equation}
\Lambda_{(0,1)} = {1\over\sqrt{2}}\begin{bmatrix} 0 & 1 & 0 &0  \\ 1 & 0 & 0 & 0 \\ 0 & 0 & 0 & 1 \\0 & 0 & 1 & 0 \end{bmatrix},\quad \Lambda_{(0,2)} = {1\over\sqrt{2}}\begin{bmatrix} 0 & -i & 0 &0  \\ i & 0 & 0 & 0 \\ 0 & 0 & 0 & -i \\0 & 0 & i & 0 \end{bmatrix}, \quad \Lambda_{(0,3)} ={1\over\sqrt{2}} \begin{bmatrix} 1 & 0 & 0 &0  \\ 0 & -1 & 0 & 0 \\ 0 & 0 & 1 & 0 \\0 & 0 & 0 & -1 \end{bmatrix},\nonumber\\
\end{equation} 
\begin{equation}
\Lambda_{(1,0)} = {1\over\sqrt{2}}\begin{bmatrix} 0 & 0 & 1 &0  \\0 & 0 & 0 & 1 \\ 1 & 0 & 0 & 0 \\0 & 1 & 0 & 0 \end{bmatrix},\quad \Lambda_{(2,0)} = {1\over\sqrt{2}}\begin{bmatrix} 0 & 0 & -i &0  \\0 & 0 & 0 & -i \\ i & 0 & 0 & 0 \\0 & i & 0 & 0 \end{bmatrix}, \quad \Lambda_{(3,0)} ={1\over\sqrt{2}} \begin{bmatrix} 1 & 0 & 0 &0  \\0 & 1 & 0 & 0 \\ 0 & 0 & -1 & 0 \\0 & 0 & 0 & -1\end{bmatrix},\nonumber
\end{equation} 
\begin{equation}
\Lambda_{(1,1)} = {1\over\sqrt{2}}\begin{bmatrix} 0 & 0 & 0 &1  \\0 & 0 & 1 & 0 \\ 0 & 1 & 0 & 0 \\1 & 0 & 0 & 0 \end{bmatrix},\quad \Lambda_{(1,2)} = {1\over\sqrt{2}}\begin{bmatrix} 0 & 0 & 0 &-i  \\0 & 0 & i & 0 \\ 0 & -i & 0 & 0 \\i & 0 & 0 & 0 \end{bmatrix}, \quad \Lambda_{(1,3)} ={1\over\sqrt{2}} \begin{bmatrix} 0 & 0 & 1 &0  \\0 & 0 & 0 & -1 \\ 1 & 0 & 0 & 0 \\0 & -1 & 0 & 0\end{bmatrix},\nonumber\\
\end{equation}
\begin{equation}
\Lambda_{(2,1)} = {1\over\sqrt{2}}\begin{bmatrix} 0 & 0 & 0 &-i  \\0 & 0 & -i & 0 \\ 0 & i & 0 & 0 \\i & 0 & 0 & 0 \end{bmatrix},\quad \Lambda_{(2,2)} = {1\over\sqrt{2}}\begin{bmatrix} 0 & 0 & 0 &-1  \\0 & 0 & 1 & 0 \\ 0 & 1 & 0 & 0 \\-1 & 0 & 0 & 0 \end{bmatrix}, \quad \Lambda_{(2,3)} ={1\over\sqrt{2}} \begin{bmatrix} 0 & 0 & -i &0  \\0 & 0 & 0 & i \\ i & 0 & 0 & 0 \\0 & -i & 0 & 0\end{bmatrix},\nonumber
\end{equation} 
\begin{equation}
\Lambda_{(3,1)} = {1\over\sqrt{2}}\begin{bmatrix} 0 & 1 & 0 &0  \\1 & 0 & 0 & 0 \\ 0 & 0 & 0 & -1 \\0 & 0 & -1 & 0 \end{bmatrix},\quad \Lambda_{(3,2)} = {1\over\sqrt{2}}\begin{bmatrix} 0 & -i & 0 &0  \\i & 0 & 0 & 0 \\ 0 & 0 & 0 & i \\0 & 0 & -i & 0 \end{bmatrix}, \quad \Lambda_{(3,3)} ={1\over\sqrt{2}} \begin{bmatrix} 1 & 0 & 0 &0  \\0 & -1 & 0 & 0 \\ 0 & 0 & -1 & 0 \\0 & 0 & 0 & 1\end{bmatrix},\label{15gen}
\end{equation} 
which we can compare to the generators (\ref{F15}), obtained from the (non-tensorial) determination (\ref{rgeneratorsN}).

\subsection{Product states}

We want now to exploit the above tensorial basis to characterize the product states directly in terms of their representative vectors within the Bloch sphere. We recall that in quantum mechanics a  product state is the state of a multipartite system that can be expressed in the form: 
\begin{equation}
D\equiv D^{A_1}\otimes D^{A_2}\otimes\cdots \otimes D^{A_{n-1}}\otimes D^{A_n}.
\label{product state}
\end{equation}
We write $D\equiv D({\bf r})={1\over N}(\mathbb{I} + c_N \,{\bf r}\cdot {\bf \Lambda})$, and $D^{A_i}\equiv D({\bf r}^{A_i})={1\over N_i}(\mathbb{I} + c_{N_i} \,{\bf r}^{A_i}\cdot {\bf \Lambda}^{A_i})$, $i=1,\dots,n$. To obtain a more compact notation, we define the $N^2$-dimensional  vector $\mathcal{R}\equiv(r_0,{\bf r})^\top$, where $r_0\equiv{1\over\sqrt{N-1}}$. Similarly, we define the $N_i^2$-dimensional vectors $\mathcal{R}^{A_i}\equiv(r_0^{A_i},{\bf r}^{A_i})^\top$, where $r_0^{A_i}\equiv{1\over\sqrt{N_i-1}}$, $i=1,\dots, n$. We also define the $N^2$-dimensional matricial vector $\mathcal{L}=({\Lambda}_0,{\bf \Lambda})^\top$, where $\Lambda_0\equiv\Lambda_{(0,\dots,0)}= \sqrt{2\over N}\,\mathbb{I}$, and the $N_i^2$-dimensional matricial vectors $\mathcal{L}^{A_i}=({\Lambda}_0^{A_i},{\bf \Lambda}^{A_i})^\top$, where ${\Lambda^{A_i}_0}\equiv \sqrt{2\over N_i}\,\mathbb{I}^{A_i}$, $i=1,\dots,n$, so that we can write:  
\begin{equation}
D\equiv D(\mathcal{R})= {c_N\over N}\, \mathcal{R}\cdot \mathcal{L},
\label{compact}
\end{equation}
and similarly:
\begin{equation}
D^{A_i}\equiv D(\mathcal{R}^{A_i})= {c_{N_i}\over N_i}\, \mathcal{R}^{A_i}\cdot \mathcal{L}^{A_i},\quad i=1,\dots,n.
\label{compact}
\end{equation}

If  $D$ is a product state, we have: 
\begin{equation}
 \mathcal{R}\cdot \mathcal{L} = 2^{1-n\over 2} d_{N_1,\dots,N_n}\,\mathcal{R}^{A_1}\cdot \mathcal{L}^{A_1}\otimes \cdots \otimes \mathcal{R}^{A_n}\cdot \mathcal{L}^{A_n},\quad d_{N_1,\dots,N_n}\equiv \sqrt{(N_1-1)\cdots (N_n-1)\over N-1}, 
\end{equation} 
or more explicitely:
\begin{equation}
\mathcal{R}\cdot \mathcal{L} = d_{N_1,\dots,N_n} \sum_{j_1=0}^{N_1^2-1} \cdots \sum_{j_n=0}^{N_n^2-1}\mathcal{R}^{A_1}_{j_1}\cdots \mathcal{R}^{A_n}_{j_n} \left(2^{1-n\over 2} \, \Lambda^{A_1}_{j_1}\otimes\cdots\otimes  \Lambda^{A_n}_{j_n}\right).
\end{equation} 
In view of  (\ref{tensor-generators}), we thus obtain:
\begin{equation}
\mathcal{R}_{(j_1,\dots,j_n)}=d_{N_1,\dots,N_n}\, \mathcal{R}^{A_1}_{j_1}\cdots  \mathcal{R}^{A_n}_{j_n}, \quad j_i=0,\dots N_i, \quad i=1,\dots, n.
\label{product state-Bloch}
\end{equation} 

Equality (\ref{product state-Bloch}) expresses what a product state is, in the Blochean formalism. Let us express it using the Bloch vectors ${\bf r}\in B_1(\real^{N^2-1})$ and ${\bf r}^{A_i}\in B_1(\real^{N_i^2-1})$, $i=1,\dots, n$. Defining $d_{N_i}\equiv \sqrt{N_i-1\over N-1}$, $i=1,\dots,n$, it follows from (\ref{product state-Bloch}) that the first $\sum_{i=1}^n (N_i^2-1)$ components of ${\bf r}$ are:
\begin{eqnarray}
&&r_{(j,0,\dots, 0)}=d_{N_1}\, r_{j}^{A_1},\quad  j=1,\dots,N_1^2-1,\\
&&r_{(0,j,0,\dots, 0)}=d_{N_2}\, r_{j}^{A_2},\quad  j=1,\dots,N_2^2-1,\\
&&\quad\vdots\nonumber\\
&&r_{(0,\dots, 0,j)}=d_{N_n}\, r_{j}^{A_n},\quad j=1,\dots,N_n^2-1.
\label{product state-Bloch1}
\end{eqnarray} 
These correspond to the one-entity contributions. Then, we have the two-entity contributions, which are of the form: 
\begin{equation}
r_{(0,\dots, j_k,\dots, j_\ell, \dots,0)}=d_{N_k, N_\ell}\, r_{j_k}^{A_k}r_{j_\ell}^{A_\ell},\quad j_k=1,\dots,N_k^2-1, \quad  j_\ell=1,\dots,N_\ell^2-1,
\label{product state-Bloch two state}
\end{equation} 
where we have defined $d_{N_k, N_\ell}\equiv \sqrt{(N_k-1)(N_\ell-1)\over N-1}$. Then, we have the three-entity contributions, and so on, up to the $n$-entity contribution: 
\begin{equation}
r_{(j_1,j_2,\dots, j_n)}= d_{N_1,\dots,N_n} \, r_{j_1}^{A_1}r_{j_2}^{A_2}\cdots r_{j_n}^{A_n},\quad j_1=1,\dots,N_1^2-1,\,\, \dots,\,\,  j_n=1,\dots,N_n^2-1.
\label{product state-Bloch n state}
\end{equation}

Bringing the above into a more compact notation, we observe that the vector ${\bf r}$ can be written as the direct sum:
\begin{equation}
{\bf r}= \underbrace{\bigoplus_{(i)} d_{N_i}{\bf r}^{A_i}}_{\rm 1-sector} \underbrace{\bigoplus_{(i,j)}  d_{N_i, N_j} {\bf r}^{A_i A_j}}_{\rm 2-sector} \underbrace{\bigoplus_{(i,j,k)}  d_{N_i, N_j,N_k} {\bf r}^{A_i A_jA_k}}_{\rm 3-sector}\, \cdots\,  \underbrace{\bigoplus_{\phantom{(i,j)}} d_{N_1,\dots,N_n} {\bf r}^{A_1\cdots A_n}}_{n-{\rm sector}},
\label{direct sum-n}
\end{equation} 
where the first direct sum is over the ${\binom n1}=n$ ways to chose one index among $n$ indices, and corresponds to the one-entity contributions, the second direct sum is over  the $\binom n2$  ways to chose two indices among $n$ indices, and corresponds to the two-entity contributions, the third direct sum is over the $\binom n3$  ways to chose three indices among $n$ indices, and corresponds to the three-entity contributions, and so on, up to the last term, which corresponds to the ${\binom nn} =1$ way to chose $n$ indices among $n$ indices, and corresponds to the $n$-entity contribution. The vectors ${\bf r}^{A_i}$ belong to the spheres $B_{1}(\real^{N_i^2-1})$; the vectors ${\bf r}^{A_i A_j}$, whose components are:
\begin{equation}
r^{A_i A_j}_{(k,\ell)}= r_{k}^{A_i}r_{\ell}^{A_j},\quad k=1,\dots,N_i^2-1, \quad \ell =1,\dots,N_j^2-1,
\end{equation}
belong to the spheres $B_{1}(\real^{(N_i^2-1)(N_j^2-1)})$; the vectors ${\bf r}^{A_i A_jA_k}$, whose components are:
\begin{equation}
r^{A_i A_jA_k}_{(k,\ell,m)}= r_{k}^{A_i}r_{\ell}^{A_j}r_{m}^{A_k},\quad k=1,\dots,N_i^2-1, \quad \ell =1,\dots,N_j^2-1, \quad  m =1,\dots,N_k^2-1,
\end{equation}
belong to the spheres $B_{1}(\real^{(N_i^2-1)(N_j^2-1) (N_k^2-1)})$, and so on, with the last vector ${\bf r}^{A_1\cdots A_n}$ in the direct sum (\ref{direct sum-n}), with components: 
\begin{equation}
r^{A_1\cdots A_n}_{(j_1,\dots, j_n)}= r_{j_1}^{A_1}\cdots r_{j_n}^{A_n},\quad j_i=1,\dots,N_i^2-1, \quad i =1,\dots,n,
\end{equation}
belonging to the sphere $B_{1}(\real^{\prod_{i=1}^n (N_i^2-1)})$. 

In other terms, a product state of a multipartite system, made of $n$ entities, can be represented in the Bloch sphere by a vector ${\bf r}$ which is the (dimensionally weighted) direct sum of vectors describing the different sub-entities, belonging to their respective Bloch spheres, plus vectors describing the 2-entity correlations, the 3-entity correlations, and so on, up to the $n$-entity correlations. What is important to observe, however, is that all the one-entity vectors are independent from one another, and that all the vectors in the other sectors are fully determined by them, as one would expect from a product state, which defines a condition where all the entities are separated, i.e., where the whole is equal to the sum of its parts.

\subsection{Product measurements in bipartite systems}
\label{Product measurements}

To keep the discussion simple and avoid a too heavy notation, in the following we will only consider joint entities formed by two sub-entities (bipartite systems). Then, we have: $D\equiv D^{A}\otimes D^{B}$, and (\ref{direct sum-n}) simply becomes: 
\begin{equation}
{\bf r}= d_{N_A}{\bf r}^{A}\oplus d_{N_B}{\bf r}^{B}\oplus d_{N_A,N_B}{\bf r}^{AB},
\label{tripartite-vector}
\end{equation}
where ${\bf r}^{A}\in B_{1}(\real^{N_A^2-1})$, ${\bf r}^{B}\in B_{1}(\real^{N_B^2-1})$, and ${\bf r}^{AB}\in B_{1}(\real^{(N_A^2-1)(N_B^2+1)})$. In other terms, a bipartite system in a product state can be represented in the extended Bloch model (if the tensorial basis of generators is used) by the tripartite vector (\ref{tripartite-vector}), where the third vector in the direct sum is fully determined by the first two, in accordance with the concept of a product state, where the knowledge of the two one-entity states (the points in the sub-spheres $B_{1}(\real^{N_A^2-1})$ and $B_{1}(\real^{N_B^2-1})$), fully determines the point in  $B_{1}(\real^{(N_A^2-1)(N_2^2-1)})$, and consequently the point in $B_1(\real^{N^2-1})$, i.e., the state of the total system.

A product measurement is described by an observable of the form: $O=O^{A}\otimes O^{B}$. Introducing the spectral decompositions: $O^{A}=\sum_{i=1}^{N_A} o_i^{A} P_i^{A}$ and $O^{B}=\sum_{j=1}^{N_B} o_j^{B} P_j^{B}$, we can write: $O=\sum_{i=1}^{N_A}\sum_{j=1}^{N_B}o_i^{A}o_j^{B} P_i^{A}\otimes P_j^{B}$. If $D({\bf r})$ is a product state, i.e., $D({\bf r}) = D({\bf r}^{A})\otimes D({\bf r}^{B})$, the probabilities for the transitions $D({\bf r})\to P_i^{A}\otimes P_j^{B}$ are given by:
\begin{eqnarray}
{\cal P}(D({\bf r})\to P_i^{A}\otimes P_j^{B})&=&{\rm Tr}\, (D({\bf r}^{A})\otimes D({\bf r}^{B}))(P_i^{A}\otimes P_j^{B})= {\rm Tr}\, D({\bf r}^{A})P_i^{A}\otimes D({\bf r}^{B})P_j^{B}\nonumber\\
&=&{\rm Tr}\, D({\bf r}^{A})P_i^{A}\, {\rm Tr}\, D({\bf r}^{B})P_j^{B}= {\cal P}(D({\bf r}^{A})\to P_i^{A})\, {\cal P}(D({\bf r}^{B})\to P_j^{B}).
\label{productprobability}
\end{eqnarray}
In other terms, the transition probabilities for a bipartite system subjected to a product measurement  are simply the products of the transition probabilities for the sub-systems, subjected to the corresponding one-entity measurements. 

Let us explain how this factorization of the probabilities can be understood when we represent the product measurement within the extended Bloch sphere. We know that the representative vector ${\bf r}$ is of the tripartite form (\ref{tripartite-vector}). The same is obviously true for the vectors representative of the eigenstates, associated with the vertices of the measurement $(N-1)$-simplex ($N=N_AN_B$), as they are also product states: 
\begin{equation}
{\bf n}_{ij}= d_{N_A}{\bf n}_{i}^{A}\oplus d_{N_B}{\bf n}_j^{B}\oplus d_{N_A,N_B}{\bf n}_{ij}^{AB},\quad i=1,\dots,N_A,\quad j=1,\dots,N_B,
\label{tripartite-vector-eigenstate}
\end{equation}
where the ${\bf n}_{i}^{A}$, $i=1,\dots,N_A$, are the vertex vectors of the $(N_A-1)$-simplex associated with  $O^{A}$, and the  ${\bf n}_{j}^{B}$, $j=1,\dots,N_B$, are the vertex vectors of the $(N_B-1)$-simplex associated with  $O^{B}$. This means that the $(N-1)$-simplex associated with the vertex vectors ${\bf n}_{ij}$, when projected onto the sub-ball  $B_{1}(\real^{N_A^2-1})$, associated with the first $N_A^2-1$ components of the vector (the first vector in the direct sum), will become a $(N_A-1)$-simplex, of a rescaled size (because of the dimensional factor $d_{N_A}$), and the same is true when the ${\bf n}_{ij}$ are projected onto the sub-ball  $B_{1}(\real^{N_B^2-1})$.

This means that in the description of the measurement we can consider two equivalent processes. The first one is the process associated with the full $(N-1)$-simplex $\triangle_{N-1}$, with the point particle associated with ${\bf r}$ orthogonally ``falling'' onto it, and the associated membrane disintegrating and collapsing toward one of the outcome eigenstates ${\bf n}_{ij}$. In accordance with the general theory (see Sec.~\ref{General measurement} and \cite{AertsSassoli2014c}), we know that such dynamics will produce exactly the probabilities (\ref{productprobability}). 

 The second process is to consider the measurement from the perspective of the two sub-entities. Indeed, when  the point particle associated with ${\bf r}$ orthogonally ``falls'' onto $\triangle_{N-1}$, it follows from (\ref{tripartite-vector}) and (\ref{tripartite-vector-eigenstate}) that the point particles associated with ${\bf r}^{A}$ and ${\bf r}^{B}$ will also orthogonally ``fall'' onto their respective sub-simplexes $\triangle_{N_A-1}$ and $\triangle_{N_B-1}$, respectively. Also, considering that ${\bf r}^{A}$ and ${\bf r}^{B}$ are independent vectors within (\ref{tripartite-vector}), in the sense that they can be varied independently from one another (which is not the case for the third ``correlation'' vector ${\bf r}^{AB}$, whose components depend on those of ${\bf r}^{A}$ and ${\bf r}^{B}$), we can consider two separate membranes associated with $\triangle_{N_A-1}$ and $\triangle_{N_B-1}$, and their collapses, in whatever order. 
 
Obviously, this double process, of the two sub-membranes working one independently from the other, can perfectly mimic the functioning of the full $(N-1)$-dimensional membrane associated with $\triangle_{N-1}$, and will produce exactly the same statistics of outcomes. In other terms, when product measurements are performed on product states, the membrane mechanism naturally decomposes into two simpler sub-membrane mechanisms, associated with the one-entity measurements. And of course, the same holds true for multipartite product states and measurements.

\subsection{Separable states (are they really?)} 

A less restrictive condition than that of product states is that of \emph{separable states}, that is, states that are interpreted, in the standard formalism, as statistical mixtures of product states: 
\begin{equation}
D({\bf r})=\sum_\mu p_\mu\, D({\bf r}^{A}_{\mu})\otimes D({\bf r}^{B}_{\mu}),
\end{equation}
where the positive numbers $p_\mu$ obey: $\sum_\mu p_\mu =1$. Instead of (\ref{tripartite-vector}), the representative vector in the $(N^2-1)$-dimensional Bloch sphere is now given by: 
\begin{equation}
{\bf r}=\sum_\mu p_\mu\,  (d_{N_A}{\bf r}^{A}_\mu\oplus d_{N_B}{\bf r}^{B}_\mu\oplus d_{N_A,N_B}{\bf r}^{AB}_\mu) = 
d_{N_A}\overline{\bf r}^{A}\oplus d_{N_B}\overline{\bf r}^{B}\oplus d_{N_A,N_B}\overline{\bf r}^{AB}
\label{trip-sep}
\end{equation}
where we have defined: $\overline{\bf r}^{A}\equiv \sum_\mu p_\mu{\bf r}^{A}_\mu$, $\overline{\bf r}^{B}\equiv \sum_\mu p_\mu{\bf r}^{B}_\mu$, and $\overline{\bf r}^{AB}\equiv \sum_\mu p_\mu{\bf r}^{AB}_\mu$. So, we still have a tripartite representation, but this time not only $\overline{\bf r}^{A}$ is not independent of $\overline{\bf r}^{B}$, as their components both include the parameters $p_\mu$, but also the components of the correlation vector $\overline{\bf r}^{AB}$ are not anymore a simple product of the components of the sub-entity vectors, i.e., $\overline{r}^{AB}_{(i,j)}\neq \overline{r}^{B}_i\overline{r}^{B}_j$.

In other terms, for a separable but non-product state, the tripartite decomposition (\ref{trip-sep}) is more involved, as one cannot anymore deduce the coordinates of the point in $B_{1}(\real^{(N_A^2-1)(N_2^2-1)})$ from the coordinates of the points in $B_{1}(\real^{N_A^2-1})$ and $B_{1}(\real^{N_B^2-1})$, nor one can move the point in $B_{1}(\real^{N_A^2-1})$ independently of the point in $B_{1}(\real^{N_B^2-1})$. This means that when we perform a product measurement, the process will not be governed anymore by separate processes associated with the two sub-entities, i.e., the collapsing mechanism of the full membrane $\triangle_{N-1}$ cannot be decomposed into two separate and independent collapsing mechanisms, associated with the lower dimensional membranes $\triangle_{N_A-1}$ and $\triangle_{N_B-1}$.

Indeed, if we consider two separate collapses, their transition probabilities are given by ${\cal P}(D(\overline{\bf r}^{A})\to P_i^{A})=\sum_\mu p_\mu {\cal P}(D({\bf r}_\mu^{A}\to P_i^{A})$ and ${\cal P}(D(\overline{\bf r}^{B})\to P_j^{B})=\sum_\mu p_\mu {\cal P}(D({\bf r}_\mu^{B}\to P_j^{B})$, so that the joint probability for both transitions would be the product:  $\sum_\mu \sum_{\nu} p_\mu p_{\nu} {\cal P}(D({\bf r}_\mu^{A}\to P_i^{A}){\cal P}(D({\bf r}_{\nu}^{B}\to P_j^{B})$, which evidently cannot be equal to the correct quantum transition probability: 
\begin{equation}
{\cal P}(D({\bf r})\to P_i^{A}\otimes P_j^{B})=\sum_\mu p_\mu  {\cal P}(D({\bf r}^{A}_\mu\to P_i^{A})\, {\cal P}(D({\bf r}^{B}_\mu\to P_j^{B}),
\end{equation}
derived from the mechanism of the full $\triangle_{N-1}$ membrane. The reason for this ``non-separability of the separable states'' is of course that in the extended Bloch representation separable states are not interpreted as classical mixtures, but as pure states, so that their measurements cannot be reduced to mixtures of measurements performed on different product states. Note that in quantum information theory the situation of separable states that can produce genuine quantum correlations, and therefore do not describe an actual separation of the sub-entities, is known as \emph{quantum discord}~\cite{OllivierZurek2001,HendersonVedral2001}.

\subsection{Entangled states}

We consider now a non-product, non-separable state $|\psi\rangle = a_1\, e^{i\alpha_1}|\psi^A\rangle\otimes |\phi^B\rangle + a_2\, e^{i\alpha_2}|\phi^A\rangle\otimes |\psi^B\rangle$. Defining:
\begin{eqnarray}
&&D({\bf r})=|\psi\rangle\langle \psi| = {1\over N}\left(\mathbb{I} +c_N\, {\bf r}\cdot\mbox{\boldmath$\Lambda$}\right), \quad  D({\bf r}^A)=|\psi^A\rangle\langle \psi^A|={1\over N_A}\left(\mathbb{I} +c_{N_A}\, {\bf r}^A\cdot\mbox{\boldmath$\Lambda$}^A\right),  \\
&&D({\bf s}^A)=|\phi^A\rangle\langle \phi^A|={1\over N_A}\left(\mathbb{I} +c_{N_A}\, {\bf s}^A\cdot\mbox{\boldmath$\Lambda$}^A\right),\quad 
D({\bf r}^B)=|\psi^B\rangle\langle \psi^B|={1\over N_B}\left(\mathbb{I} +c_{N_B}\, {\bf r}^B\cdot\mbox{\boldmath$\Lambda$}^B\right),\\
&&D({\bf s}^B)=|\phi^B\rangle\langle \phi^B|={1\over N_B}\left(\mathbb{I} +c_{N_B}\, {\bf s}^B\cdot\mbox{\boldmath$\Lambda$}^B\right),
\label{entanglement-operators}
\end{eqnarray}
 we can write: 
\begin{equation}
D({\bf r})=a_1^2\, D({\bf r}^{A})\otimes D({\bf s}^{B}) + a_2^2\, D({\bf s}^{A})\otimes D({\bf r}^{B}) + I,
\label{entanglement-operator}
\end{equation}
where the interference contribution is given by:
\begin{eqnarray}
I&=&a_1a_2\, e^{i(\alpha_1-\alpha_2)}\left(|\psi^A\rangle \otimes  |\phi^{B}\rangle\right) \left(\langle \phi^{A}|\otimes \langle \psi^{B}|\right) + {\rm c.c.}\nonumber\\
&=& a_1a_2\, e^{-i\alpha}|\psi^A\rangle\langle \phi^{A}|\otimes |\phi^{B}\rangle\langle \psi^{B}| + {\rm c.c.},\quad \alpha \equiv \alpha_2-\alpha_1,
\label{interference-contribution}
\end{eqnarray}
and of course, the Hermitian traceless matrix $I$ can also be associated with a representative vector ${\bf r}^{\rm int}\in \real^{N^2-1}$, defined by: $I \equiv I({\bf r}^{\rm int})={c_N\over N}  {\bf r}^{\rm int}\cdot\mbox{\boldmath$\Lambda$}$.

Let us calculate explicitly the components of ${\bf r}$ in terms of the components of ${\bf r}^{A}$, ${\bf s}^{A}$, ${\bf r}^{B}$, ${\bf s}^{B}$, ${\bf r}^{\rm int}$, of the weights $a_1$ and $a_2$, and of the relative phase $\alpha$. For this, we have to make a choice for the $N_A^2-1$ generators $\Lambda_i^A$ of $SU(N_A)$, and for the $N_B^2-1$ generators $\Lambda_j^B$ of $SU(N_B)$. The natural one is to consider that the first two generators are: 
\begin{equation}
\Lambda^{A}_1 = |\psi^{A}\rangle\langle \phi^{A}| + |\phi^{A}\rangle\langle \psi^{A}|, \quad 
\Lambda^{A}_2 =-i(|\psi^{A}\rangle\langle \phi^{A}| - |\phi^{A}\rangle\langle \psi^{A}|),
\label{choicegenerators1}
\end{equation}
\begin{equation}
\Lambda^{B}_1 = |\psi^{B}\rangle\langle \phi^{B}| + |\phi^{B}\rangle\langle \psi^{B}|, \quad 
\Lambda^{B}_2 =-i(|\psi^{B}\rangle\langle \phi^{B}| - |\phi^{B}\rangle\langle \psi^{B}|).
\label{choicegenerators2}
\end{equation}
For the following $N_A-1$ generators of $SU(N_A)$, we chose them to be: $\Lambda_{n+2}^{A}= W_n^{A}$,  $n=1,\dots,N_A-1$. Similarly, for the following $N_B-1$ generators of $SU(N_B)$, we chose them to be: $\Lambda_{n+2}^{B}= W_n^{B}$,  $n=1,\dots,N_B-1$,  where the $W_n^{A}$ and $W_n^{B}$ are the diagonal matrices defined in (\ref{W}). The remaining generators can be taken for instance according to (\ref{rgeneratorsN}), for a given choice of the bases, but hereafter we will not need to specify them.  

In view of (\ref{components}), the components of ${\bf r}^{A}$ are given by: $r^A_i=e_{N_A}{\rm Tr}\, D({\bf r}^A)\Lambda^A_i, \quad i=1,\dots,N_A^2-1$, where $e_{N_A}\equiv {N_A\over 2c_{N_A}}$. But since $D({\bf r}^A)=|\psi^A\rangle\langle \psi^A|$, with the choice (\ref{choicegenerators1}) we clearly have that $r^A_1=r^A_2=0$, and for the same reason also the other components associated with the non-diagonal generators $\Lambda_i^{A}$, for $i=N_A+2,\dots, N_A^2-1$, are zero. Regarding the components associated with the generators $\Lambda_{n+2}^{A}= W_n^{A}$,  we observe that  ${\rm Tr}\, |\psi^A\rangle\langle \psi^A| \Lambda_{n+2}^{A}=\sqrt{2\over n(n+1)}$, for $n=1,\dots,N_A-1$, or equivalently: ${\rm Tr}\, |\psi^A\rangle\langle \psi^A| \Lambda_{i}^{A} =\sqrt{2\over (i-2)(i-1)}$, for $i=3,\dots,N_A+1$. This means that the components of ${\bf r}^{A}$ are: 
\begin{equation}
{\bf r}^{A} =e_{N_A}(0,0,1, {1\over\sqrt{3}},{1\over\sqrt{6}},{1\over\sqrt{10}},\dots, c_{N_A}^{-1},0\dots,0)^\top \in B_1(\real^{N_A^2-1}),
\label{bra}
\end{equation}
where $e_{N_A}=\sqrt{N_A\over 2(N_A-1)}$. In the same way, since $D({\bf s}^A)=|\phi^A\rangle\langle \phi^A|$, for the components of the vector ${\bf s}^{A}$ we find: 
\begin{equation}
{\bf s}^{A} =e_{N_A}(0,0,-1, {1\over\sqrt{3}},{1\over\sqrt{6}},{1\over\sqrt{10}},\dots, c_{N_A}^{-1},0\dots,0)^\top \in B_1(\real^{N_A^2-1}),
\label{bsa}
\end{equation}
where the difference in sign in the third component comes from the fact that $\Lambda_3^A=W_1^A=|\psi^A\rangle\langle \psi^A|-|\phi^A\rangle\langle \phi^A|$. For the same reasons, we also find that:
\begin{equation}
{\bf r}^{B} =e_{N_B}(0,0,1, {1\over\sqrt{3}},{1\over\sqrt{6}},{1\over\sqrt{10}},\dots, c_{N_B}^{-1},0\dots,0)^\top \in B_1(\real^{N_B^2-1}),
\label{brb}
\end{equation}
\begin{equation}
{\bf s}^{B} =e_{N_B}(0,0,-1, {1\over\sqrt{3}},{1\over\sqrt{6}},{1\over\sqrt{10}},\dots, c_{N_B}^{-1},0\dots,0)^\top \in B_1(\real^{N_B^2-1}).
\label{bsb}
\end{equation}

We now calculate the components of the vector ${\bf r}^{\rm int}$, which are given by: $r^{\rm int}_i=e_{N}{\rm Tr}\, I({\bf r}^{\rm int})\Lambda_i$, $i=1,\dots,N^2-1$. Using the tensorial determination (\ref{tensor-generators}), we can write more explicitely: 
\begin{equation}
r^{\rm int}_{(i,j)}={e_N\over \sqrt{2}}{\rm Tr}\, I \Lambda^{A}_i\otimes \Lambda^{B}_j.
\end{equation}
We then observe that:
\begin{eqnarray}
{\rm Tr}\, I \Lambda^{A}_i\otimes \Lambda^{B}_{j} &=& a_1a_2\, e^{-i\alpha} {\rm Tr}\, \left(|\psi^A\rangle\langle \phi^{A}|\Lambda^{A}_i\right) \otimes  \left(|\phi^{B}\rangle\langle \psi^{B}| \Lambda^{B}_{j}\right) + {\rm c.c.}\nonumber\\
 &=&a_1a_2\, e^{-i\alpha} {\rm Tr}\,|\psi^A\rangle\langle \phi^{A}|\Lambda^{A}_i \,\,{\rm Tr}\, |\phi^{B}\rangle\langle \psi^{B}| \Lambda^{B}_{j} + {\rm c.c.}\nonumber\\
 &=& a_1a_2\, e^{-i\alpha} \langle \phi^{A}|\Lambda^{A}_i|\psi^A\rangle \langle \psi^{B}| \Lambda^{B}_{j}|\phi^{B}\rangle + {\rm c.c.}
\end{eqnarray}
Clearly, ${\rm Tr}\, I \Lambda^{A}_0\otimes \Lambda^{B}_{j}={\rm Tr}\, I \Lambda^{A}_i\otimes \Lambda^{B}_{0}=0$, for all 
 $i=1,\dots,N_A^2-1$, and  $j=1,\dots,N_B^2-1$, which means that ${\bf r}^{\rm int}$ does not contribute to the first $(N_A^2-1)+(N_B^2-1)$ components of ${\bf r}$. In view of (\ref{choicegenerators1}) and (\ref{choicegenerators2}), we also have: 
\begin{equation}
 {\rm Tr}\, I \Lambda^{A}_1\otimes \Lambda^{B}_{1} =  a_1a_2\, e^{-i\alpha} \langle \phi^{A}|\Lambda^{A}_1|\psi^A\rangle \langle \psi^{B}| \Lambda^{B}_{1}|\phi^{B}\rangle + {\rm c.c.}= 2 a_1a_2\, \cos \alpha.
\end{equation}
\begin{equation}
 {\rm Tr}\, I \Lambda^{A}_2\otimes \Lambda^{B}_{2} =  a_1a_2\, e^{-i\alpha} \langle \phi^{A}|\Lambda^{A}_2|\psi^A\rangle \langle \psi^{B}| \Lambda^{B}_{2}|\phi^{B}\rangle + {\rm c.c.}= 2 a_1a_2\, \cos \alpha.
\end{equation}
\begin{equation}
 {\rm Tr}\, I \Lambda^{A}_1\otimes \Lambda^{B}_{2} =  a_1a_2\, e^{-i\alpha} \langle \phi^{A}|\Lambda^{A}_1|\psi^A\rangle \langle \psi^{B}| \Lambda^{B}_{2}|\phi^{B}\rangle + {\rm c.c.}= -2 a_1a_2\, \sin \alpha.
\end{equation}
\begin{equation}
 {\rm Tr}\, I \Lambda^{A}_2\otimes \Lambda^{B}_{1} =  a_1a_2\, e^{-i\alpha} \langle \phi^{A}|\Lambda^{A}_2|\psi^A\rangle \langle \psi^{B}| \Lambda^{B}_{1}|\phi^{B}\rangle + {\rm c.c.}= 2 a_1a_2\, \sin \alpha,
\end{equation}
and of course, the trace of $I$ with all the other generators is zero. Thus, the only non-null components of ${\bf r}^{\rm int}$ are: 
\begin{eqnarray}
 r^{\rm int}_{(1,1)} &=& e_N \sqrt{2}\, a_1a_2\, \cos \alpha,\quad  \quad r^{\rm int}_{(2,2)} = e_N \sqrt{2}\, a_1a_2\, \cos \alpha,\\
r^{\rm int}_{(1,2)} &=& -e_N  \sqrt{2}\, a_1a_2\, \sin \alpha,  \quad \quad r^{\rm int}_{(2,1)} = e_N \sqrt{2}\, a_1a_2\, \sin \alpha.\label{int-contr1}
 \label{int-contr2}
\end{eqnarray}

We now have all we need to determine the components of  ${\bf r}$. Observing that the first two terms in (\ref{entanglement-operator}) are of the ``mixture'' kind, we can use (\ref{trip-sep}) and write: 
\begin{equation}
{\bf r} =  d_{N_A}\overline{\bf r}^{A}\oplus d_{N_B}\overline{\bf r}^{B}\oplus d_{N_A,N_B}\overline{\bf r}^{AB} +{\bf r}^{\rm int},
\label{trip-sep-int}
\end{equation}
and in view of (\ref{bra}), (\ref{bsa}), and (\ref{brb}), (\ref{bsb}), we have: 
\begin{equation}
\overline{\bf r}^{A} =e_{N_A}(0,0,a_1^2-a_2^2, {1\over\sqrt{3}},{1\over\sqrt{6}},{1\over\sqrt{10}},\dots, c_{N_A}^{-1},0\dots,0)^\top \in B_1(\real^{N_A^2-1}),
\label{overline-ra}
\end{equation}
\begin{equation}
\overline{\bf r}^{B} =e_{N_B}(0,0,a_2^2-a_1^2, {1\over\sqrt{3}},{1\over\sqrt{6}},{1\over\sqrt{10}},\dots, c_{N_B}^{-1},0\dots,0)^\top\in B_1(\real^{N_B^2-1}).
\label{overline-rb}
\end{equation}
To specify the components of $\overline{\bf r}^{AB}$, one needs to make a choice regarding the order for the double index $(i,j)$, $i=1,\dots,N_A^2-1$, $j=1,\dots,N_B^2-1$. If we choose the order: $(1,1)$, $(2,2)$, $(1,2)$, $(2,1)$, $(3,3)$, $(1,3)$, $(2,3)$, $(3,2)$, $(3,1)$, $(4,4)$, $(1,4)$, $(2,4)$, $(3,4)$, $(4,3),\dots$, and consider that: $\overline{r}^{AB}_{(i,j)}=a_1^2\, r_i^As_j^B + a_2^2\, s_i^A r_j^B$, we obtain:
\begin{equation}
\overline{\bf r}^{AB} =e_{N_A}e_{N_B}(0,0,0,0, -1,0,0,0,0, {1\over 3},0,0, {a_1^2-a_2^2\over\sqrt{3}},{a_2^2-a_2^2\over\sqrt{3}},\dots)^\top\in B_1(\real^{(N_A^2-1)(N_B^2-1)}).
\label{overline-rab}
\end{equation}
Finally, the components of the interference contribution are: 
\begin{equation}
{\bf r}^{\rm int} =e_{N}\sqrt{2}\, a_1a_2\, (\underbrace{0,\dots,0,}_{N_A^2-1\,{\rm terms}}\underbrace{0,\dots,0,}_{N_B^2-1\,{\rm terms}}\cos \alpha,\cos \alpha,-\sin \alpha,\sin \alpha, 0,\dots,0)^\top\in \real^{N^2-1}.
\label{comp-int}
\end{equation}

To recapitulate, observing that $({\bf r}-{\bf r}^{\rm int})\perp {\bf r}^{\rm int}$, we have found  that the entangled state (\ref{entanglement-operator}) is represented in the Bloch sphere by the sum (\ref{trip-sep-int}), which is the sum of two orthogonal vectors, the first one being of the ``separable'' kind, and the second one being the term that truly distinguishes an entangled state from a separable state (which however, as we have seen in the previous section, is not truly separable, if interpreted as a pure state). We can also observe that ${\bf r}^{\rm int}$ does not contribute to the sectors describing the individual entities, as is clear that its first $(N_A^2-1)+(N_B^2-1)$ components are zero. Therefore, defining the reduced vector: 
\begin{equation}
\tilde{\bf r}^{\rm int} =e_{N}\sqrt{2}\, a_1a_2\, (\cos \alpha,\cos \alpha,-\sin \alpha,\sin \alpha, 0,\dots,0)^\top\in \real^{(N_A^2-1)(N_B^2-1)},
\label{comp-int-reduced}
\end{equation}
we can also write (\ref{trip-sep-int}) in the more compact form:
\begin{equation}
{\bf r} =  d_{N_A}\overline{\bf r}^{A}\oplus d_{N_B}\overline{\bf r}^{B}\oplus \left( d_{N_A,N_B}\overline{\bf r}^{AB} +\tilde{\bf r}^{\rm int}\right).
\label{trip-sep-int-reduced}
\end{equation}

\subsection{The $N_A=N_B=2$ case}
\label{TheN=2case}

Not to unnecessarily complicate the discussion, we consider the simple situation of two entangled qubits. Then, $N_A=N_B=2$, $e_{N_A}=e_{N_B}=1$, $e_{N}=\sqrt{2\over 3}$, $d_{N_A}=d_{N_B} =d_{N_A,N_B} ={1\over\sqrt{3}}$, and (\ref{trip-sep-int}) becomes: 
\begin{equation}
{\bf r} =  {1\over\sqrt{3}}\, \overline{\bf r}^{A}\oplus \overline{\bf r}^{B}\oplus \overline{\bf r}^{AB} +{\bf r}^{\rm int}.
\label{trip-sep-int-qubit}
\end{equation}
Equations (\ref{overline-ra}) and (\ref{overline-rb}) reduce to: 
\begin{equation}
\overline{\bf r}^{A} =(0,0,a_1^2-a_2^2)^\top, \quad \overline{\bf r}^{B} =(0,0,a_2^2-a_1^2)^\top = -\overline{\bf r}^{A}.
\label{overline-ra-and-rb-qubit}
\end{equation}
The correlation vector $\overline{\bf r}^{AB}$ is 9-dimensional, with a single constant non zero component:
\begin{equation}
\overline{\bf r}^{AB} =(0,0,0,0, -1,0,0,0,0)^\top,
\label{overline-rab-qubit}
\end{equation}
and the $15$-dimensional interference vector becomes: 
\begin{equation}
{\bf r}^{\rm int} ={2\over \sqrt{3}}\, a_1a_2\, (0,0,0,0,0,0, \cos \alpha,\cos \alpha,-\sin \alpha,\sin \alpha,0,0,0,0,0)^\top,
\label{comp-int-qubit}
\end{equation}
so that the full expression for ${\bf r}$ is: 
\begin{eqnarray}
&&{\bf r} =  {1\over\sqrt{3}}(0,0,a_1^2-a_2^2)^\top \oplus (0,0,a_2^2-a_1^2)^\top \oplus (0,0,0,0, -1,0,0,0,0)^\top\nonumber\\
&& \quad\quad + \,\, {2\over \sqrt{3}}\, a_1a_2\, (0,0,0,0,0,0, \cos \alpha,\cos \alpha,-\sin \alpha,\sin \alpha,0,0,0,0,0)^\top.
\label{trip-sep-int-qubit}
\end{eqnarray}

Let us consider, as we did in Sec.~\ref{Entanglement},  a product observble $O^A\otimes O^B$, such that the eigenstates of $O^{A}$ are $|\psi^A\rangle$ and $|\phi^{A}\rangle$, and the eigenstates of $O^{B}$ are $|\psi^B\rangle$ and $|\phi^{B}\rangle$, so that when $O^A\otimes O^B$ is measured only the transitions $D({\bf r})\to D({\bf r}^{A})\otimes D({\bf s}^{B})$ and $D({\bf r})\to D({\bf s}^{A})\otimes D({\bf r}^{B})$ can take place, with probabilities $a_1^2$ and $a_2^2$, respectively. We denote ${\bf n}_{\psi\phi}$  the vertex vector representative of the eigenstate $D({\bf r}^{A})\otimes D({\bf s}^{B})$, and ${\bf n}_{\phi\psi}$ the vertex vector representative of the eigenstate $D({\bf s}^{A})\otimes D({\bf r}^{B})$. Their explicit coordinates are obtained by simply setting $a_2=0, a_1=1$, and $a_1=0, a_2=1$, in (\ref{trip-sep-int-qubit}), respectively:
\begin{eqnarray}
&&{\bf n}_{\psi\phi} =  {1\over\sqrt{3}}(0,0,1)^\top \oplus (0,0,-1)^\top \oplus (0,0,0,0, -1,0,0,0,0)^\top\nonumber \\
&& {\bf n}_{\phi\psi} =  {1\over\sqrt{3}}(0,0,-1)^\top \oplus (0,0,1)^\top \oplus (0,0,0,0, -1,0,0,0,0)^\top.
\label{trip-eigenstates}
\end{eqnarray}
When $O^A\otimes O^B$ is measured, the point particle representative of the state orthogonally ``falls'' onto the measuring 3-simplex (a tetrahedron), and as we have seen in Sec.~\ref{Entanglement}, it lands exactly on the edge between ${\bf n}_{\psi\phi}$ and ${\bf n}_{\phi\psi}$. It immediately follows from (\ref{trip-sep-int-qubit}) and (\ref{trip-eigenstates}) that the on-simplex vector is: ${\bf r}^\parallel = {\bf r}-{\bf r}^{\rm int}$, i.e., that the interference term is precisely the contribution that is perpendicular to the simplex, i.e., ${\bf r}^\perp={\bf r}^{\rm int}$, so that we can write: 
 \begin{equation}
{\bf r}^\parallel = a_1^2\, {\bf n}_{\psi\phi} + a_2^2\, {\bf n}_{\phi\psi}.
\end{equation}

Different from the description of Sec.~\ref{Entanglement}, and thanks to the tensorial basis of generators that we have adopted, we can now contemplate the unfolding of the measurement process also from the viewpoint of the one-entity sub-systems, described by the two vectors $\overline{\bf r}^{A}$ and $\overline{\bf r}^{B}$, given in (\ref{overline-ra-and-rb-qubit}). These vectors are located on the 1-simplex associated with the North-South axis of their respective 3-dimensional Bloch sphere, and by construction they are perfectly anti-correlated. If the $A$-point particle moves upward, the $B$-point particle moves downward, and vice versa. This means that once the point particle representative of the state of the entangled entity is on the edge of the measurement 3-simplex, and the associated elastic substance disintegrates, thus bringing the  particle either to ${\bf n}_{\psi\phi}$ or to ${\bf n}_{\phi\psi}$, from the perspective of the two sub-systems their representative particles will move as if they would experience a measurement produced by an elastic band stretched between $(0,0,1)^\top$ and $(0,0,-1)^\top$, but with perfectly correlated outcomes. 

\begin{figure}[!ht]
\centering
\includegraphics[scale =.6]{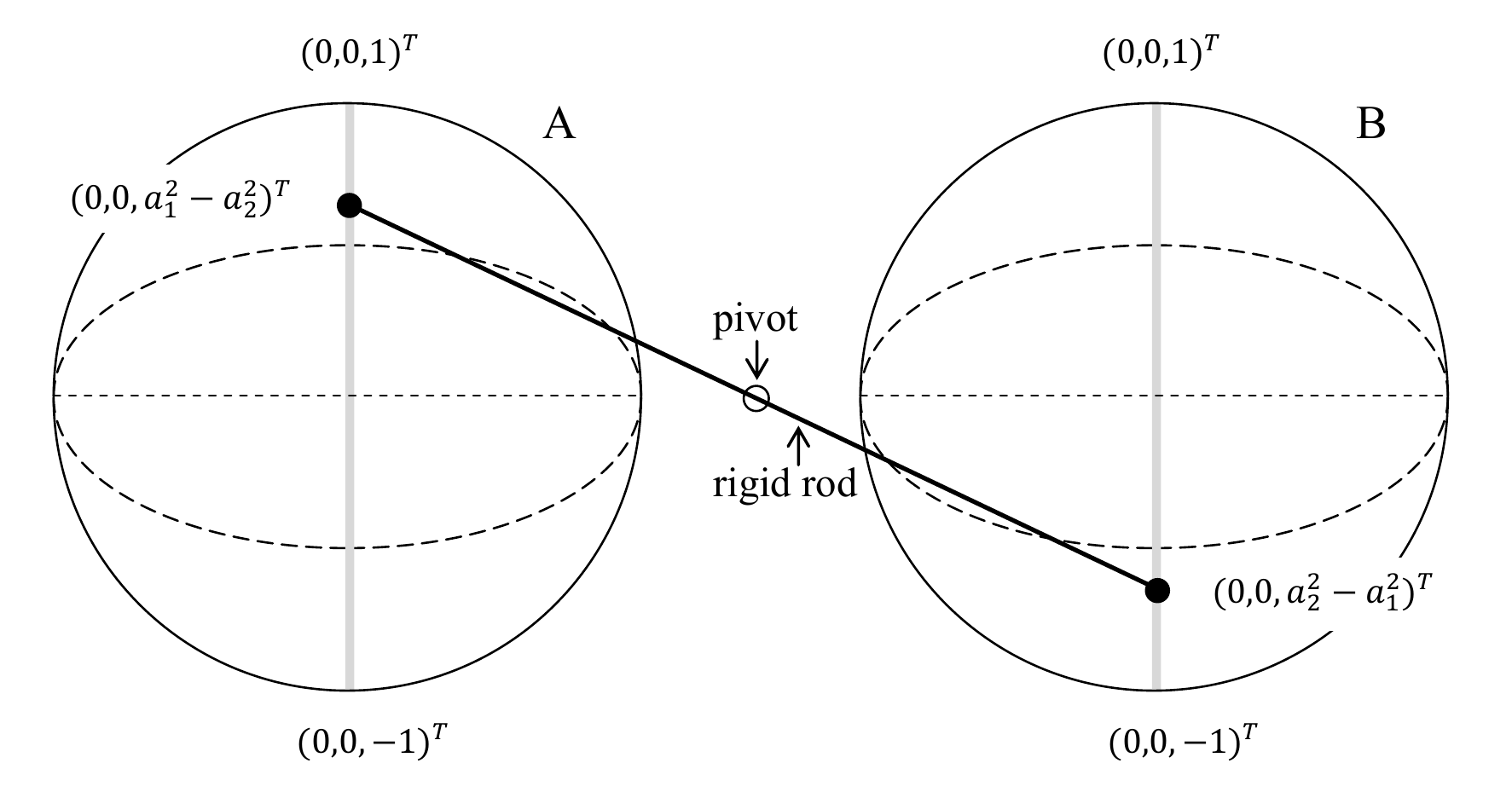}
\caption{When the two particles are connected through a rigid extendable rod, the measurement of  $O^A$, followed by the measurement of $O^B$ (or vice versa) is equivalent to the measurement of $O^A\otimes O^B$, when the bipartite system is in the entangled state (\ref{entanglement}) and the product observable $O^A\otimes O^B$ is such that $|\psi^A\rangle$ and $|\phi^{A}\rangle$ are eigenstates of $O^{A}$, and $|\psi^B\rangle$ and $|\phi^{B}\rangle$ are eigenstates of $O^{B}$.
\label{figure8}}
\end{figure}

One can understand the process by imagining that the two point particles are connected by an invisible extendable rod (see Fig.~\ref{figure8}), and that the measurement of $O^A$ (resp., $O^B$) is first performed, and that when the particle is drawn to one of the anchor points, because of the action of the rod, the other particle is drawn to the opposite point. Then, $O^B$ (resp., $O^A$) is also measured, but since the entity is in an eigenstate, nothing happens. We can then easily check that the probabilities obtained in this way are the same as those produced by the 3-simplex measurement. Indeed, being the $A$-point particle located at position $a_1^2-a_2^2$, along an elastic band stretched between points $-1$ and $1$, it is clear that the probability to be drawn to point $1$ is: ${1\over 2}(a_1^2-a_2^2 -(-1))=a_1^2$, and the probability to be drawn to point $-1$ is: ${1\over 2}(1-(a_1^2-a_2^2))=a_2^2$. Similarly, and compatibly, the $B$-point particle being located at position $a_2^2-a_1^2$, along an elastic band stretched between points $-1$ and $1$, the probability to be drawn to point $1$ is: ${1\over 2}(a_2^2-a_1^2 -(-1))=a_2^2$, and the probability to be drawn to point $-1$ is: ${1\over 2}(1-(a_2^2-a_1^2))=a_1^2$.

\subsection{More general product measurements}

In Sec.~\ref{Entanglement}, and in the previous section, we have only considered a product measurement $O^A\otimes O^B$ such that its eigenstates are precisely the orthogonal states that form the entangled state. For instance, in the case of a spin measurement, this corresponds to the situation where the spin observable $\sigma_3\otimes\sigma_3$ is measured on a state $|\psi\rangle = a_1 e^{i\alpha_1}|+\rangle\otimes |-\rangle  + a_2 e^{i\alpha_2}|-\rangle\otimes |+\rangle$, with $ |+\rangle$ and $ |-\rangle$ the eigenstates of $\sigma_3$. But what about more general situations, like those intervening in the tests of Bell's inequalities, which involve the measurements of spin observables of the form $\sigma_{\bf n}\otimes\sigma_{\bf m}$, with ${\bf n}$ and ${\bf m}$ two arbitrary directions in space? Can we still obtain a simple description of the measurement, from the viewpoint of the individual sub-systems? 

To answer this question, we now consider a more general product observable, still denoted $O^A\otimes O^B$, with no specific assumptions on its eigenstates. For simplicity, we limit our discussion to the $N_A=N_B=2$ case. We write: $O^A= o^A_+P_{+}^A+ o^A_-P_{-}^A$ and $O^B= o^B_+P_{+}^B+ o^B_-P_{-}^B$. To each projection operator we can associate a specific unit vector in the (here 3-dimensional) Bloch sphere: $P_{\pm}^A\equiv P({\bf n}_\pm^A)={1\over 2}(\mathbb{I} + {\bf n}_\pm^A\cdot\mbox{\boldmath$\Lambda$})$, and similarly: $P_{\pm}^B \equiv P({\bf n}_\pm^B)={1\over 2}(\mathbb{I} + {\bf n}_\pm^B\cdot\mbox{\boldmath$\Lambda$})$, where $\mbox{\boldmath$\Lambda$}=(\sigma_1,\sigma_2,\sigma_3)^\top$,  ${\bf n}_+^A =-{\bf n}_-^A$, and ${\bf n}_+^B =-{\bf n}_-^B$. The eigenstates of $O^A\otimes O^B$ are the four projections: $P_{\sigma\rho}^{AB}\equiv P({\bf n}^A_\sigma)\otimes P({\bf n}^B_\rho) = {1\over 4}(\mathbb{I} + \sqrt{6}\,{\bf n}_{\sigma\rho}^{AB}\cdot\mbox{\boldmath$\Lambda$})$, $\sigma,\rho\in\{+,-\}$, where the unit vectors ${\bf n}_{\sigma\rho}^{AB}$ belong to the 15-dimensional Bloch sphere, and the generators are given by (\ref{15gen}). Then, all four transitions can take place in a measurement, as the point particle in the entangled state (\ref{entanglement})-(\ref{entanglement-operator}) will now not necessarily ``fall'' onto one of the edges of the measurement 3-simplex, in the first deterministic phase of the measurement, but will generally ``fall'' onto one of its interior points. 

The eigenstates being product states, we know they admit the tripartite representation (\ref{tripartite-vector-eigenstate}), which here reduces to: 
\begin{equation}
{\bf n}_{\sigma\rho}^{AB}= {1\over\sqrt{3}}{\bf n}_{\sigma}^{A}\oplus {\bf n}_{\rho}^{B}\oplus {\bf n}_{\sigma\rho}^{AB},\quad \sigma,\rho\in\{+,-\},
\label{tripartite-vector-eigenstate-qubits}
\end{equation}
and different from the previous situation, the interference contribution of the entangled state (\ref{comp-int-qubit}) is not orthogonal to the measurement simplex. Also, the point particles representative of the sub-entities do not lie onto the 1-simplexes associated with the one-entity measurements $O^A$ and $O^B$, unless of course $a_1=a_2$, as in this case they would be located exactly at the center of their respective sub-spheres (this would be so if the bipartite system is in a singlet state, in the case of a spin measurement).  

However, in view of (\ref{trip-sep-int-reduced}), we have that the point particle representative of the bipartite system is associated with the vector:
\begin{equation}
{\bf r} ={1\over\sqrt{3}}\, \overline{\bf r}^{A}\oplus \overline{\bf r}^{B}\oplus (\overline{\bf r}^{AB} +\sqrt{3}\, \tilde{\bf r}^{\rm int}).
\label{trip-sep-int2}
\end{equation}
If we write ${\bf r}={\bf r}^\parallel + {\bf r}^\perp$, with ${\bf r}^\perp$ the component orthogonal to the measurement simplex $\triangle_3$, associated with $O^A\otimes O^B$, and ${\bf r}^\parallel$ the on-simplex component, then, because of the direct sum structure of ${\bf r}$, we necessarily have: 
\begin{equation}
{\bf r}^\parallel ={1\over\sqrt{3}}\, \overline{\bf r}^{A\parallel}\oplus \overline{\bf r}^{B\parallel}\oplus (\overline{\bf r}^{AB} +\sqrt{3}\, \tilde{\bf r}^{\rm int})^\parallel,
\label{trip-sep-int-parallel}
\end{equation}
where $\overline{\bf r}^{A\parallel}$ is the vector obtained by orthogonally projecting $ \overline{\bf r}^{A}$ onto the measurement 1-simplex associated with $O^A$, and $\overline{\bf r}^{B\parallel}$ is the vector obtained by orthogonally projecting $ \overline{\bf r}^{B}$ onto the measurement 1-simplex associated with $O^B$. In other terms, as we already observed in Sec.~\ref{Product measurements}, the deterministic process through which the point particle, in the 15-dimensional Bloch sphere, orthogonally moves onto the 3-dimensional simplex, also corresponds to the deterministic processes through which the point particles representative of the two sub-systems, orthogonally move onto their respective 1-dimensional simplexes. 

At this point, the natural question is: Can we obtain the transition probabilities ${\cal P}({\bf r}\to {\bf n}_{\sigma\rho}^{AB})$ by considering only the one-entity measurements in the sub-spheres, in a sequential way, if we connect the process through a rod-like structure, as we did in the previous section? To answer this question, we observe that: 
\begin{eqnarray}
{\cal P}({\bf r}\to {\bf n}_{\sigma\rho}^{AB})&=& {\rm Tr}\,(a_1^2\, D({\bf r}^{A})\otimes D({\bf s}^{B}) + a_2^2\, D({\bf s}^{A})\otimes D({\bf r}^{B}) + I)P({\bf n}^A_\sigma)\otimes P({\bf n}^B_\rho)\\
&=& a_1^2\,{\cal P}({\bf r}^A\to {\bf n}_{\sigma}^{A}){\cal P}({\bf s}^B\to {\bf n}_{\rho}^{B}) +a_2^2\, {\cal P}({\bf s}^A\to {\bf n}_{\sigma}^{A}){\cal P}({\bf r}^B\to {\bf n}_{\rho}^{B})+{\rm Tr}\,I P({\bf n}^A_\sigma)\otimes P({\bf n}^B_\rho).\nonumber
\end{eqnarray}
The third term, which comes from the interference contribution (\ref{interference-contribution}), depends not only on $a_1$ and $a_2$, but also on the relative phase $\alpha$. However, the one-entity vectors $ \overline{\bf r}^{A}$ and $ \overline{\bf r}^{B}$, only contain information about $a_1$ and $a_2$, via their third component -- see (\ref{overline-ra-and-rb-qubit}) -- but no information about $\alpha$, as the only vector in (\ref{trip-sep-int2}) varying with $\alpha$ is $\tilde{\bf r}^{\rm int}$. Therefore, even if we connect the two vectors $ \overline{\bf r}^{A}$ and $ \overline{\bf r}^{B}$, and perform sequential measurements, we will never succeed in deriving the probability produced by the disintegration of the full 3-simplex.

In the special situation considered in Sec.~\ref{Entanglement} and Sec.~\ref{TheN=2case}, the rod mechanism was however sufficient to do the job. This was the case because the interference term was not contributing to the measurement, as it was perpendicular to the measurement simplex. It is therefore natural to ask if the above mentioned ``rod mechanism'' would nevertheless be able to produce the correct values of the transition probabilities in the situations where the interference contribution would be zero. To answer this question, we need to find an example of a situation of this kind.  

Consider a product spin-measurement $\sigma_{{\bf n}_{+}^{A}}\otimes\sigma_{{\bf n}_{+}^{B}}$, performed on a singlet state ($a_1=a_2={1\over\sqrt{2}}$, $\alpha =\pi$): $|\psi_S\rangle = {1\over\sqrt{2}}( |{\bf n}\rangle\otimes |-{\bf n}\rangle - |-{\bf n}\rangle\otimes |{\bf n}\rangle)$, with ${\bf n}$ a unit vector. To place ourselves in a situation where the interference contribution is zero, we remember that a singlet state, being of zero spin, is a rotationally invariant state. This means that the vector ${\bf n}$ is arbitrary, and we are free to choose it as is best for us. For the specific choice ${\bf n}={\bf n}_{+}^{A}= -{\bf n}_{-}^{A}$, we have $|\psi_S\rangle = {1\over\sqrt{2}}( |{\bf n}_{+}^{A}\rangle\otimes |{\bf n}_{-}^{A}\rangle - |{\bf n}_{-}^{A}\rangle\otimes |{\bf n}_{+}^{A}\rangle)$. Then, the interference contribution (\ref{interference-contribution}) becomes: $I= -{1\over 2}\, |{\bf n}_{+}^{A}\rangle\langle {\bf n}_{-}^{A}|\otimes |{\bf n}_{-}^{A}\rangle\langle {\bf n}_{+}^{A}| + {\rm c.c.}$, so that:
\begin{eqnarray}
{\rm Tr}\,I P({\bf n}^A_\sigma)\otimes P({\bf n}^B_\rho)&=& -{1\over 2}\, {\rm Tr}\, |{\bf n}_{+}^{A}\rangle\langle {\bf n}_{-}^{A}|{\bf n}^A_\sigma\rangle\langle {\bf n}^A_\sigma| \otimes |{\bf n}_{-}^{A}\rangle\langle {\bf n}_{+}^{A}|{\bf n}^B_\rho\rangle\langle {\bf n}^B_\rho|   + {\rm c.c.}\nonumber\\
&=& -{1\over 2}\, \delta_{-,\sigma}\langle {\bf n}_{+}^{A}|{\bf n}^B_\rho\rangle \,{\rm Tr}\,|{\bf n}_{+}^{A}\rangle\langle {\bf n}^A_\sigma| \,{\rm Tr}\, 
 |{\bf n}_{-}^{A}\rangle \langle {\bf n}^B_\rho| + {\rm c.c.}\nonumber\\
 &=&-{1\over 2}\, \delta_{-,\sigma}\delta_{+,\sigma}\langle {\bf n}_{+}^{A}|{\bf n}^B_\rho\rangle \, {\rm Tr}\, 
 |{\bf n}_{-}^{A}\rangle \langle {\bf n}^B_\rho| + {\rm c.c.} = 0.
\end{eqnarray}
The interference contribution being zero, the transition probabilities become: 
\begin{eqnarray}
{\cal P}({\bf r}\to {\bf n}_{\sigma\rho}^{AB}) &=& {1\over 2} \,{\cal P}({\bf n}_{+}^{A}\to {\bf n}_{\sigma}^{A}){\cal P}({\bf n}_{-}^{A}\to {\bf n}_{\rho}^{B}) +{1\over 2}\, {\cal P}({\bf n}_{-}^{A}\to {\bf n}_{\sigma}^{A}){\cal P}({\bf n}_{+}^{A}\to {\bf n}_{\rho}^{B}) \nonumber\\
&=& {1\over 2}\,\delta_{+,\sigma}{\cal P}({\bf n}_{-}^{A}\to {\bf n}_{\rho}^{B}) +{1\over 2}\,\delta_{-,\sigma}{\cal P}({\bf n}_{+}^{A}\to {\bf n}_{\rho}^{B}), \quad \sigma,\rho \in\{+,-\},
\label{transition-prob-4}
\end{eqnarray}
or, more explicitly:
\begin{eqnarray}
&&{\cal P}({\bf r}\to {\bf n}_{++}^{AB}) = {1\over 2} \,{\cal P}({\bf n}_{-}^{A}\to {\bf n}_{+}^{B}),\quad 
{\cal P}({\bf r}\to {\bf n}_{--}^{AB}) = {1\over 2} \,{\cal P}({\bf n}_{+}^{A}\to {\bf n}_{-}^{B}),\nonumber \\
&&{\cal P}({\bf r}\to {\bf n}_{+-}^{AB}) = {1\over 2} \,{\cal P}({\bf n}_{-}^{A}\to {\bf n}_{-}^{B}),\quad 
{\cal P}({\bf r}\to {\bf n}_{-+}^{AB}) = {1\over 2} \,{\cal P}({\bf n}_{+}^{A}\to {\bf n}_{+}^{B}).
\label{transition-prob-4-bis}
\end{eqnarray}

\begin{figure}[!ht]
\centering
\includegraphics[scale =.6]{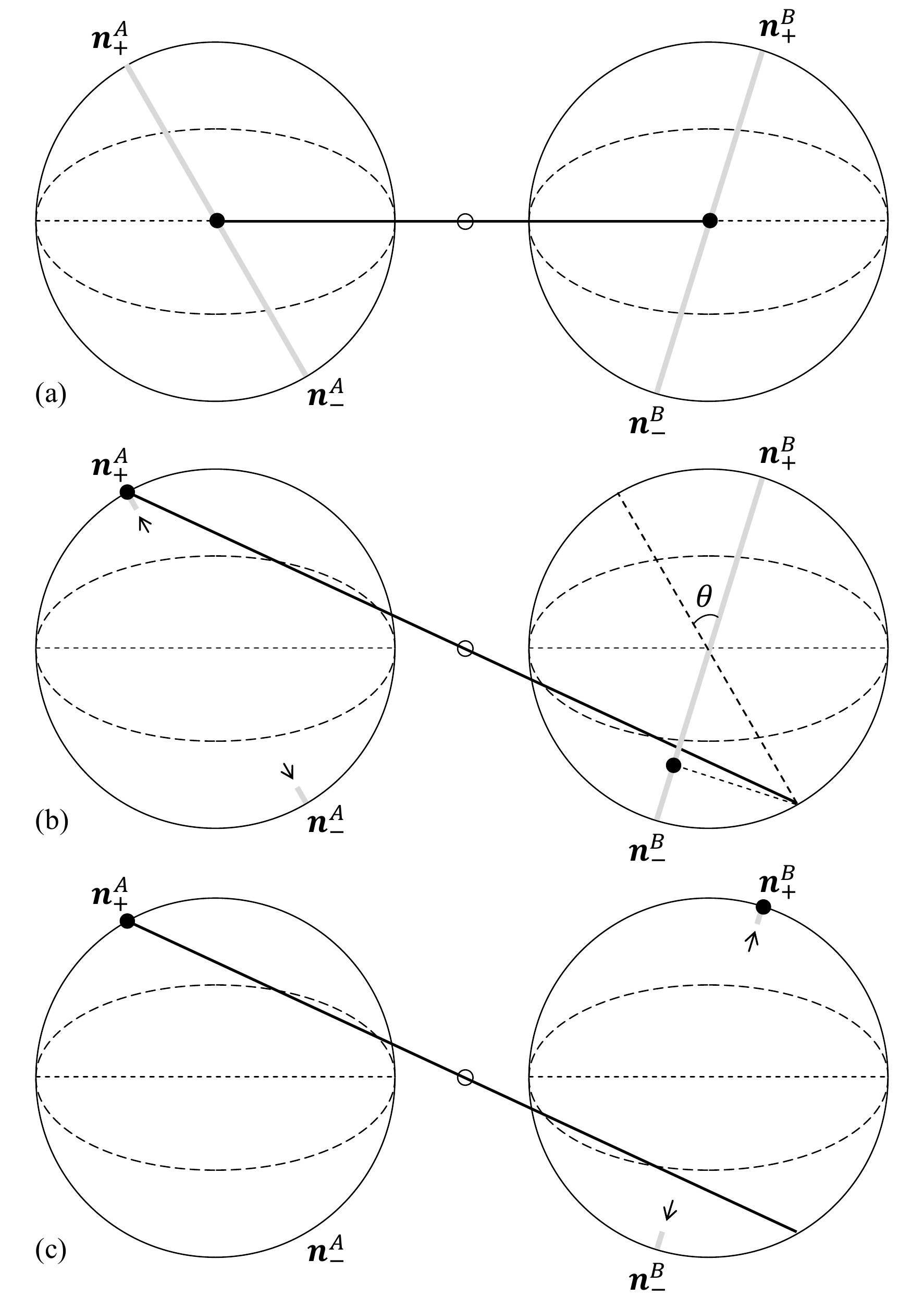}
\caption{The measurement of $\sigma_{{\bf n}_{+}^{A}}\otimes\sigma_{{\bf n}_{+}^{B}}$, when the bipartite system is in a singlet state, is equivalent to the measurement of  $\sigma_{{\bf n}_{+}^{A}}$ followed by the measurement of $\sigma_{{\bf n}_{+}^{B}}$, when the two particles are connected through an extendable rigid rod: (a) initially the two particles are at the center of their respective spheres; (b) following the measurement of $\sigma_{{\bf n}_{+}^{A}}$, the $A$-particle is drawn either to ${\bf n}_{+}^{A}$ or to ${\bf n}_{-}^{A}$, with equal probability; here the outcome is ${\bf n}_{+}^{A}$; because of the rod-connection, the $B$-particle is forced to acquire the opposite position ${\bf n}_{-}^{A}$. Then, the rod-connection is disabled, and the $B$-particle orthogonally ``falls'' onto the elastic band associated with $\sigma_{{\bf n}_{+}^{B}}$; (c) the $B$-particle is finally drawn either to ${\bf n}_{-}^{B}$, or to ${\bf n}_{+}^{B}$ (here ${\bf n}_{+}^{B}$), with probabilities ${1\over 2}(1+\cos\theta)$ and ${1\over 2}(1-\cos\theta)$, respectively.
\label{figure9}}
\end{figure}

Let us show that it is possible to reproduce these probabilities by means of sequential measurements, performed on the two sub-entities, if we assume that a rigid extendable rod connects the two point particles representative of the individual states (see Fig.~\ref{figure9}). For a singlet state, we know that $\overline{\bf r}^{A} =(0,0,0)^\top$ and $\overline{\bf r}^{B} =(0,0,0)^\top$, i.e., that the point particles representative of the two sub-entities are located exactly at the center of their spheres. This means that if we perform the measurement  $\sigma_{{\bf n}_{+}^{A}}$ on the first entity, both outcomes can occur with equal probability ${1\over 2}$. If the outcome is ${\bf n}_{-}^{A}$ (resp., ${\bf n}_{+}^{A}$),  because of the connection through the rigid rod, the entity in the second sphere will reach the opposite position ${\bf n}_{+}^{A}$ (resp., ${\bf n}_{-}^{A}$). Then, assuming that following the first measurement the rod-connection is disabled, performing the second measurement $\sigma_{{\bf n}_{+}^{B}}$, on the second entity, we will obtain the outcome  ${\bf n}_{\rho}^{B}$, with probability ${\cal P}({\bf n}_{+}^{A}\to {\bf n}_{\rho}^{B})$ (resp., ${\cal P}({\bf n}_{-}^{A}\to {\bf n}_{\rho}^{B})$). Finally, considering the joint probability of these two sequential measurements, we exactly obtain the quantum mechanical values (\ref{transition-prob-4})-(\ref{transition-prob-4-bis}). And of course, the same statistics of outcomes will be obtained if one performs $\sigma_{{\bf n}_{+}^{B}}$ as the first measurement and  $\sigma_{{\bf n}_{+}^{A}}$ as the second one. 

A few remarks are in order. One could object that the measurement performed using the rod mechanism is not equivalent to the quantum measurement, because it is not a coincidence measurement, but a sequential measurement. However, it should be considered that coincidence measurements are idealized measurements, in the sense that their simultaneity is always only approximate, and never exact, in real experimental situations. In other terms, strictly speaking also coincidence measurements are, in practice, sequential measurements, very close in time. What is important is that the statistics of outcomes does not depend on the order of the two connected sequential measurements. In that respect, we also observe that the quantum mechanical probabilities (\ref{transition-prob-4-bis}) are precisely formed by the product of the probabilities of two one-entity  processes, which means that the very quantum formalism already suggests that the coincidence measurement should in fact be interpreted as two sequential (although connected) measurements. 

Having said that, it is certainly also possible to imagine a more complicate mechanisms, different from that exemplified by the two elastic bands and the rod, that would allow for a description of the product measurement in a purely coincident way, for instance using electric-like forces and charges, but this of course would give rise to a very complicated dynamics, which is not of particular interest. 

From (\ref{transition-prob-4-bis}), we can easily calculate the expectation value:
\begin{eqnarray}
E({\bf n}_{+}^{A},{\bf n}_{+}^{B}) &=& {\cal P}({\bf r}\to {\bf n}_{++}^{AB}) +{\cal P}({\bf r}\to {\bf n}_{--}^{AB})-{\cal P}({\bf r}\to {\bf n}_{+-}^{AB})-{\cal P}({\bf r}\to {\bf n}_{-+}^{AB})\nonumber\\
&=& {1\over 4}(1-\cos\theta) + {1\over 4}(1-\cos\theta)- {1\over 4}(1+\cos\theta)- {1\over 4}(1+\cos\theta)\nonumber\\
&=& -\cos\theta = - {\bf n}_{+}^{A}\cdot {\bf n}_{+}^{B},
\label{average}
\end{eqnarray}
which of course corresponds to the well-known quantum average $\langle\psi_S|\sigma_{{\bf n}_{+}^{A}}\otimes\sigma_{{\bf n}_{+}^{B}}|\psi_S\rangle$. Now, since we haven't used the interference contribution (\ref{interference-contribution}) to calculate (\ref{average}), and that (\ref{average}) can be used to violate Bell's inequalities, one may be tempted to conclude that separable states would be able to violate the latter, which we know to be false. To solve this apparent contradiction, we have to observe that although the singlet state, as we observed, is rotationally invariant, the same is not true for its separable and interference components, if taken individually. This means that when we vary the vector ${\bf n}$ used to mathematically describe $|\psi_S\rangle$, its separable and interference parts will both change, but their changes will exactly compensate, so maintaining the singlet state unchanged (apart from a global phase factor). The specific choice ${\bf n}={\bf n}_{+}^{A}$, is precisely that which brings the interference contribution to zero, thus transferring all the information about the entanglement to the separable part of the state. This separable part, however, now depends on that specific choice, which in turn depends on the measurement which is carried out. 

In other terms, a product spin-measurement on a singlet state will deliver the same statistics of outcomes if performed on a separable, non-entangled state, provided we connect first the orientation of the spin states in the ``mixture'' with the orientation of one of the two Stern-Gerlach apparatuses. More precisely, the measurement of $\sigma_{{\bf n}_{+}^{A}}\otimes\sigma_{{\bf n}_{+}^{B}}$ will give the same results if performed on $|\psi_S\rangle$, or on ${1\over 2}(D({\bf n}_{+}^{A})\otimes D(-{\bf n}_{+}^{A}) + D(-{\bf n}_{+}^{A})\otimes D({\bf n}_{+}^{A}))$, or ${1\over 2}(D({\bf n}_{+}^{B})\otimes D(-{\bf n}_{+}^{B}) + D(-{\bf n}_{+}^{B})\otimes D({\bf n}_{+}^{B})$. These two product states, as we said, are no more rotationally invariant, in accordance with the fact that although the two one-entity states $\overline{\bf r}^{A} =(0,0,0)^\top$ and $\overline{\bf r}^{B} =(0,0,0)^\top$, being at the center of their respective spheres, are rotationally invariant when considered separately, if connected through the rod (which mimics the effect of the third correlation component $\overline{\bf r}^{AB}$) they will lose their rotational invariance. The presence of the rod is however not sufficient to break the rotational invariance of the two sub-entity states. For this, the two entities need to be subjected to their respective elastic measurements, which will create a direction by bringing them to a point at the surface of their spheres, with the rod correlating their movements, but also with the rod being disabled at the end of the process, in accordance with the fact that the outcome state is a product state. 

It is worth observing that one of the consequences of the above analysis is that a macroscopic mechanistic entity can be easily constructed and put in a state which is equivalent to a singlet spin state, and is therefore able to violate Bell's inequalities with exactly the same $2\sqrt{2}$ numerical  value (which corresponds to the maximal violation obtainable in quantum mechanics~\cite{Cirel1980}). This was already observed years ago by one of us~\cite{Aerts1991}, and in fact macroscopic objects can easily violate Bell's inequality with all possible numerical values, also beyond Cirel'son's bond~\cite{Aerts1984,AertsBroekaert2000,Sassoli2013a,Massimiliano2013b,Massimiliano2014}. Of course, the construction of interconnected macroscopic objects mimicking the behavior of entangled microscopic entities is only possible here because spin-${1\over 2}$ entities live in a 3-dimensional Bloch sphere of states, still representable within our 3-dimensional Euclidean theater, and because we have considered a very special entangled state, the singlet state, which has the remarkable property of being rotationally invariant, thus allowing us to ``push'' all the information about the entanglement into the separable part of the state. For more general states, and measurement situations, such a simple three-dimensional representation is not expected to hold anymore, in accordance with the fact that spin entities are genuine non-spatial entities~\cite{AertsSassoli2014e}.

\section{Completed quantum mechanics}
\label{Completed quantum mechanics}

As its name indicates, our extended Bloch model, and the associated hidden-measurement interpretation, constitute a \emph{completed version of quantum mechanics}. This is so because it allows for a full description of the measurement process, and for a non-circular derivation of the Born rule~\cite{AertsSassoli2014c}. As we have seen, this can be done by completing the standard state space of a quantum entity by also adding the operator-states, i.e., the density matrices, to be also interpreted as pure states. In other terms, the completed quantum mechanics of the extended Bloch model considers that not only the states at the surface of the Bloch sphere, but also those in its interior, can represent pure states. 

As we want to emphasize in this section, this completion of the states allows to solve a paradox, which was clearly formulated some years ago by one of us~\cite{Aerts2000}. A possible solution to the paradox was already given at that time, and is the same solution that we are going to propose here. However, thanks to the new general representation of multipartite systems, given in Sec.~\ref{Multipartite systems}, and the recent discovery that general quantum measurements can be fully described within the extended Bloch representation~\cite{AertsSassoli2014c}, the solution acquires today a much higher plausibility.  But let us start by formulating the paradox. Following~\cite{Aerts2000}, we have the two general physical principles: 
 
 \vspace{0.25cm}
\noindent {\bf General physical principle 1} \emph{A physical entity $S$ is said to exists, at a given moment, if and only if it is in one and only one pure state, at that moment.}
 \vspace{0.25cm}
 
 \noindent {\bf General physical principle 2} \emph{Let $S$ be a joint physical entity formed by two sub-entities $S^A$ and $S^B$. $S$ exists, at a certain moment, if and only if $S^A$ and $S^B$ exist, at that moment.}
 \vspace{0.25cm}
  
\noindent These two principles, however natural and self-evident, are incompatible with the following two principles of standard quantum mechanics (SQM): 

 \vspace{0.25cm}
\noindent {\bf SQM principle 1} \emph{Let $S$ be a physical entity, with Hilbert space ${\cal H}$. Each ray-state (i.e., vector-state) of ${\cal H}$ is a pure state of $S$, and all the pure states of $S$ are of this form.}

 \vspace{0.25cm}
\noindent {\bf SQM principle 2} \emph{Let $S^A$ and $S^B$ be two physical entities, with Hilbert spaces ${\cal H}^A$ and ${\cal H}^B$, respectively. The states of the joint quantum entity $S$, formed by the two sub-entities $S^A$ and $S^B$, are described in the tensor product Hilbert
space ${\cal H}^A\otimes {\cal H}^B$. The two sub-entities $S^A$ and $S^B$ are in the ray-states $|\psi^A\rangle\in {\cal H}^A$ and  $|\phi^B\rangle\in {\cal H}^B$, if and only if the joint entity $S$ is in the ray-state $|\psi^A\rangle\otimes|\phi^B\rangle$.}
 \vspace{0.25cm}

\noindent The reason why the above four principles cannot all be satisfied together is of course the existence of non-product states. Indeed, if $S$ is a joint entity in the state (\ref{entanglement}), being a ray-state (i.e., a vector-state, in the terminology of this article), from the \emph{SQM principle 1} we know that it is a pure state of $S$. Then, according to the \emph{General physical principle 1}, we know that the entity exists, and by the  \emph{General physical principle 2} we also know that the two sub-entities $S^A$ and $S^B$ exist. But then, by the \emph{SQM principle 1}, we have that  $S^A$ and $S^B$ must be in ray-states, and according to the \emph{SQM principle 2} $S$ has to be in a product state, which is a contradiction. 

Of course, the perception of the above difficulty was present since the discovery of entangled states, for instance when Schr{\oe}dinger emphasized that when two quantum entities are in an entangled state only the properties of the pair appear to be defined, whereas the individual properties of each one of the two sub-entities that have formed the pair remain totally undefined~\cite{Schroedinger1935}. But it is also true that most of the attention, in earlier and subsequent investigations, went to the `non-local' properties of entangled states, so that the above paradox has often been overlooked, although partly mentioned in some texts (see~\cite{Fraassen1991}, Sec.~7.3, and the references therein). 

Facing this incompatibility of the above four principles, a possible (and usual) strategy is that of considering that the \emph{General physical principle 2} cannot have general validity, in the sense that when the joint entity (the bipartite system) is in an entangled state, the sub-entities simply, and literally, would cease to exist, in the same way that, for example, two droplets of water would cease to exist when fused into a single bigger droplet. This strategy, however, is not fully consistent when considering entangled microscopic entities. Indeed, the two composing entities do not completely disappear in the entanglement, as there are properties associated with the pair that remain actual. 

For instance, when two electrons are in an entangled state, we are still in the presence of two masses, which can easily be separated by a large spatial distance. So, the entanglement is neither a situation where the two masses are completely fused together, nor a situation where a spatial connection would bond the two entities together, making it difficult to spatially separate them (as it would be the case in a chemical bond). Also, in case of entangled spins  in a singlet state, we know that they are always perfectly anticorrelated, and the property of ``being anticorrelated'' is clearly meaningful only if we are in the presence of two entities. 

In other terms, it is not fully consistent to affirm that in an entangled state the composing entity would completely cease to exist. Also, the \emph{General physical principle 2} is very close to a tautology, and it is very difficult to imagine a situation where it would not apply. Even the above example of two droplets of water fused together cannot be considered as a counterexample, as we are not really allowed to describe the bigger droplet, once formed, as the combination of two actual sub-droplets. But fortunately, there are no reasons to abandon the \emph{General physical principle 2}, as the extended Bloch model already contains a way out of the paradox. Indeed, according to it, the \emph{SQM principle 1} is simply incomplete, and needs to be replaced by the following completed quantum mechanics (CQM) principle: 

 \vspace{0.25cm}
\noindent {\bf CQM principle 1} \emph{Let $S$ be a physical entity, with Hilbert space ${\cal H}$. Each operator-state (i.e., density matrix) of ${\cal H}$ is a pure state of $S$, and all the pure states of $S$ are of this form.}
 \vspace{0.25cm}

When it was initially formulated, fifteen years ago, the \emph{CQM principle 1} was conjectured on the basis of the existence of the elastic sphere-model (the Bloch model for only two-outcomes) and of the rod-model example providing a description of coincidence measurements on singlet states. Other works were also available, going in the same direction, like the proof, within axiomatic approaches, that standard quantum mechanics was unable to describe all possible situations of a joint entity formed by two sub-entities~\cite{Aerts1982}, or the promising results by Coecke, showing that hidden-measurement models for higher dimensional quantum entities could also be worked out~\cite{Coecke1995a}, and that the states in a tensor product could be realized by introducing correlations on the different component states~\cite{Coecke1995b,Coecke1996}.

On the other hand, within the now available extended Bloch representation, not only the \emph{CQM principle 1} naturally follows from the formalism, as operator-states are needed to describe the measurements and derive the Born rule, but also, the model allows to exactly determine what are the states of the two sub-entities, given the state of the joint entity. Indeed, as we have seen in Sec.~\ref{Multipartite systems}, and according to (\ref{trip-sep-int}), if the joint entity is in a state described by the vector  ${\bf r}$, within the Bloch sphere $B_1(\real^{N^2-1})$, it can always be written in the form ${\bf r} =  d_{N_A}\overline{\bf r}^{A}\oplus d_{N_B}\overline{\bf r}^{B}\oplus d_{N_A,N_B}\overline{\bf r}^{AB} +{\bf r}^{\rm int}$, where $\overline{\bf r}^{A}\in B_1(\real^{N_A^2-1})$ is the state of the sub-system $S^A$, and $\overline{\bf r}^{B}\in B_1(\real^{N_B^2-1})$ is the state of the sub-system $S^B$, with the two vectors $\overline{\bf r}^{AB}$ and ${\bf r}^{\rm int}$ describing the correlations between the two sub-entities, which cannot be deduced from the states of the sub entities, in accordance with the philosophical principle that the whole is greater than the sum of its parts (so that the states of the parts cannot fully determine the state of the whole).

\section{Conclusion}
\label{Conclusion}

In the present work we have continued our exploration of the extended Bloch representation~\cite{AertsSassoli2014c,AertsSassoli2014d,AertsSassoli2014e}, which is a candidate for a completed version of quantum theory, in which operator-states (density matrices) also play the role of pure states, in the description of the measurement processes, as well as in the characterization of multipartite systems. Let us summarize below the results that we have obtained. 

Following an introduction to the fundamentals of the extended Bloch formalism, we have described states that are a superposition of two orthogonal states, and shown that the interference effects they give rise result from the different possible orientations of the Bloch vector within the Bloch sphere (along circles of latitudes), with respect to the measurement simplex, with the no-interference condition corresponding to the situation where the point particle representative of the state lands exactly at the center of the simplex. When applied to entangled states, which are a special case of superposition states, the approach reveals that the perfect correlations they subtend results from the fact that the point particle exactly lands on the edge of the measurement simplex. 

We have also explored the situation of a superposition of three orthogonal states, finding that the no-interference condition still corresponds to the situation where the point particle lands exactly at the center of the simplex, the interference effects being always the consequence of the different possible orientations of the Bloch vector with respect to the measurement simplex. However, since for $N>2$ the shape of the convex set of states, within the Bloch sphere, is more complicated, superposition states are then associated with more complicated geometries, as different relative phases (and therefore different circles) intervene in the determination of the probabilities. 

 We have also considered general multipartite systems, introducing a suitable tensorial determination of the $SU(N)$ generators, which allowed us to partition the Bloch sphere of the total system into a direct sum of sub-spheres, representing the different sub-systems, plus their correlations. This allowed us to analyze the structure and behavior of product, separable and entangled states, in relation to their measurements. We have observed that when the state is a product state,  a product measurement factorizes into separate measurements, which can performed independently of one another, within the respective sub-entity's spheres. This, however, is not anymore possible if the state is separable, but non-product, as in this case the different one-entity measurements cannot be carried each independently of the other. And this means that separable states, in the Blochean completed quantum mechanics, are not truly such.

We also observed that the difference between separable and entangled states is the presence of an additional interference term, which is responsible for the violation of Bell's inequalities. Also in this case, a fortiori, product measurements cannot be decomposed into separated sub-measurements. However, we have shown that in special circumstances, like that of a spin coincidence measurement on a single state, it is possible to describe the correlation induced by the entanglement as the result of the action of an extendable rigid rod, correlating the outcomes of the two one-entity measurements, in their respective sub-spheres. 

But although the simple rigid rod mechanism doesn't work in the general situation, our analysis shows that an entangled state can always be understood as a condition in which two sub-entities are in a well defined state, but are also interconnected. This explains why the state of a joint entity generally allows to determine the states of the sub-entities, but that the converse is not true. Indeed, different types of interconnections can give rise to different states of the joint entity, with the sub-entities possibly remaining in the same states (for instance, by varying $\alpha$ in (\ref{comp-int-qubit})). 

Coincidence measurements can therefore be understood as processes during which such interconnections between the entangled sub-entities are severed, bringing them in a situation of true separation. The breaking of the connection between the two sub-entities, like the breaking of an elastic band, is a process that creates correlations that were only potentially present prior to the measurement. Therefore, the extended Bloch formalism strongly suggests that the violation of Bell's inequalities is due to the presence of so-called \emph{correlations of the second kind}~\cite{Aerts1991, Sassoli2013a}, i.e., correlations that were not already present before the measurement (these are the \emph{correlations of the first kind}), but are created by the measurement itself. 

We have seen that for spin measurements on a singlet state, a macroscopic classical laboratory situation, using two spheres and a rigid rod, can perfectly reproduce the quantum correlations, and the corresponding violation of Bell's inequalities. This is due to the specificity of the singlet state (which is rotationally invariant) and the fact that for spin-${1\over 2}$ entities the Bloch sphere is three-dimensional, and therefore can be fully represented in our Euclidean theater. But in the general situation, a mechanistic laboratory situation using macroscopic interconnected entities will not be able to simulate all the correlations, in the different possible measurements, as quantum entanglement is a \emph{non-spatial} form of interconnection, i.e., a connection not through our Euclidean space, but through a space of higher dimensionality, comparable to that of the Bloch sphere of the joint system. 
 
Finally, as emphasized in the last section of the article, and in its very title, the extended Bloch formalism allowed us to propose a solution to a paradox, that of the interpretation of entangled states, which according to the standard quantum formalism would be self-contradictory states, formed by non-existing sub-entities which nevertheless would possess properties. The solution of the paradox results from the simple observation that any state of a joint entity can be written in the direct sum form (\ref{trip-sep-int}), which specifies at any moment what are the states of the sub-entities, and what part of the total state describes their potential correlations (i.e., their non-spatial interconnection). This means that entangled joint entities exist because the sub-entities which form them also exist, as they are always in a well-defined pure state, generally represented by an operator-state, in accordance with the completed quantum mechanics principle 1.


\begin{thebibliography}{}

\bibitem{Poincare1892} H. Poincar\'e, \emph{The\'eorie Mathematique de la Lumi\`ere}, Gauthiers-Villars, Paris, Vol. 2 (1892).
\bibitem{Bloch1946} F. Bloch, ``Nuclear induction,'' Phys. Rev. 70, 460--474 (1946).
\bibitem{Aerts1986} D. Aerts, ``A possible explanation for the probabilities of quantum mechanics,'' Journal of Mathematical Physics 27, 202--210 (1986).
\bibitem{Aerts1987} D. Aerts, ``The origin of the non-classical character of the quantum probability model.'' In: Information, Complexity, and Control in Quantum Physics, eds. A. Blanquiere et al, Springer-Verlag, Berlin (1987).
\bibitem{AertsSassoli2014c} D. Aerts and M. Sassoli de Bianchi, ``The extended Bloch representation of quantum mechanics and the hidden-measurement solution to the measurement problem,'' Annals of Physics 351,  975--1025 (2014).
\bibitem{Aerts1998b} D. Aerts, ``The entity and modern physics: the creation-discovery view of reality.'' In: \emph{Interpreting Bodies: Classical and Quantum Objects in Modern Physics}, ed. Castellani, E. Princeton Unversity Press, Princeton (1998).
\bibitem{Aerts1999} D. Aerts, ``The stuff the world is made of: Physics and reality,'' pp. 129--183. In: \emph{The White Book of `Einstein Meets Magritte'}, Edited by Diederik Aerts, Jan Broekaert and Ernest Mathijs, Kluwer Academic Publishers, Dordrecht, 274 pp. (1999).
\bibitem{Gleason1957} A. M. Gleason, ``Measures on the closed subspaces of a Hilbert space,'' J. Math. Mech. 6, 885--893 (1957). 
\bibitem{Kochen1967} S. Kochen and E. P. Specker, ``The problem of hidden variables in quantum mechanics,'' J. Math. Mech. 17, 59--87 (1967).
\bibitem{Aertsetal1997} D. Aerts, B. Coecke, B. D. Hooghe and F. Valckenborgh, ``A mechanistic macroscopic physical entity with a three-dimensional Hilbert space description,'' Helv. Phys. Acta 70, 793 (1997).
\bibitem{Coecke1995aa} B. Coecke, ``Hidden measurement representation for quantum entities described by finite dimensional complex Hilbert spaces,'' Found. Phys., 25, 1185 (1995).
\bibitem{Coecke1995bb} B. Coecke, ``Generalization of the proof on the existence of hidden measurements to experiments with an infinite set of outcomes,'' Found. Phys. Lett., 8, 437 (1995).
\bibitem{Arvind1997} Arvind, K. S. Mallesh and N. Mukunda, ``A generalized Pancharatnam geometric phase formula for three-level quantum systems,'' J. Phys. A 30, 2417 (1997)
\bibitem{Kimura2003} G. Kimura``The Bloch vector for $N$-level systems,'' Phys. Lett. A 314, 339 (2003).
\bibitem{Byrd2003} M. S. Byrd and N. Khaneja, ``Characterization of the positivity of the density matrix in terms of the coherence vector representation,'' Phys. Rev. A 68, 062322 (2003).
\bibitem{Kimura2005} G. Kimura and A. Kossakowski, ``The Bloch-vector space for N-level systems -- the spherical-coordinate point of view,'' Open Sys. Information Dyn. 12, 207 (2005).
\bibitem{Bengtsson2006} I. Bengtsson and K. \.{Z}yczkowski, \emph{Geometry of  Quantum States: An Introduction to Quantum Entanglement}, Cambridge University Press, Cambridge (2006).
\bibitem{Bengtsson2013} I. Bengtsson and K. \.{Z}yczkowski, ``Geometry of the set of mixed quantum states: An apophatic approach,'' pp. 175--197. In: \emph{Geometric Methods in Physics, XXX Workshop 2011, Trends in Mathematics}, Springer (2013).
\bibitem{AertsSassoli2014d} D. Aerts and M. Sassoli de Bianchi, ``Many-measurements or many-worlds? A dialogue,'' Foundations of Science (2014). DOI: 10.1007/s10699-014-9382-y.
\bibitem{AertsSassoli2014e} D. Aerts and M. Sassoli de Bianchi, ``Do spins have directions?'' arXiv:1501.00693 [quant-ph] (2014).
\bibitem{Hughston1993} L. P. Hughston, R. Jozsa and William K. Wootters,``A complete classification of quantum ensembles having a given density matrix,'' Physics Letters A 183, 14--18 (1993).
\bibitem{Hioe1981} F. T. Hioe, J. H. Eberly, ``$N$-level coherence vector and higher conservation laws in quantum optics and quantum mechanics,'' Phys. Rev. Lett. 47, 838--841 (1981).
\bibitem{Alicki1987} R. Alicki, K. Lendi, \emph{Quantum Dynamical Semigroups and Application}, Lecture Notes in Physics Vol. 286, Springer-Verlag, Berlin (1987).
\bibitem{Mahler1995} G. Mahler, V. A. Weberruss, \emph{Quantum Networks}, Springer, Berlin (1995).
\bibitem{Cirel1980} B. S. Cirel'son, ``Quantum generalizations of BI,'' Lett. Math. Phys., 4, 93 (1980).
\bibitem{OllivierZurek2001} H. Ollivier and W. H. Zurek, ``Quantum discord: A measure of the quantumness of correlations,'' Phys. Rev. Lett. 88, 017901 (2001).
\bibitem{HendersonVedral2001} L. Henderson and V. Vedral, ``Classical, quantum and total correlations,'' J. Phys. A 34, 6899 (2001).
\bibitem{Aerts1991} D. Aerts, ``A mechanistic classical laboratory situation violating the Bell inequalities with $2\sqrt{2}$, exactly `in the same way' as its violations by the EPR experiments,'' Helv. Phys. Acta 64, 1--23 (1991).
\bibitem{Aerts1984} D. Aerts, ``The missing element of reality in the description of quantum mechanics of the EPR paradox situation,'' Helv. Phys. Acta, 57, 421--428 (1984).
\bibitem{AertsBroekaert2000} D. Aerts, S. Aerts, J. Broekaert and  L. Gabora, ``The violation of Bell inequalities in the macroworld,'' Found. Phys., 30, pp. 1387-1414 (2000).
\bibitem{Sassoli2013a} M. Sassoli de Bianchi, ``Using simple elastic bands to explain quantum mechanics: a conceptual review of two of Aerts' machine-models,'' Centr. Eur. J. Phys. 11, 147--161 (2013).
\bibitem{Massimiliano2013b} M. Sassoli de Bianchi, ``Quantum dice,'' Annals of Physics 336, 56--75 (2013).
\bibitem{Massimiliano2014} M. Sassoli de Bianchi, ``A remark on the role of indeterminism and non-locality in the violation of Bell's inequality,'' Annals of Physics 342, 133--142 (2014).
\bibitem{Aerts2000}  D. Aerts, ``The description of joint quantum entities and the formulation of a paradox," Int. J. Theor. Phys. 39, 485--496 (200).
\bibitem{Schroedinger1935} E. Schr{\oe}dinger, Naturwissenschaftern 23, 807 (1935). English translation: John D. Trimmer, Proceedings of the American Philosophical Society, 124, 323 (1980). Reprinted in: J. A. Wheeler and W. H. Zurek (Eds.), \emph{Quantum Theory and Measurement} (Princeton University Press, Princeton, 1983) 152.
\bibitem{Fraassen1991} B. C. Van Fraassen,  \emph{Quantum Mechanics: An Empiricist View}, Oxford University Press, Oxford, New York, Toronto (1991).
\bibitem{Aerts1982} D. Aerts, ``Description of many physical entities without the paradoxes encountered in quantum mechanics,'' Found. Phys. 12, 1131--1170 (1982).
\bibitem{Coecke1995a} B. Coecke,  ``Representation for pure and mixed states of quantum physics in Euclidean space,'' Int. J. Theor. Phys. 34, 1165 (1995).
\bibitem{Coecke1995b} B. Coecke, ``Representation of a spin-1 entity as a joint system of two spin-1/2 entities on which we introduce correlations of the second kind,'' Helv. Phys. Acta 68, 396 (1995).
\bibitem{Coecke1996} Coecke, B. ``Superposition states through correlations of the second kind,'' Int. J. Theor. Phys. 35, 1217 (1996).


\end{thebibliography}
\end{document}